\documentclass{JFM-FLM_Au}

\lefttitle{M. V. Chubynsky et al.}
\righttitle{Dynamics of fluid-fluid displacements: A spectral approach}

\title{Dynamics of fluid-fluid displacements in a model rough fracture beyond the quasistatic limit: A spectral approach}

\author{Mykyta V. Chubynsky\aff{1}, {Jordi Ort{\'\i}n}\aff{2}, Marco Dentz\aff{1} \and Ran Holtzman\aff{1}}

\affiliation{\aff{1}Institute of Environmental Assessment and Water Research (IDAEA), Spanish National Research Council (CSIC), Barcelona, Spain
\aff{2}
Departament de F\'{\i}sica de la Mat{\`e}ria Condensada, 
Universitat de Barcelona, 
Barcelona, Spain, 
and
University of Barcelona Institute of Complex Systems (UBICS), Barcelona, Spain
}

\corresau{Mykyta Chubynsky, chubynsky@gmail.com; Ran Holtzman, ran.holtzman@csic.es}

\begin{document}

\maketitle

\begin{abstract}
In fluid-fluid displacements in porous and fractured materials, viscous friction in microscale interfacial (Haines) jumps is intimately linked to macroscale energy dissipation and the associated pressure-saturation (retention) hysteresis. The mode of control (flow rate vs. pressure) and the driving rate (from quasistatic to finite) can substantially affect hysteresis. Despite the significance of hysteresis and dissipation in various technological and natural processes, a quantitative understanding of the link between the micro- and macro-scales, and of the impact of the inherent heterogeneity of porous media, remains elusive. An ``imperfect'' Hele-Shaw cell of variable aperture is a simple model system which allows to study all these  in details. However, simulating fluid-fluid interface evolution in heterogeneous media, even in such a simple system, is computationally prohibitive, as multiple length and time scales need to be resolved simultaneously. We develop here a spectral computational approach for interface evolution and energy dissipation and validate it via comparison to computational fluid dynamics simulations and experiments. Computational efficiency is demonstrated by following interface evolution in a cell with a single ``defect'', as well as with random roughness; in both, disparate length scales lead to nontrivial dynamics over many orders of magnitude in time. We also show theoretically that, in the quasi-static limit, while viscous forces during Haines jumps fully account for the energy dissipated between consecutive metastable equilibria, viscosity does not change the total dissipated \emph{amount}, merely the relaxation \emph{time}.
Our approach provides a stepping stone towards upscaling of fluid-fluid flows in porous media.
\end{abstract}


\section{Introduction}
The flow of multiple fluid phases in porous materials is an important topic in a multitude of natural and engineered applications. One inherent intriguing aspect is its history-dependent nature, typically described in terms of pressure--saturation hysteresis and the corresponding energy dissipation \citep{Albers2014,Cueto2016,HoltzmanCommPhys2020,Holtzman2023}. 
Hysteresis and dissipation substantially affect the rate and pathways by which one fluid displaces another,
which is crucial for planning and controlling processes such as soil irrigation and evaporation, carbon dioxide sequestration, and underground hydrogen storage \citep{Sahimi2011}. 

In idealised systems like a Hele-Shaw (HS) cell of uniform aperture or a cylindrical capillary of uniform cross section, the interfacial dynamics can be obtained analytically for pressure-driven flows \citep{CLOTET_JCIS2012,Fries2008}. 
In such homogeneous media, hysteresis {(history-dependence of the interface position given the inlet pressure)} and dissipation do not emerge in the \emph{quasistatic} driving limit---when the {fluid is} driven at zero
rate (an idealised limit that cannot be fully exercised in practice). 
That is, history-dependence and dissipation are {present} only at finite driving rates (rate of changing inlet pressure).
In contrast, displacements in heterogenous porous systems display hysteresis and dissipation even in the {quasistatic} driving limit,
because the heterogeneity (in conduit geometry, e.g. pore sizes) gives rise to multiple equilibrium interfacial configurations, and energy is dissipated in out-of-equilibrium interfacial jumps between consecutive metastable equilibria. These so-called Haines jumps are triggered whenever the driving pressure pushes the two-fluid configuration beyond its metastability limit \citep{Haines1930}.

Phenomenological models of hysteresis, based on the collective behaviour of independent or interacting hysteretic units (hysterons), rely on the link between interfacial jumps and energy losses \citep{Bertotti_V1,Cueto2016,Helland_WRR2021}. However, it is not clear what hysterons represent at microscopic scales, and thus how to establish a connection between model parameters and actual physical and geometrical quantities and obtain quantitative predictions without fitting of parameter values.
Progress has been made using mildly heterogeneous confined systems that provide a physical understanding of the interfacial dynamics \citep{deGennes1986,Rubio1989,Glass1998,Soriano2002,Geromichalos2002,Glass2003,Meakin2009,Planet2009,Santucci2011,CLOTET_JCIS2012,Clotet2014}. 
The convoluted geometry of the fluid-fluid interfaces in three-dimensions (3D) makes quasi-two-dimensional (2D) systems an attractive model system. 
A prominent example of such media is the 2D array of ``pillars'' (cylindrical objects of various radii), which can be easily manufactured and studied experimentally by standard microfluidics techniques \citep{Ferrari2015,BORGMAN2019}.

Recently, in particular, the ``imperfect Hele-Shaw cell'' (IHSC; Fig. \ref{fig:fig1_schematic}) has allowed a thorough investigation of the pore-scale processes underlying Haines jumps and hysteresis \citep{Planet2020,HoltzmanCommPhys2020,Holtzman2023,Lavi_PRF_2023,Holtzman2024}.
The IHSC is an idealised model for an open fracture of random aperture, where fluids move slowly enough (in the Stokes regime) for flows to be quantified via Darcy's law, and fluid-fluid interfaces are distorted by capillary forces arising from aperture variations. 
IHSCs prove to be highly useful in understanding instabilities in quasi 2D systems, and a promising stepping stone towards understanding 3D flows in disordered media [e.g. \citet{Orozco2026TwoPhase}].
%
%
Using IHSCs composed of single, pair, and multiple randomly-distributed mesa-shaped heterogeneities (``defects''), we have studied both experimentally and numerically individual and collective capillary jumps, and related them to their macroscopic manifestation as pressure-saturation hysteresis and energy dissipation \citep{Planet2020,HoltzmanCommPhys2020,Holtzman2023,Lavi_PRF_2023,Holtzman2024}; 
these works, however, are limited to quasistatic driving [except \citet{Lavi_PRF_2023} which treats a single constriction].

In practice processes are never truly quasistatic, as both driving (changing pressure or fluid volume) and the consequent relaxation (from these perturbations) occur over a finite time scale, and, in particular, Haines jumps are dynamic by nature [even if the driving were truly quasi-static, the interface jumps would occur at high speeds \citep{Moebius2012,Berg2013}]. 
Dynamics of individual Haines jumps in quasistatic driving can provide insights into the various time scales, from which one can deduce when the quasistatic approximation is appropriate \citep{Maloy1992,Moura_EPL_2017,Reis_JFM_2026}.
Furthermore, finite-rate displacements go through different paths (in terms of interfacial configurations) than quasistatic displacements, as the system progressively departs from mechanical equilibrium as rates increase. 
At finite driving rates, viscous dissipation occurs both in the intermittent jumps triggered by the medium heterogeneities and in the continuous displacements of the fluid phases \citep{Moebius2012,Berg2013}. 
The question arises as to whether these two contributions — stemming from viscous friction in perfect cells and from Haines jumps in disordered cells — can simply be added, e.g. to quantify energy dissipation.

The driving mode of displacement also plays a significant role. 
In flow-rate-controlled displacements the global mass per unit time is conserved, while it is not in pressure-controlled displacements. 
The corresponding displacement paths can thus differ significantly. 
For example, Haines jumps are suppressed and pressure-saturation hysteresis disappears in volume-controlled quasistatic displacements in a vertical capillary tube of varying cross section \citep{Nepal2026}. 
This is a consequence of the fact that the interface between the two fluids has only one degree of freedom, the height $h(t)$.
While in IHSC the interface $h(x,t)$ presents additional degrees of freedom (many different configurations for a given invaded area), differences between the two driving modes may still be expected. How controlling the volume instead of the pressure impacts hysteresis is an interesting open question [noting earlier works examining related aspects, e.g. burst dynamics in monolayer of beads experiments \citep{Moura_EPL_2017}, and pressure-saturation relationship from a phenomenological compartment model \citep{Cueto2016}].

Simulating the dynamics of the displacements at finite rates in large disordered media is computationally prohibitive.
Computational fluid dynamics models (CFD; also called ``direct methods'') discretising the Navier-Stokes equations are capable of resolving two-phase flows at finite rates in complex pore geometries, however, at a computational cost that limits simulations to very small domains,  insufficient to describe complex heterogeneous pore geometry or generate the large number of realizations required for upscaling \citep{ramstad2019pore,Zhao2019,Giudici2023GNMvsDNS}. 
The computational cost of CFD remains prohibitive even when considering 2D systems, or averaging over the third dimension \citep{Krishna_JFM_2025}. 
Pore network models require much less computational resources, but they use a simplified pore geometry that cannot resolve the actual interface configuration \citep{Hughes2001,Ferer2011,Giudici2023GNMvsDNS}.

The main purpose of the present work is to provide a computational tool that makes it possible to predict and analyse two-phase displacements in IHSCs, at high accuracy and computational efficiency, for both pressure and flow-rate driving. 
The proposed approach is applicable to all scenarios discussed thus far. Here, we benchmark its performance against interfacial behaviours observed in IHSCs of diverse geometries and under varying conditions.


The paper is organised as follows. In Sec.\ \ref{sec:HS}, we derive the fundamental equations for the IHSC, and use the Hele-Shaw formulation to explicitly show that viscous (frictional) forces during Haines jumps fully account for the energy dissipated between consecutive metastable equilibria in quasistatic displacements. Our analysis also provides a non-intuitive result: while the energy dissipation is viscous in origin, the total dissipation in a Haines jump is viscosity-independent.
Sec. \ref{sec:spectral} presents a novel spectral approach, formulated based on the hydrodynamic equations for the bulk flow in the IHSC and kinematic and dynamic boundary conditions for the interface dynamics. The approach tracks the dynamics of the Fourier modes into which the interface shape is decomposed, providing both high accuracy and high computational efficiency. 
Sec.\ \ref{sec:validation} validates the model by comparing it for the simple case of a single (isolated) ``mesa" defect against experiments and high-fidelity CFD simulations.
Next, we consider two examples illustrating the promising power of the spectral approach. First, in Sec.~\ref{sec:relaxation}, we study in more detail the single defect case, where, despite its apparent simplicity, our numerical and analytical studies with the spectral approach reveal complex, multistage behaviour of a Haines jump that would be extremely challenging to study using traditional approaches.
Then, in Sec.\ \ref{sec:disorder} we simulate a large disordered IHSC with random roughness (proxy of a rough fracture). 
Finally, the conclusions are drawn in Sec.\ \ref{sec:conclusions}.


\section{Flow and energy dissipation in an imperfect Hele-Shaw cell} 
\label{sec:HS}



\begin{figure}
    \centering    \includegraphics[width=0.5\linewidth]{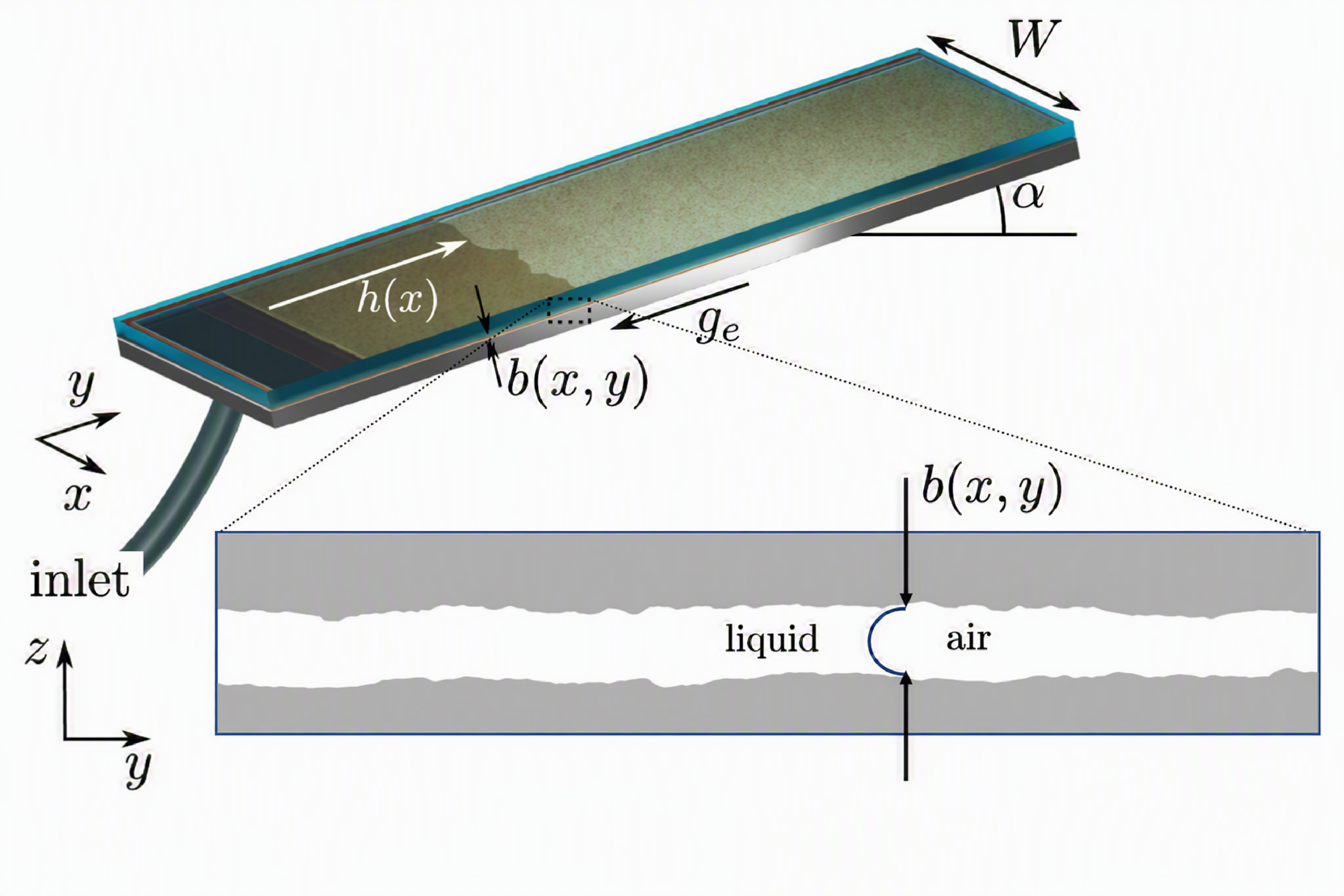}
\caption{
Schematic illustration of the model system: An imperfect Hele-Shaw cell with randomly varying aperture $b(x,y) = b_0 \pm \delta b(x,y)$. 
The cell is tilted to provide an effective gravity $g_e=g\sin\alpha$.
}
    \label{fig:fig1_schematic}
\end{figure}

We consider an IHSC in which one inner surface is flat, and the other is rough such that the aperture (inter-surface distance) $b(x,y)$ varies in space. Here $x$ is horizontal and $y$ is directed along the surface upwards (Fig. \ref{fig:fig1_schematic}). 
The cell is filled with two fluids, one fully wetting with viscosity $\mu$ and density $\rho$ (which we will refer to as `liquid') and the other nonwetting with comparatively negligible density and viscosity (`air'). The interface between the two fluids is assumed to be given by a single-valued function of $x$, $h(x,t)$, such that liquid fills the space $y<h(x,t)$, namely disallowing interface disconnections and trapping. 

We first recall the basic equations describing the above system and introduce two types of boundary conditions corresponding to pressure and flow-rate control (Sec.~\ref{subsec:HS-equations}). As considered previously, when the controlling inlet pressure is constant or the flow rate is zero, it is possible to write down an equation for the shape of the interface in equilibrium; in the quasistatic limit, the interface may follow the equilibrium shape that now changes slowly, except during Haines jumps (Sec.~\ref{subsec:equil}). Starting from an equation for the energy dissipation rate, we then show that the total viscous dissipation amount during a Haines jump in the quasistatic limit is viscosity-independent (Sec.~\ref{subsec:viscosity-independence}). A simplifying assumption, important, in particular, for the derivation of the spectral approach, is made in Sec.~\ref{subsec:b_0}. This assumption is used to derive an equation for the ``instantaneous equilibrium'' interface shape, generalising the concept of equilibrium to nonzero flow rates (Sec.~\ref{subsec:instant}). The concepts of true and instantaneous equilibria are important in the derivation of the spectral approach in Sec.~\ref{sec:spectral}, as perturbations of the interface shape with respect to an equilibrium are considered. An expression for the dissipation rate involving only the evolution of the interface shape is derived in Sec.~\ref{subsec:dissipation_rate}; this is used later with the spectral approach (Sec.~\ref{subsec:dissip_spectral}) and also shows explicitly that the dissipation amount matches the work of external forces. Finally, an expression for the amount of dissipation in a Haines jump is derived and compared to previous work in Sec.~\ref{subsec:qe-p-driving}.

\subsection{Governing equations}\label{subsec:HS-equations}
We simplify the problem to two dimensions (2D) --- the $x-y$ plane --- by neglecting velocity gradients along the cell compared to those in the transverse direction, as well as transverse pressure gradients {compared to longitudinal ones} \citep{Homsy1987}. 
%
%
Assuming Stokes flow, we get the following expression for the 2D flow velocity (averaged in the $z$ direction) in the liquid phase (Darcy's law):
\begin{equation}
    \mathbf{u}(x,y)=-\frac{b^2(x,y)}{12\mu}(\nabla p(x,y)+\rho g_e \hat{y}).\label{eq:u}
\end{equation}
Here $p(x,y)$ is the pressure in the liquid phase, $g_e=g\sin\alpha$ is the effective gravity ($\alpha$ is the tilt angle), and $\hat{y}$ is the unit vector along the $y$ direction. 
The continuity equation is
\begin{equation}
    \nabla\cdot[b(x,y)\mathbf{u}(x,y)]=0.\label{eq:cont}
\end{equation}
These two equations are valid for $y<h(x,t)$. At the interface [$y=h(x,t)$], the dynamic boundary condition is Young-Laplace law, which states that
\begin{equation}
    p_a - p =\gamma\left(\frac{\pi}{4}\kappa(x,t)+\frac{2}{b(x,h(x,t))}\right),\label{eq:pBC}
\end{equation}
where $\gamma$ is the surface tension, 
$\kappa$ is the in-plane curvature of the interface, and the expression in the parentheses is an approximation for the sum of the principal curvatures of the interface assuming complete wetting~\citep{Park/Homsy:1984}. 
From here on, we take the non-wetting phase pressure $p_a$ to be atmospheric, and set $p_a=0$.
Hereafter, for brevity we do not explicitly write the functional dependence of these variables on $x$, $y$, and $t$, except where it is essential.

The interface moves according to the kinematic boundary condition,
\begin{equation}
    \frac{\partial h}{\partial t}=u_y-u_x\frac{\partial h}{\partial x}.\label{eq:kinem}
\end{equation}
The boundaries at the sides, $x=0$ and $x=W$, are impermeable ($u_x=0$), enforced via $\partial p/\partial x=0$. 

We will be considering throughout this paper two types of boundary conditions for the inlet (bottom boundary, $y=0$): (i) pressure-controlled, setting the pressure at the inlet, $p(y=0)=P$; and (ii) flow-rate-controlled, setting the inflow speed, $u_y(y=0)=V$ {(note that we specify $u_y$ \emph{everywhere} along the inlet, as specifying only the \emph{total} flow rate would leave the problem underdetermined)}. The latter {boundary condition} also includes the case of a closed inlet ($V=0$).
%
%

The above relies on a number of assumptions, conditions and approximations:  
\begin{itemize}

    \item {We assume a single interface with a shape that is a single-valued function of $x$. While this limitation is not inherent in the approach, it simplifies the formulation; it is also necessary for the spectral approach we develop in Sec.~\ref{sec:spectral} in any case.}
    
    \item Further, we assume negligible interface deformation by gravity, valid when $\ell_c\gg b$, where
    \begin{equation}
        \ell_c=\left(\frac{\pi\gamma}{4\rho g_e}\right)^{1/2}
    \end{equation}
    is the capillary length \citep{deGennes2004}.
    This allows us to use the expression for the interface curvature in Eq.~(\ref{eq:pBC}).
    
    \item The tangential speed on the {side walls at $x=0$ and $x=W$} cannot be specified in this formulation, meaning that there is slip at the side walls (which, however, only affects flow up to distances of the order of $b$).
    
    \item The wetting properties of the side walls are not an input.
    The wall wettability affects the meniscus only up to a distance of $\sim\ell_c$ from the walls \citep{deGennes2004}, which is typically small ($\ell_c\ll W$). 
    This is further considered in Sec.~\ref{sec:spec_evol}.
    
    \item Complete wetting of the top and bottom surfaces of the cell is assumed [while noting that generalisation to partial wetting with a fixed contact angle is straightforward, we do not consider dynamic contact angle variations, including hysteresis of the dynamic contact angle between imbibition and drainage displacement directions \citep{PahlavanPRL2015}].
    
    \item Inertial effects, which may be important at very early times of fast Haines jumps \citep{Fries2008}, are not included. 
    
    

    \item Equation~(\ref{eq:u}) assumes that in the Laplacian $(\partial^2/\partial x^2+\partial^2/\partial y^2+\partial^2/\partial z^2)\mathbf{u}$ the last term dominates.
    
    \item Finally, 
    we do not account for 3D wetting effects, including thin residual coating layers of the wetting liquid along the solid walls left in the drained region during imbibition, which modify the kinematic and dynamic boundary conditions to account for volume conservation and enhanced curvature and viscous stresses at the interface \citep{Park/Homsy:1984,Reinelt1985,Morrow2023}, as well as instabilities caused by entrainment of the less-wetting fluid \citep{Levache2014,Odier2017} and corner films \citep{Chauvet2009} which may appear along mesa defects and possibly other locations where there are abrupt changes of aperture.

\end{itemize}
Despite these assumptions, the approach applies to a wide range of conditions of interest, confirmed, for example, by the good agreement with experiments discussed in Sec.~\ref{sec:imbdra}. 

\subsection{Equilibria and quasistatic driving}
\label{subsec:equil}

Before introducing dynamics, we recall the equation for the equilibrium shape of the interface and the concept of quasistatic driving.

For the \textit{\textbf{pressure control}} case with $P={\rm const}$, we introduce a quantity measuring the imbalance between the pressure at the interface [Eq.~(\ref{eq:pBC})], the pressure at the inlet, $P$, and the gravity term, $\rho g_e h$,
\begin{equation}
    p_e(x)=\rho g_e h(x)-\gamma\left(\frac{\pi}{4}\kappa(\{h(x)\})+\frac{2}{b(x,h(x))}\right)-P.
    \label{eq:pe}
\end{equation}
The system is in equilibrium if there is a balance for all $x$, i.e.,
\begin{equation}
    p_e(x)\equiv 0.\label{eq:equilP}
\end{equation}
Indeed, the solution of Eqs.~(\ref{eq:u})--(\ref{eq:kinem}) in the liquid phase then gives
\begin{eqnarray}
    p &=&P-\rho g_e y,\\
    \mathbf{u}&=&0,\\
    \partial h/\partial t&=&0,
\end{eqnarray}
and the interface does not move. Depending on the details of $b(x,y)$, the equilibrium condition (\ref{eq:equilP}) may have multiple solutions $h_{\rm eq}(x)$.
If Eq.~(\ref{eq:equilP}) is not satisfied, it implies pressure imbalance that produces flow and interface motion.

In the quasistatic limit, changes in the inlet pressure $P$ in time are applied slowly such that the system remains in equilibrium [Eq.~(\ref{eq:equilP}) is satisfied] at all times.
This limit was considered in previous publications~\citep{HoltzmanCommPhys2020,Holtzman2023}.
{Note that the factor $\pi/4$ in front of the curvature term in Eq.~(\ref{eq:pe}), correcting for 3D curvature \citep{Park/Homsy:1984}, was not included in the 2D model in \citet{HoltzmanCommPhys2020,Holtzman2023} (except where quantitative agreement with experimental data was sought).} 
The existence of multiple solutions may give rise to hysteresis (history dependence) of the interface evolution with $P$, even in this quasistatic limit. At a particular $P$ a solution branch may terminate [$h_{\rm eq}(x)$ may become non-real], with the interface necessarily undergoing a Haines jump to a different equilibrium solution, on a time scale
so short that the change of $P$ during the jump can be neglected.

For \textit{\textbf{flow-rate control}} with $V=0$ (a \textit{\textbf{closed inlet}}), the equilibrium condition is still given by Eq.~(\ref{eq:equilP}), however, $P$ is no longer a known constant but an unknown determined by the volume of the liquid (conserved in this case).

%
%
\subsection{Energy dissipation: Rate, amount, and the role of viscosity}
\label{subsec:viscosity-independence}

Out of equilibrium, the moving liquid dissipates energy. The dissipation rate $R$ is equal to the work per unit time of viscous forces, which in the overdamped limit is equal {in magnitude and opposite in sign} to 
the work per unit time of all other forces. 
Using Eq.~(\ref{eq:u}), we get
\begin{equation}
    |R|=-\int b(\mathbf{u}\cdot[\nabla p+g_e\rho\hat{y}])\; dA
    =12\mu\int\frac{u^2}{b} \; dA,
    \label{eq:Rgen}
\end{equation}
where
integration is over the area of the cell filled with liquid. 
{(There are different conventions for the sign of energy dissipation; here we use $| \; |$ to remove the ambiguity.)}
%
To use this expression, the full velocity field $\mathbf{u}(x,y)$ must be known (e.g.\ from numerical simulations of the Hele-Shaw equations).
{Equation~(\ref{eq:Rgen}) can also be obtained directly as the work per unit time of viscous forces, using the parabolic (Poiseuille) approximation of the transverse velocity profile, from which Eq.~(\ref{eq:u}) is derived (see Appendix \ref{sec:dissip_direct}).}


In the quasistatic limit, outside of Haines jumps---as the liquid velocity tends to zero---the dissipation rate $|R|$ approaches zero as well. 
Moreover, considering a process in which the interface moves between two given equilibrium configurations without Haines jumps, in the quasistatic limit the \textit{total amount} of dissipation tends to zero as well. 
This is because, even though in this limit the duration of the process (inversely proportional to the interface speed) tends to infinity, it does so more slowly than the dissipation rate (proportional to the square of the interface speed) tends to zero. Thus, in this limit one only needs to consider dissipation during Haines jumps.

Notably, even though energy dissipation in the system we consider is viscous in nature, 
the total amount of dissipation in a Haines jump under quasistatic driving
does not depend on the liquid viscosity $\mu$. 
This is true in both the pressure-controlled case where $P$ can be considered constant during the jump, and the flow-rate-controlled case with $V\to 0$.
We demonstrate this rather nonintuitive statement by considering systems that only differ in liquid viscosity, all starting with the same initial interface profile $h(x)$. 
The interface will move through the same configurations, 
but on different (viscosity-dependent) time scales, i.e.,
\begin{equation}
    h(x,t;\mu)=h(x,t/[\mu/\mu_0];\mu_0),
\end{equation}
where $\mu_0$ is an arbitrary reference viscosity.
This means that it is possible to consider a process where such systems not only start with the same interface profile, but also end with the same interface profile though at different times ($\propto\mu$).
Further, the velocity fields at the corresponding times in the evolution process are likewise the same, only being scaled by the viscosity, i.e.,
\begin{equation}
    \mathbf{u}(x,y,t;\mu)=(\mu/\mu_0)^{-1}\mathbf{u}(x,y,t/[\mu/\mu_0];\mu_0).
\end{equation}
%
%
Therefore, the total dissipation in this process is the same in all systems. Indeed, Eq.~(\ref{eq:Rgen}) gives for the dissipation rate
\begin{eqnarray}
|R(t;\mu)|&=&12\mu\int_{x=0}^W\int_{y=0}^{h(x,t/[\mu/\mu_0];\mu_0)}\frac{\left[(\mu/\mu_0)^{-1}u(x,y,t/[\mu/\mu_0];\mu_0)\right]^2}{b(x,y)} \; dx \, dy\nonumber\\
&=&\frac{\mu_0}{\mu}|R(t/[\mu/\mu_0];\mu_0)|,
\end{eqnarray}
and the total dissipation is
\begin{eqnarray}
|\Psi(\mu)|&=&\int_{0}^{(\mu/\mu_0)t_0}|R(t;\mu)| \; dt\nonumber\\
   &=&\frac{\mu_0}{\mu}\int_{0}^{(\mu/\mu_0)t_0}|R(t/[\mu/\mu_0];\mu_0)|(\mu/\mu_0)\; d(t/[\mu/\mu_0])\nonumber\\
   &=&|\Psi(\mu_0)|.
\end{eqnarray}
Here $t_0$ is defined as the duration of the process for {$\mu=\mu_0$}. Thus, the total dissipation amount is viscosity-independent. 
A particular case of such a process is a Haines jump during quasistatic driving, where $t_0 \rightarrow \infty$ because the relaxation to the static equilibrium takes infinite time.
Intuitively, a more viscous liquid dissipates less per unit time (while the viscosity is higher, the flow speed has a stronger effect as the dissipation rate is quadratic in the speed), yet the process takes longer by the same factor.

The viscosity independence of the total dissipation is natural, because in the Stokes limit where Eq.\ (\ref{eq:u}) was derived 
the contribution of the kinetic energy to the energy balance is negligible, 
and the total dissipation simply equals the work of all forces other than viscous friction. This work only depends on the initial and final interface shapes.
This is discussed in more detail in Secs.~\ref{subsec:dissipation_rate} ({and the associated} Appendix \ref{sec:demo_energy_diss_visc}) and \ref{subsec:qe-p-driving} below. {Moreover, as long as the liquid in the initial and final states is at rest, this} makes the result valid in presence of inertia and dynamic wetting effects, which are neglected in the above consideration.
%

\subsection{Simplifying assumption: Computing permeability from mean aperture ($b \approx b_0$)}
\label{subsec:b_0}



A simplifying assumption made throughout this paper (unless stated otherwise), which is essential for our derivation of the spectral approach, is that 
the variations of the cell aperture $b(x,y)$ are small compared to the mean aperture $b_0$, such that we neglect their effects on the permeability [Eq.~(\ref{eq:u})] and continuity 
[Eq.~(\ref{eq:cont})].
That is, we approximate Eqs.~(\ref{eq:u}) and (\ref{eq:cont}) by replacing $b$ with $b_0$:
\begin{eqnarray}
    \mathbf{u}&\approx& -\frac{b_0^2}{12\mu}(\nabla p+g_e\rho\hat{y}),
    \label{eq:uappr}\\
    \nabla\cdot\mathbf{u}&\approx&0.
    \label{eq:contappr}
\end{eqnarray} 
We stress that the effect of the non-uniformity of $b(x,y)$ on capillary pressure is explicitly included.
This may appear inconsistent, since all three contributions due to variations of $b$ are of the same order (linear) in $\delta b=b-b_0$ in the equations they enter.

To justify this apparent inconsistency, we note that while the capillary pressure directly affects the equilibrium interface shape, the permeability and continuity only influence the \emph{rate} at which this equilibrium is approached.
%
%
{A simple analogue with one degree of freedom is 
    ${ds}/{dt}=-s+\epsilon +\epsilon {ds}/{dt}$,
where {$t$ and $s$ are dimensionless variables and $\epsilon$ is a small dimensionless parameter}. Formally, both perturbation terms are $O(\epsilon)$. However, the first shifts the {stationary state (i.e., the value of $s$ in the limit $t\to\infty$)} by $O(\epsilon)$, while the second only affects the rate at which the {stationary state} is approached.
For {initial ($t=0$)} perturbations of the {stationary state} of $O(\epsilon)$), $ds/dt$ is $O(\epsilon)$ {at all $t>0$}, and thus the second perturbation term is $O(\epsilon^2)$ with a negligible effect in the first-order approximation.

Finally, note we use $b=b_0$ in Eq.~(\ref{eq:Rgen}) to calculate the dissipation.

\subsection{Instantaneous equilibrium for finite flow rates}
\label{subsec:instant}

The concept of equilibrium can be generalised to the \textit{\textbf{flow-rate control}} case with $V\ne 0$. We define
\begin{equation}
    p_{eV}(x)=\rho g_{eV}h(x)-\gamma\left(\frac{\pi}{4}\kappa(\{h(x)\})+\frac{2}{b(x,h(x))}\right)-\Pi(t),
    \label{eq:peV}
\end{equation}
where $\Pi(t)$ is chosen so that the average over $x$ of $p_{eV}$ is zero, and the modified effective gravity is
\begin{equation}
 g_{eV}=g_e\left(1+\frac{12\mu V}{\rho g_e b_0^2}\right).
 \label{eq:geV}
\end{equation}
While there is, obviously, no true equilibrium in this case (since the problem is not stationary, unlike in Sec.~\ref{subsec:equil}), if
\begin{equation}
p_{eV}(x)=0,
\label{eq:equilFR}
\end{equation}
then in the approximation (\ref{eq:uappr})--(\ref{eq:contappr}) the solution is
\begin{eqnarray}
    p&=&\Pi(t)-\rho g_{eV}y,\\
    \mathbf{u}&=&V\hat{y},\\
    \partial h/\partial t&=&V,
\end{eqnarray}
i.e., the interface simply translates at speed $V$ with a zero instantaneous rate of shape change. This is a kind of ``instantaneous equilibrium'' state [in fact, the interface shape persists indefinitely if the cell aperture is $y$-independent, $b=b(x)$]. 
Equation~(\ref{eq:equilP}) which is valid for $V=0$ is, of course, a particular case of Eqs.~(\ref{eq:peV})--(\ref{eq:equilFR}) with $\Pi(t)=P={\rm const}$ and no approximation needed. Instantaneous equilibria can be considered for any non-stationary problems with an arbitrary time dependence of $P$ or $V$. 
Our spectral approach (Sec.~\ref{sec:spectral}) is based on considering perturbations of either true or instantaneous equilibrium profiles. Note that for $V<-\rho g_e b_0^2/(12\mu) <0$, $g_{eV}$ becomes negative, leading to the Saffman-Taylor instability; this case will not be considered in this paper.

Finally, note that the arguments of Sec.~\ref{subsec:b_0} about the validity of the approximation of Eqs.~(\ref{eq:uappr})--(\ref{eq:contappr}) do not apply to flow-rate control with $V\ne 0$, since the equilibrium condition (\ref{eq:equilFR}) itself relies on this approximation and is no longer exact. 
In earlier studies on scaling laws for interface roughness in a horizontal cell (without gravity), \citet{Paune2002, Paune2003} concluded that the approximation of Eqs.~(\ref{eq:uappr})--(\ref{eq:contappr}) applies to most conditions of practical interest.
The validity of the spectral approach 
is further justified by our tests (Sec.~\ref{sec:imbdra}). 

\subsection{Computing the dissipation rate based on interface dynamics}
\label{subsec:dissipation_rate}

The dissipation rate in Eq.~(\ref{eq:Rgen}) requires knowledge of the flow speed in the bulk. 
It is useful (in particular, for the spectral approach of Sec.~\ref{subsec:dissip_spectral}) to obtain an expression which does not require this {and only uses the knowledge of interface dynamics}. With the simplifying assumptions of Sec.~\ref{subsec:b_0}, Gauss theorem can be used to transform the area integral in the first equality of Eq.~(\ref{eq:Rgen}) into a contour integral along the boundaries of the liquid domain (the inlet, the side walls and the interface), which gives
\begin{equation}
|R(t)|\approx -b_0\int_{x=0}^W p_e(x,t)\frac{\partial h(x,t)}{\partial t}\, dx,
\label{eq:Rfinal}
\end{equation}
where
$p_e$ is given by Eq.~(\ref{eq:pe}) and for flow-rate control $P$ is the inlet pressure $p(x,y=0,t)$ averaged along the inlet. The details of the derivation are given in Appendix~\ref{sec:demo_energy_diss_visc}. Equation~(\ref{eq:Rfinal}) can be interpreted straightforwardly as the power of external forces acting on the liquid volume (gravity, capillary and inlet pressure; see Appendix \ref{sec:demo_energy_diss_visc}).

As expected, for pressure control the dissipation rate vanishes in the quasistatic limit outside Haines jumps ($p_e\to 0$). However, for flow-rate control note that
Eq.~(\ref{eq:Rfinal}) contains $p_e$ [Eq.~(\ref{eq:pe})] and not $p_{eV}$ [Eq.~(\ref{eq:peV})]. 
Dissipation does not vanish for $p_{eV}=0$, as the liquid is still moving, albeit at a uniform speed $V$.

\subsection{Dissipation in Haines jumps for quasistatic pressure driving}
\label{subsec:qe-p-driving}

It is instructive to compare our derivation of the energy dissipated with that in the model in \citet{Holtzman2023} for quasistatic pressure driving.
In \citet{Holtzman2023}, the energy dissipation in a Haines jump (between two static equilibria) was calculated from the knowledge of the initial and final interface configurations alone, not of the details of the dynamics in between. 
The authors subdivided the change in the interface height during the jump $\Delta h(x)$ into elementary steps $\delta h_k(x)$, $\sum_k \delta h_k(x)=\Delta h(x)$, small enough such that during each step $p_e(x,t)$ can be considered constant [equal to $p_e^k(x)$]. The total energy dissipation per unit aperture of the cell during the jump is then \citep{Holtzman2023}:
%
\begin{equation}
    |\psi|=-\sum_k \int_0^W p_e^k(x)\delta h_k(x) \; dx.
    \label{eq:PsiQS2}
\end{equation}
In this derivation, the specific way the subdivision into elementary steps is done does not affect the computed dissipation, as long as the initial and the final interface profiles are correct.
{Note that in Ref.~\citet{Holtzman2023} dissipation per unit aperture is denoted by $\Psi$, which here is used for the \emph{total} dissipation (not per unit aperture)}.


On the other hand, from Eq.~(\ref{eq:Rfinal})
the energy per unit aperture width dissipated in a Haines jump, between two equilibrium configurations, is:
\begin{eqnarray}
    |\psi|=\frac{1}{b_0}\int_{t_i}^{t_f} |R(t)| \; dt&=&-\int_{x=0}^W\int_{t=t_i}^{t_f} p_e(x,t)\frac{\partial h(x,t)}{\partial t} \; dx \, dt\nonumber\\
    &\approx&-\sum_k \int_0^W p_e^k(x)\delta h_k(x) \; dx.
    \label{eq:Psispec}
\end{eqnarray}
Here $t_i$ and $t_f$ are the initial and final times of the jump, respectively ($t_f \rightarrow \infty$, since it takes an infinitely long time to reach the exact equilibrium), and in the last step the time integral is discretised as a sum. This is the same expression as Eq.~(\ref{eq:PsiQS2}) obtained for the quasistatic model, except $\delta h_k(x)$ are now specifically steps of the actual interface evolution during the Haines jump, rather than arbitrary subdivisions of the jump. This equivalence is yet another confirmation of the match between work of non-viscous forces and viscous dissipation. 
%
{It also suggests that, in effect, \citet{Holtzman2023} used the approximation that $b=b_0$ everywhere but in the capillary pressure, as done here.}

\section{Spectral approach for interface dynamics}
\label{sec:spectral}

As discussed above, deviations of the interface shape from equilibrium lead to flow resulting in interface motion. It is tempting then to consider formulating a partial differential equation (PDE) for the shape of the interface $h(x,t)$, describing this relaxation process towards equilibrium. 
An attempt to propose such a PDE is discussed in the Supplemental Information (SI), where we show that there are fundamental reasons why the PDE approach for $h(x,t)$, 
fails, {regardless of the specific form of the PDE}. 
The interface dynamics is nonlocal in space, and thus governed by an integro-differential equation~\citep{Ganesan1998,Alvarez-Lacalle_PRE_2001,Paune2002,Paune2003}, which is computationally costly to solve. 
The difficulty is that perturbations of the interface with different length scales relax at different rates, {with the specific form of the dependence not producible by any PDE}.
Fourier expansion of a perturbation of the interface allows consideration of these different length scales separately. This is the essence of the spectral approach presented here.

\subsection{Evolution of an interface perturbed from equilibrium}
\label{sec:spec_evol}

Consider an interface slightly perturbed from a given (instantaneous) equilibrium configuration, 
\begin{equation}
    h(x,t)=h_{\rm eq}(x,t)+\delta h(x,t).
\end{equation}

\subsubsection{Pressure control}
From Eqs.~(\ref{eq:uappr}) and (\ref{eq:contappr}) we get
\begin{equation}
    p(x,y,t)=P(t)-\rho g_e y+\delta p(x,y,t),
    \label{eq:deltap}
\end{equation}
where we allow an arbitrary (i.e. not necessarily infinitely slow) change of the inlet pressure $P$, and $\delta p$ satisfies Laplace equation, 
\begin{equation}
     \nabla^2 \delta p=0\label{eq:Laplace}
\end{equation}
with the following boundary conditions: (i) $\delta p=0$ at $y=0$; (ii) $\partial(\delta p)/\partial x=0$ at $x=0$ and $x=W$; and (iii) $p$ satisfies Young-Laplace law [Eq.~(\ref{eq:pBC})] at the interface:
\begin{equation}
    \delta p(h(x,t),t)=p_e(x,t).
    \label{eq:pint}
\end{equation}

The general solution satisfying Eq.~(\ref{eq:Laplace}) with these boundary conditions can be represented as a Fourier series in $x$,
\begin{equation}
    \delta p(x,y,t)=\sum_{n=0}^{\infty} p_n(t)\sinh(k_ny)\cos (k_nx)/\sinh(k_n h_0),
    \label{eq:dp}
\end{equation}
where
\begin{equation}
    k_n=\frac{\pi n}{W}.
    \label{eq:kn}
\end{equation}
$h_0$ is (at this point) arbitrary, and $p_n(t)$ must be chosen to satisfy Eq.~(\ref{eq:pint}). Matching this last boundary condition is particularly simple if the interface is nearly flat,
\begin{equation}
    h\approx h_0,
    \label{eq:h0}
\end{equation}
as in this case
\begin{equation}
    p_e(x,t)\approx \delta p(x,h_0,t)=\sum_{n=0}^{\infty} p_n(t)\cos (k_nx).
    \label{eq:dpi}
\end{equation}
Thus, $p_n$ are given by the Fourier transform of $p_e$. Also, in this approximation and assuming that $|h'|\ll 1$ (using the prime throughout the paper to denote the $x$-derivative), from the kinematic boundary condition (\ref{eq:kinem}) and Eq.~(\ref{eq:uappr}),
\begin{equation}
    \frac{\partial h(x,t)}{\partial t}\approx u_y(x,h_0,t)=-\frac{b_0^2}{12\mu}\left.\frac{\partial [\delta p(x,y,t)]}{\partial y}\right|_{y=h_0}.
    \label{eq:dhdt}
\end{equation}
If we expand the interface profile $h(x,t)$ in cosines in $x$, without any additional terms,
\begin{equation}
h(x,t)=\sum_{n=0}^{\infty} h_n(t)\cos (k_nx),
    \label{eq:hfour}
\end{equation}
then
\begin{equation}
    \left.\frac{\partial h}{\partial x}\right|_{x=0}=\left.\frac{\partial h}{\partial x}\right|_{x=W}=0,
    \label{eq:90deg}
\end{equation}
which, in effect, corresponds to the contact angle of $90^{\circ}$ at the side walls. Given that deviations of the actual contact angle at the side walls from this value are expected to only affect the interface shape
very close to the sides [of the order of $\ell_c$ \citep{deGennes2004}], 
we assume that Eq.~(\ref{eq:hfour}) is valid in all our considerations.
Moreover, this is also consistent with Eqs.~(\ref{eq:u})--(\ref{eq:kinem}) in the sense that, while generally these equations do not enforce a particular value of the side wall contact angle, \textit{if} Eq.~(\ref{eq:90deg}) holds at $t=0$, then it remains valid at all later times. 
Indeed, if at a given time Eq.~(\ref{eq:90deg}) holds, then at a side wall
\begin{equation}
    \frac{\partial^2 h}{\partial x \partial t}=\left.\frac{\partial u_y}{\partial x}\right|_{y=h}\approx -\frac{b_0^2}{12\mu}\frac{\partial^2 p}{\partial x \partial y}=0,
\end{equation}
where, first, the kinematic condition (\ref{eq:kinem}), then the approximation (\ref{eq:uappr}) (which becomes exact if the cell is perfect near the wall, $b=b_0$), and, finally, the impenetrability condition {at the side walls} $\partial p/\partial x=0$ are used. This match between the side wall boundary conditions in Eqs.~(\ref{eq:u})--(\ref{eq:kinem}) and in the approach of this section is what allows the comparison in Sec.~\ref{sec:relax_test}, despite the cell not being wide enough for the side wall effects to be negligible ($W/\ell_c$ = 10).
Using Eqs.~(\ref{eq:dp}) and (\ref{eq:dhdt}) and approximation~(\ref{eq:h0}) gives
\begin{eqnarray}
    \frac{dh_n}{dt}&=&-\frac{b_0^2}{12\mu}k_n\coth(k_n h_0)p_n,\ n\ne 0,\label{eq:dh_vs_p}\\
    \frac{dh_0}{dt}&=&-\frac{b_0^2}{12\mu h_0}p_0.
    \label{eq:dh0_vs_p0}
\end{eqnarray}
Note that the natural choice for $h_0$, as used in Eq.~(\ref{eq:dpi}) and the right-hand side of Eq.~(\ref{eq:dh_vs_p}), is the mean value (the zeroth Fourier component) of $h$, so the notation is consistent. Equations~(\ref{eq:dh_vs_p}) and (\ref{eq:dh0_vs_p0}) directly relate the evolution of the Fourier components of the interface shape to the Fourier components of the pressure imbalance. Evolution of the $n\ne 0$ modes preserves the area occupied by the liquid, and only evolution of $h_0$ changes it.


\subsubsection{Flow-rate control}
Using the approximations in Eqs.~(\ref{eq:uappr})--~(\ref{eq:contappr}), we get
\begin{equation}
    p(x,y,t)=\Pi(t)-\rho g_{eV}y+\delta p(x,y,t),
    \label{eq:deltapV}
\end{equation}
where $\Pi(t)$ is the same as in Eq.~(\ref{eq:peV}). 
Here, as in the pressure-controlled case, $\delta p$ satisfies the Laplace equation~(\ref{eq:Laplace}), and the boundary conditions at $x=0$ and $x=W$ (side walls) are $\partial(\delta p)/\partial x=0$. 
However, here at $y=0$ we have $\partial(\delta p)/\partial y=0$, and
\begin{equation}
    \delta p(h(x,t),t)=p_{eV}(x,t).
    \label{eq:pintV}
\end{equation}
The solution is
\begin{equation}
    \delta p(x,y,t)=\sum_{n=0}^{\infty} p_n(t)\cosh(k_ny)\cos (k_nx)/\cosh(k_n h_0),
    \label{eq:dpV}
\end{equation}
and for a nearly flat interface ($h\approx h_0$),
\begin{equation}
    p_{eV}(x,t)\approx \delta p(x,h_0,t)=\sum_{n=1}^{\infty} p_n(t)\cos (k_nx)
    \label{eq:dpiV}
\end{equation}
(note $p_0=0$, as by construction the mean value of $p_{eV}$ is zero). The equation for the evolution of the interface is now
\begin{equation}
    \frac{\partial h(x,t)}{\partial t}\approx u_y(x,h_0,t)=V-\frac{b_0^2}{12\mu}\left.\frac{\partial [\delta p(x,y,t)]}{\partial y}\right|_{y=h_0},
    \label{eq:dhdtV}
\end{equation}
which gives
\begin{eqnarray}
    \frac{dh_n}{dt}&=&-\frac{b_0^2}{12\mu}k_n\tanh(k_n h_0)p_n,\ n\ne 0,
    \label{eq:dh_vs_pV}\\
    \frac{dh_0}{dt}&=&V,
    \label{eq:dh0_vs_p0V}
\end{eqnarray}
where $h_n$ are the Fourier coefficients of $h$ defined, as before, via Eq.~(\ref{eq:hfour}).

\subsection{An approximate solution for interfacial dynamics}

In principle, Eqs.~(\ref{eq:dh_vs_p})--(\ref{eq:dh0_vs_p0}) [or (\ref{eq:dh_vs_pV})--(\ref{eq:dh0_vs_p0V}), for flow-rate control] can be solved using a straightforward numerical integration approach: At every step $p_e$ (or $p_{eV}$) is calculated, expanded in Fourier series, and the Fourier modes of the interface profile are advanced in time according to these equations, using a standard explicit numerical algorithm for ODEs. 
However, this would require an extremely small integration timestep $\Delta t$ to maintain numerical stability, since short-wavelength modes relax fast. Alternatively, we can increase $\Delta t$ by finding an approximate analytical solution of these sets of equations over a time period that is longer than the relaxation times of some of the modes. We establish this here by considering a time interval $\Delta t$ such that during this interval we can neglect the temporal changes in the following: $b(x,h(x,t))$ for any given $x$; the mean interface height $h_0$; and $P(t)$ in the expression for $p_e$ [Eq.~(\ref{eq:pe}); pressure control], or $\Pi(t)$ in the expression for $p_{eV}$ [Eq.~(\ref{eq:peV}); flow-rate control].

\subsubsection{Pressure control}
Bearing in mind that for $|h'|\ll 1$ the interface curvature $\kappa\approx h''$, and denoting $\Delta h(x,\Delta t)\equiv h(x,t+\Delta t)-h(x,t)$, we get from Eq.~(\ref{eq:pe}),
\begin{equation}
    p_e(x,t+\Delta t)\approx p_e(x,t)+\rho g_e\Delta h(x,\Delta t)-\frac{\pi}{4}\gamma\Delta h''(x,\Delta t).
    \label{eq:pe_perturb}
\end{equation}
The Fourier transform of this is
\begin{equation}
    p_n(t+\Delta t)=p_n(t)+\rho g_e(1+\ell_c^2 k_n^2)\Delta h_n(\Delta t),
\end{equation}
where
\begin{equation}
    \Delta h(x,\Delta t)=\sum_{n=0}^{\infty}\Delta h_n(\Delta t)\cos(k_n x).
    \label{eq:dh}
\end{equation}
Equations~(\ref{eq:dh_vs_p})--(\ref{eq:dh0_vs_p0}) then become
\begin{eqnarray}
    \frac{d[\Delta h_n(\Delta t)]}{d[\Delta t]}&=&\nonumber\\
    & &\hspace{-1.5cm}-\frac{b_0^2}{12\mu}k_n\coth[k_n h_0(t)]\left[p_n(t)+\rho g_e(1+\ell_c^2 k_n^2)\Delta h_n(\Delta t)\right],\ n\ne 0,\\
    \frac{d[\Delta h_0(\Delta t)]}{d[\Delta t]}&=&-\frac{b_0^2}{12\mu h_0(t)}\left[p_0(t)+\rho g_e\Delta h_0(\Delta t)\right].
\end{eqnarray}
The solutions of these equations are
\begin{eqnarray}
    \Delta h_n(\Delta t)&=&-\frac{p_n(t)}{\rho g_e (1+\ell_c^2 k_n^2)}[1-\exp(-\omega_p(k_n) \Delta t)],\ n\ne 0,\label{eq:Deltahn}\\
    \Delta h_0(\Delta t)&=&-\frac{p_0(t)}{\rho g_e}[1-\exp(-\omega_p(0) \Delta t)],\label{eq:Deltah0}
\end{eqnarray}
where the mode relaxation rates are
\begin{eqnarray}
    \omega_p(k)&=&\beta k(1+\ell_c^2 k^2)\coth (kh_0(t)),\label{eq:omp}\\
    \omega_p(0)&=&\frac{\beta}{h_0(t)}\label{eq:omp0}
\end{eqnarray}
[note that Eq.~(\ref{eq:omp0}) is the $k\to 0$ limit of Eq.~(\ref{eq:omp})], and
\begin{equation}
    \beta=\frac{\rho g_e b_0^2}{12\mu}
    \label{eq:beta}
\end{equation}
is a characteristic speed at which the liquid would move down due to gravity with $P=0$ and no capillary effects.

\subsubsection{Flow-rate control}
Making the same approximations as for Eq.~(\ref{eq:pe_perturb}), in the flow-rate-controlled case we get from Eq.~(\ref{eq:peV})
\begin{equation}
    p_{eV}(x,t+\Delta t)\approx p_{eV}(x,t)+\rho g_{eV}\Delta h(x,\Delta t)-\frac{\pi}{4}\gamma\Delta h''(x,\Delta t).
\end{equation}
The Fourier transform of this is
\begin{equation}
    p_n(t+\Delta t)=p_n(t)+\rho g_{eV}(1+\ell_{cV}^2 k_n^2)\Delta h_n(\Delta t),
\end{equation}
where the meaning of $\Delta h_n$ is the same as above [Eq.~(\ref{eq:dh})], and the modified capillary length is
\begin{equation}
    \ell_{cV}=\left(\frac{\pi\gamma}{4\rho g_{eV}}\right)^{1/2}.
\end{equation}
Equations~(\ref{eq:dh_vs_pV})--(\ref{eq:dh0_vs_p0V}) give
\begin{eqnarray}
    \frac{d[\Delta h_n(\Delta t)]}{d[\Delta t]}&=&\nonumber\\
    & &\hspace{-2cm}-\frac{b_0^2}{12\mu}k_n\tanh(k_n h_0(t))\left[p_n(t)+\rho g_{eV}(1+\ell_{cV}^2 k^2)\Delta h_n(\Delta t)\right],\ n\ne 0,\\
    \frac{d[\Delta h_0(\Delta t)]}{d[\Delta t]}&=&V,
\end{eqnarray}
and the solutions are
\begin{eqnarray}
    \Delta h_n(\Delta t)&=&-\frac{p_n(t)}{\rho g_{eV} (1+\ell_{cV}^2 k_n^2)}[1-\exp(-\omega_V(k_n) \Delta t)],\ n\ne 0,
    \label{eq:DeltahnV}\\
    \Delta h_0(\Delta t)&=&V\Delta t,\label{eq:Deltah0V}
\end{eqnarray}
where
\begin{equation}
    \omega_V(k)=\beta_V k(1+\ell_{cV}^2 k^2)\tanh (kh_0(t))\label{eq:omV}
\end{equation}
and
\begin{equation}
    \beta_V=\frac{\rho g_{eV} b_0^2}{12\mu}.
    \label{eq:betaV}
\end{equation}

\subsection{The algorithm for computing the interface evolution}
\label{sec:alg}


On longer time intervals over which $b(x,h(x,t))$, $P$ ($\Pi$), or $h_0$ cannot be considered constant in time, Eqs.~(\ref{eq:Deltahn})--(\ref{eq:Deltah0}) [or (\ref{eq:DeltahnV})--(\ref{eq:Deltah0V})] cannot be applied directly. Instead, we use them within a numerical algorithm, in which we advance in small enough steps $\Delta t$ such that this constancy can be assumed. The algorithm is as follows. The solution $h(x,t)$ is discretised in space using a uniform grid of size $N$ and step $\Delta x=W/N$, and in time using $\Delta t$. At time $t$, given the values of $h(x,t)$ at all mesh points, the following procedure is carried out to advance in time by $\Delta t$:

\begin{enumerate}

    \item \hspace{2pt} Calculate $p_e(x,t)$ [or $p_{eV}(x,t)$ for rate-controlled] numerically on the mesh using a discretisation of Eq.~(\ref{eq:pe}) [or Eq.~(\ref{eq:peV})].
    
    \item \hspace{2pt} Run a discrete Fourier transform procedure for $p_e(x,t)$ [or $p_{eV}(x,t)$] to obtain the Fourier components $p_n(t)$.

    \item \hspace{2pt} Use Eqs.~(\ref{eq:Deltahn})--(\ref{eq:Deltah0}) [or (\ref{eq:DeltahnV})--(\ref{eq:Deltah0V})] to calculate $\Delta h_n(\Delta t)$.

    \item \hspace{2pt} Run the inverse discrete Fourier transform on $\Delta h_n(\Delta t)$ to obtain $\Delta h(x,\Delta t)$ at all mesh points.

    \item \hspace{2pt} Advance in time, $h(x,t+\Delta t) = h(x,t) + \Delta h(x,\Delta t)$.
    
\end{enumerate}

Using highly optimised Fast Fourier Transform (FFT) routines, e.g., those in the FFTW library \citep{FFTW05}, simulations with $N\sim 10^4 - 10^5$, such as those in Sec.~\ref{sec:disorder}, can be carried out in minutes on an ordinary laptop---substantially faster than CFD \citep{ramstad2019pore,Zhao2019,Giudici2023GNMvsDNS}. For FFT, values of $N$ equal to an integer power of 2 are preferable for computational efficiency.

\subsection{Dissipation rate} 
\label{subsec:dissip_spectral}

Once the evolution of the interface shape $h(x,t)$ is known from spectral calculations, we can use Eq.~(\ref{eq:Rfinal}) to calculate the energy dissipation rate. Since the spectral approach provides equations for Fourier coefficients, it is convenient to express the result in terms of these coefficients as well. The results for the two modes of driving are given below (derivation details are provided in Appendix \ref{APPENDIX_dissip_spectral}).


\textit{\textbf{Pressure control}}:
\begin{equation}
    |R(t)|= \frac{b_0 W}{\rho g_e}\left[\omega_p(0)p_0^2(t)+\sum_{n=1}^{\infty}\frac{\omega_p(k_n)p_n^2(t)}{2(1+\ell_c^2 k_n^2)}\right].\label{eq:Rspec}
\end{equation}

\textit{\textbf{Flow-rate control}}:
\begin{equation}
    |R(t)|=\frac{12\mu WV^2h_0}{b_0}-\frac{b_0 VW}{2}\sum_{n=1}^{\infty}k_n\tanh(k_n h_0)h_n p_n+\frac{b_0 W}{2\rho g_{eV}}\sum_{n=1}^{\infty}\frac{\omega_V(k_n)p_n^2(t)}{1+\ell_{cV}^2k_n^2}.\label{eq:RspecV}
\end{equation}

It is useful to compare the equations for pressure and rate control, Eq.~(\ref{eq:RspecV}) and Eq.~(\ref{eq:Rspec}).
The first term in Eq.~(\ref{eq:RspecV}) appears, at first glance, to have no analog in Eq.~(\ref{eq:Rspec}) and, in fact, remains nonzero in ``equilibrium'' ($p_{eV}=0$). This corresponds to dissipation during advancement of the liquid at a constant uniform speed $V$. Note, however, that in a pressure-controlled situation where $P$ increases linearly in time at a finite rate, the averaged pressure imbalance $p_0$ will likewise increase linearly (in time, or, equivalently, $h_0$), with the first term of Eq.~(\ref{eq:Rspec}) then producing the same behaviour as the first term of Eq.~(\ref{eq:RspecV}). Even though $p_0$ is no longer small in such a situation, the equations for the interfacial dynamics should still remain accurate.
%
There is no analog of the second term of Eq.~(\ref{eq:RspecV}) in Eq.~(\ref{eq:Rspec}). This is because such a term ($\sim p_0 p_n h_n$) is formally of a higher order (if $p_0$ is still treated as a small parameter). It can originate from neglected higher-order terms in Eq.~(\ref{eq:dh_vs_p}), but is not expected to be dominant as long as Eq.~(\ref{eq:dh_vs_p}) remains accurate.
Finally, the last term in Eq.~(\ref{eq:RspecV}) is directly analogous to the last term in Eq.~(\ref{eq:Rspec}) (replacing $\omega_V\to\omega_p$, $\ell_{cV}\to\ell_c$ and $g_{eV}\to g_e$).

\subsection{Applicability of the approach}
\label{subsec:Applicability}


When deriving the equations underlying the spectral approach, we have made a number of assumptions. Here we discuss the associated potential pitfalls, and the tests made to ensure the validity of our approach (further described in Sec.~\ref{sec:validation}).

The first assumption was that cell aperture variations must be small, $|\delta b|\ll b_0$. Since the smaller $|\delta b|$, the flatter the interface in equilibrium (all other factors being equal), we generally expect our second assumption, $|h'|\ll 1$, to follow from the first. 
This is not guaranteed, especially in conditions approaching the Saffman-Taylor instability, when $\delta b$ not being very small also becomes potentially more problematic on its own (regardless of the flatness of the interface). 
We test this (in Sec.~\ref{sec:imbdra}) and show that the approach remains adequate even when the $|h'|\ll 1$ condition is not valid, as long as the full expression for $p_{eV}$ (not its linear approximation) is used. 
We have also assumed, in effect, that the domain on which the Laplace equation for $\delta p$ is solved is rectangular, which requires the interface to be sufficiently flat overall. 
Strictly speaking, this requires that the full amplitude of variations of the interface height $h$ is much smaller than its mean $h_0$.
On the other hand, we expect the relaxation rates to remain approximately correct even for wavelengths smaller than the total interface deformation amplitude, since the \emph{local} flatness of the interface on the scale of the wavelength is the most important factor. 

Another important aspect relevant to the applicability of our approach is the 
assumption that the change of $b(x,h(x,t))$ is negligible during the time step $\Delta t$. 
If the variation of the cell aperture is smooth, then this condition can be satisfied for small enough $\Delta t$. 
However, for surfaces with steps [e.g., ``mesa'' defects considered in \citet{Planet2020,HoltzmanCommPhys2020,Lavi_PRF_2023}], an interface that has reached the step may cross it immediately, thus, the condition cannot be satisfied for any value of $\Delta t$. 
We expect that this would not be important for small enough $\Delta t$, because even in the case where the interface is pinned at the defect's end, and the model erroneously predicts it will detach and move downstream (``overshoot''), 
this will be corrected in the next time step. 
Thus, for sufficiently small $\Delta t$, the interface will remain in the vicinity of the defect's end without becoming unstable. 
Our tests for a mesa defect (Sec.~\ref{sec:imbdra}) confirm this. 
%
In the opposite extreme case of a defect of a constant profile (where $b$ is $y$-independent, $b=b(x)$, everywhere within the part of the cell visited by the interface), 
for pressure control with $P={\rm const}$ or a closed inlet ($V=0$) the conditions are satisfied for arbitrarily large $\Delta t$. In the closed inlet case $\Pi$ remains constant in our approximation and does not enter the equations anyway.
This means the interface shape can be calculated at any time $t$ \emph{directly} (without timestepping). 
This is exemplified in Sec.~\ref{sec:relaxation}, where we exploit this to compute the interface evolution over more than ten orders of magnitude in time.

\section{Validating the spectral approach}
\label{sec:validation}


We validate our approach by comparing with high-fidelity CFD simulations and experiments, for the illustrative (idealised) case of a system composed of a single step-like ``mesa defect'' (Fig.~\ref{fig:schematic_single_defect}), 
\begin{equation}
    b(x,y)=\begin{cases}
        b_0+\delta b, & |x-W/2|<w/2\ \text{and}\ Y_1<y<Y_2\\
        b_0 & \text{otherwise}
    \end{cases}
\end{equation}
where $\delta b<0$ is a constant, and $Y_1$ and $Y_2$ are the y-coordinates of the defect egdes. Note that for a finite cell size, the mean aperture is slightly different from $b_0$, but this is neglected due to its minor impact. 

\begin{figure}
    \centering    \includegraphics[width=0.3\linewidth]{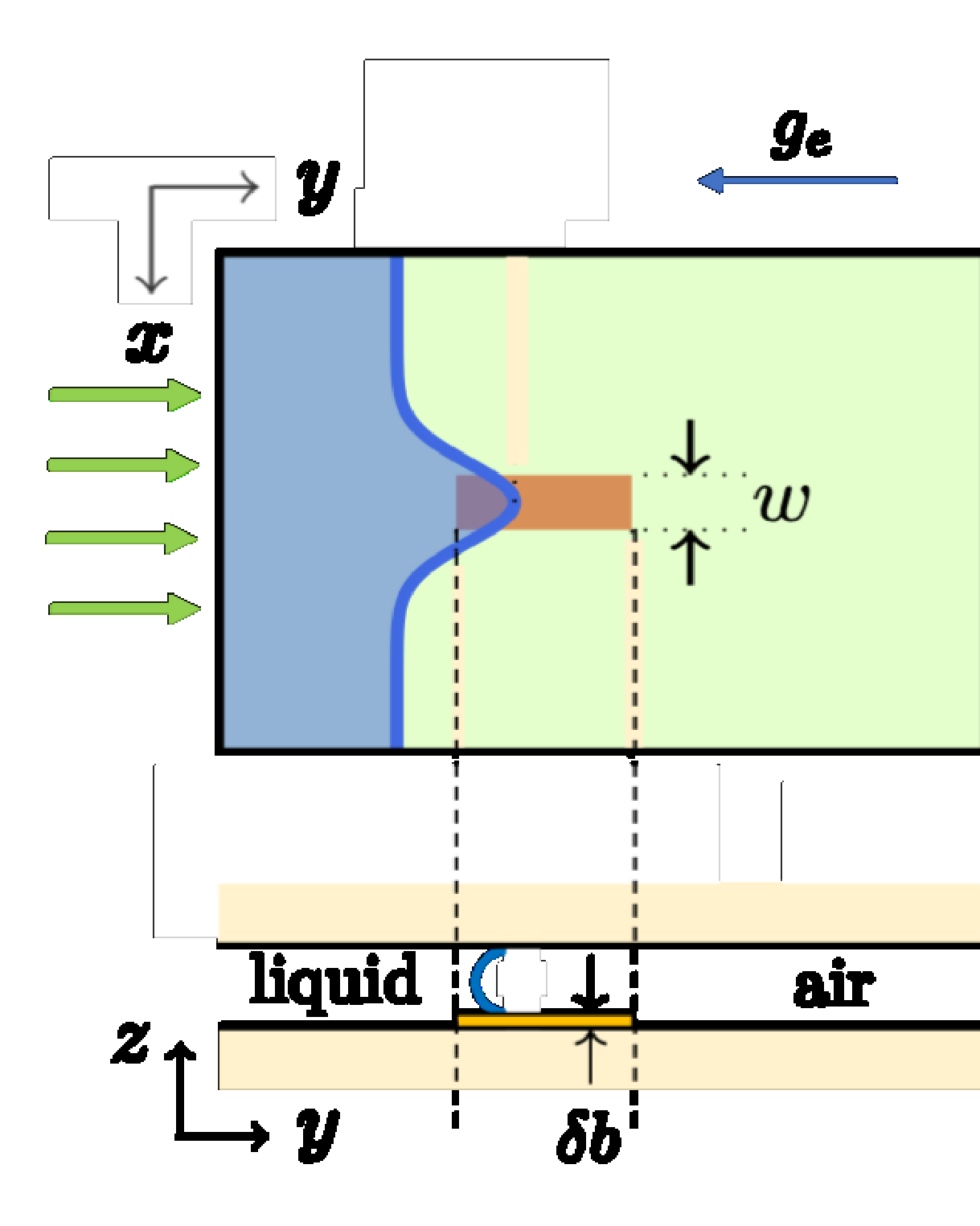}
\caption{
Schematics of idealised system consisting of IHSC with a single step-like ``mesa defect'' (of width $w$ and aperture perturbation $\delta b$).
}
    \label{fig:schematic_single_defect}
\end{figure}

We conduct two series of tests.
In the first, we consider relaxation of an initially flat interface crossing the defect, for 
(i) a closed inlet which is very far from the interface, such that the influence of that boundary is negligible; and (ii) an open inlet closer to the interface, with a fixed pressure at the inlet such that the interface would be in equilibrium initially for a cell without defect ($\delta b = 0$). 
In the second series of tests, we test the flow-rate-controlled case, for both imbibition and drainage, namely where the liquid (i) flows into the cell ($V>0$) and (ii) is drained from the cell ($V<0$).


\subsection{Relaxation of an interface perturbed from equilibrium}
\label{sec:relax_test}

For these tests, we ensure that $Y_1<h(x,t)<Y_2$ for $|x-W/2|<w/2$ at all times. In the spectral approach, we can then take
\begin{equation}
    b=b(x)=\begin{cases}
        b_0+\delta b, & |x-W/2|<w/2,\\
        b_0, & w/2<|x-W/2|<W/2.
    \end{cases}
    \label{eq:b_inf}
\end{equation}
The interface is initially flat, $h(x,0)=h^{(0)}$, and then evolves (relaxes) towards the equilibrium shape. While this situation of relaxation from a perfectly flat nonequilibrium shape is somewhat artificial, the process is very similar to a Haines jump that happens when the interface starts outside the defect ($h^{(0)}<Y_1$) in its equilibrium flat shape and then approaches the defect quasistatically ($V>0$ but very small in the flow-rate-controlled case or a very slow change of $P$ in the pressure-controlled case) and deforms after reaching the defect. In fact, within the spectral approach, the evolution of the interface is identical in these two scenarios, if in the first one $h(x,t)>h^{(0)}$ for all $x$ with $|x-W/2|<w/2$ at all $t>0$ (whether this is so depends on the details of system parameters, but, in particular, is true for sufficiently wide cells).


\subsubsection{Spectral calculations}
\label{sec:mesa_spectral}

The dependence $b=b(x)$ in Eq.~(\ref{eq:b_inf}) allows us to avoid timestepping and obtain closed-form expressions for interface evolution and the dissipation rate, which are valid at all times. To obtain these expressions, we introduce the quantity
\begin{equation}
    \mathcal{H}=\frac{2\gamma}{\rho g_e}\left(\frac{1}{b_0+\delta b}-\frac{1}{b_0}\right)\approx -\frac{2\gamma\delta b}{\rho g_e b_0^2},
    \label{eq:H}
\end{equation}
which has the meaning of the equilibrium capillary rise far inside a wide ($w\gg\ell_c$) defect relative to the level far outside the defect. Then
for the \textbf{closed inlet} case (i.e., flow-rate control with $V=0$), Fourier transforming Eq.~(\ref{eq:peV}) for $p_{eV}$ and using the result in Eq.~(\ref{eq:DeltahnV}) gives for the interface evolution 
\begin{equation}
    h(x,t)= h^{(0)}+\frac{4 \mathcal{H}}{W}\sum_{n=1}^{\infty}\frac{\sin(k_{2n}w/2)}{k_{2n}(1+\ell_c^2 k_{2n}^2)}\left\{1-\exp[-\beta k_{2n}(1+\ell_c^2 k_{2n}^2)t]\right\}\cos[k_{2n}\Delta],
    \label{eq:hmesa}
\end{equation}
where $\Delta=x-{W}/{2}$ is
the horizontal distance from the defect axis. The sum is an inverse Fourier transform and can be computed using the Fast Fourier Transform method. For the dissipation rate, from Eq.~(\ref{eq:RspecV})
\begin{equation}
    |R(t)| = \frac{8b_0\beta\rho g_e \mathcal{\mathcal{H}}^2}{W}\sum_{n=1}^{\infty}\frac{\sin^2(k_{2n}w/2)}{k_{2n}}\exp[-2\beta k_{2n}(1+\ell_c^2 k_{2n}^2)t].
    \label{eq:Rmesa}
\end{equation}
Details of these derivations are in Appendix~\ref{sec:Spectral_h_R_details}.
Obviously, {to obtain the numerical results in the plots presented in this section {and Sec.~\ref{sec:relaxation}}, the infinite sums in these equations must be truncated;}
{however, this is done in such a way that}
adding more terms would have minute effect on the results (e.g., it would not be visible in the plots).

For the \textbf{pressure-controlled} case, to avoid timestepping (for a finite inlet-interface distance), we take $h_0(t)\equiv h^{(0)}$, neglecting the change of $h_0$ in the equations for $\omega_p$. 
From Eqs.~(\ref{eq:pe}) and (\ref{eq:Deltahn})--(\ref{eq:Deltah0}), the interface evolution is 
\begin{eqnarray}
    h(x,t) &=& h^{(0)}+\frac{\mathcal{H} w}{W}\left[1-\exp\left(-\frac{\beta}{h^{(0)}}t\right)\right]\\
    & & \hspace{-1.5cm}+\frac{4 \mathcal{H}}{W}\sum_{n=1}^{\infty}\frac{\sin(k_{2n}w/2)}{k_{2n}(1+\ell_c^2 k_{2n}^2)}\left\{1-\exp[-\beta k_{2n}(1+\ell_c^2 k_{2n}^2)\coth (k_{2n}h^{(0)})t]\right\}\cos[k_{2n}\Delta].\nonumber
    \label{eq:hmesa_P_control}
\end{eqnarray}
The details are likewise provided in Appendix~\ref{sec:Spectral_h_R_details}. {An expression for the dissipation rate (when pressure is controlled) can also be derived, but since here we only test the dissipation rate in the flow-rate control case, it is not given.}

\subsubsection{CFD simulations}
The CFD data we compare with was generated by solving Eqs.~(\ref{eq:u})--(\ref{eq:kinem}) using the finite element method, with the code from Ref.~\citet{Lavi_PRF_2023} implemented in \textsc{FreeFem++} software \citep{hecht2012new}; for a detailed description 
see Appendix B of Ref.~\citet{Lavi_PRF_2023}. 
We use here a relatively coarse mesh with around 60 nodes along the interface, which improves computational efficiency with only a small reduction in accuracy.
To compare with the closed inlet case, we revise the code to change the boundary condition for the bottom boundary.

\subsubsection{Quantitative comparison}

Our spectral approach predictions closely match those of the CFD in terms of both the interface evolution (Fig.~\ref{fig:relax_compar}) and the dissipation rate (Fig.~\ref{fig:dissip_compar}), at a small fraction of the computational resources. {As an example, the computational costs of the actual runs used to produce Fig.~\ref{fig:dissip_compar} differed by a factor $\sim 10^3$ even without using FFT in the spectral code.} While code efficiency could be further optimized for both, the spectral approach is expected to remain much more efficient.
The special case of $b=b(x)$ is extremely fast to compute using the spectral approach, as no timestepping is required (and hence fewer interfacial configurations {need to be} obtained). 
In this way, our approach provides only the essential information at the times of interest, {whereas the CFD timestep is determined by the requirements of accuracy and numerical stability.}
{As a consequence, the speed-up factor also depends on the required number of data points. For this comparison, we selected the lengths $\ell_c$, $w$ and $W$ to be similar in magnitude, so the good agreement likely means that the approach is valid for any relation between these lengths, rather than just in a particular limit. 
We also chose $|\delta b|/b_0$ to be just small enough that the linear approximation is reasonable.
The values of the parameters and the detailed explanations are given in Appendix \ref{sec:relax_test_param}.}

%

Figure~\ref{fig:relax_compar} compares the interface evolution between the CFD simulation and the spectral approach, for both the closed inlet and pressure-control driving, with excellent agreement (noting small artifacts caused by the coarse mesh in the CFD). 
Even in the case we are considering here, where the length scales $W$, $w$, and $\ell_c$ are similar, nontrivial evolution of the interface spans several orders of magnitude in time. For this reason, the time points are chosen roughly uniformly on the logarithmic scale, with 5-6 profiles per decade (see caption of Fig.~\ref{fig:relax_compar}). For convenience, we introduce a time scale, $t_c\sim\ell_c/\beta$ (the significance of which will become clear in Sec.~\ref{sec:relaxation}), defining the dimensionless time
\begin{equation}
    \hat{t}=\frac{\beta t}{\ell_c};\label{eq:t-hat}
\end{equation}
the chosen time range includes $\hat{t}=1$. 
Notably, the interface evolution as a function of this dimensionless time depends on the geometry, but not on $\rho$, $g_e$ or $\mu$.
Initially, the evolution is nearly identical in the closed-inlet and pressure-controlled cases. It is driven by local flows while the total amount of liquid that has entered through the inlet is negligible even when the inlet is open; thus, the volume is conserved in both cases. At the earliest stage, the most rapid interface deformation is near the defect edges ($\Delta/\ell_c=\pm 2$), but the central part of the interface ``catches up'' eventually. The volume, of course, remains conserved at all times in the closed-inlet case; relative to that, in the pressure-controlled case the interface shifts upwards until it reaches the equilibrium position. This evolution of the interface is analysed theoretically in more detail in Sec.~\ref{sec:relaxation}.
%

\begin{figure}
    \centering

    \includegraphics[width=0.45\linewidth]{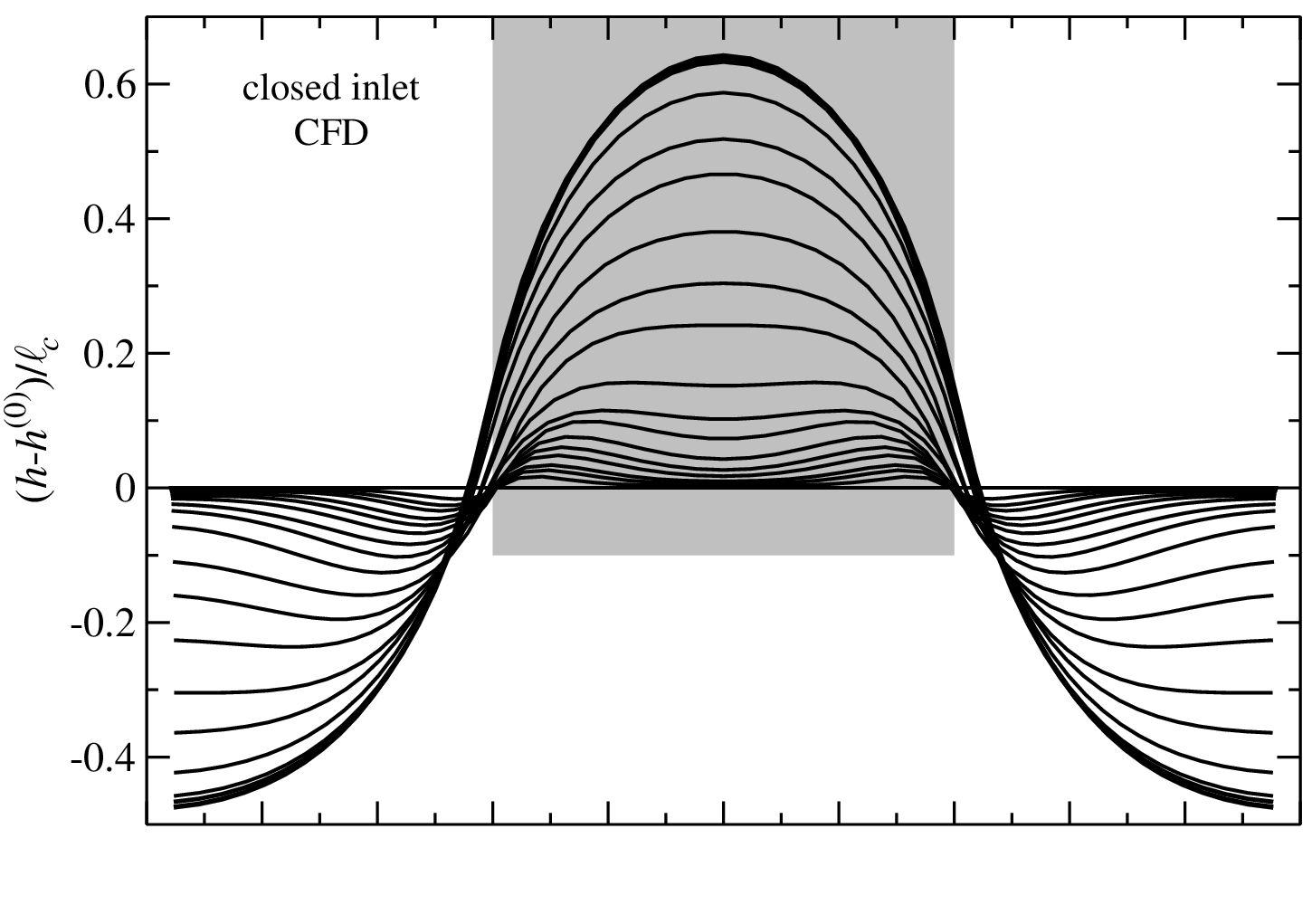}
    \includegraphics[width=0.45\linewidth]{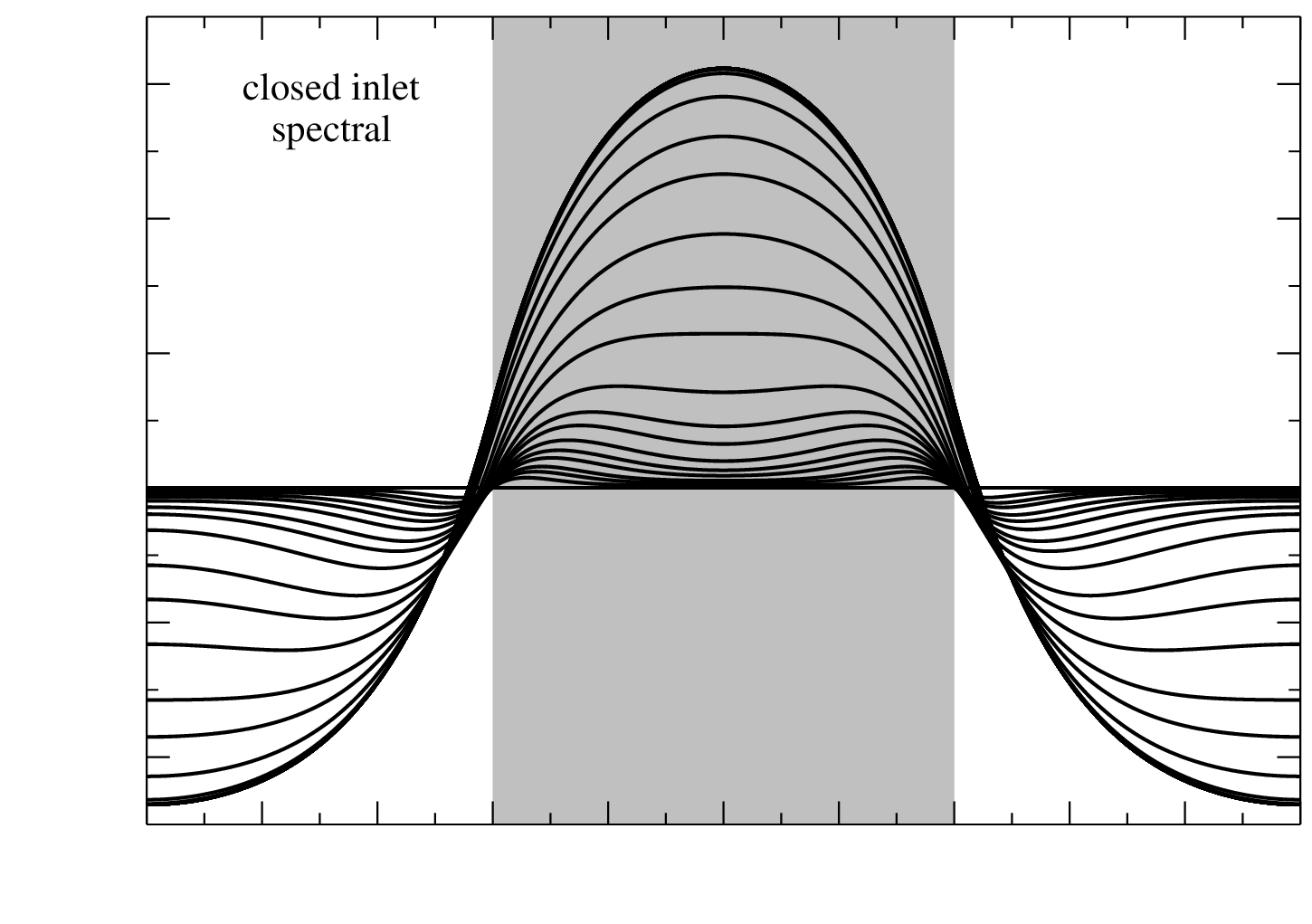}
    \includegraphics[width=0.45\linewidth]{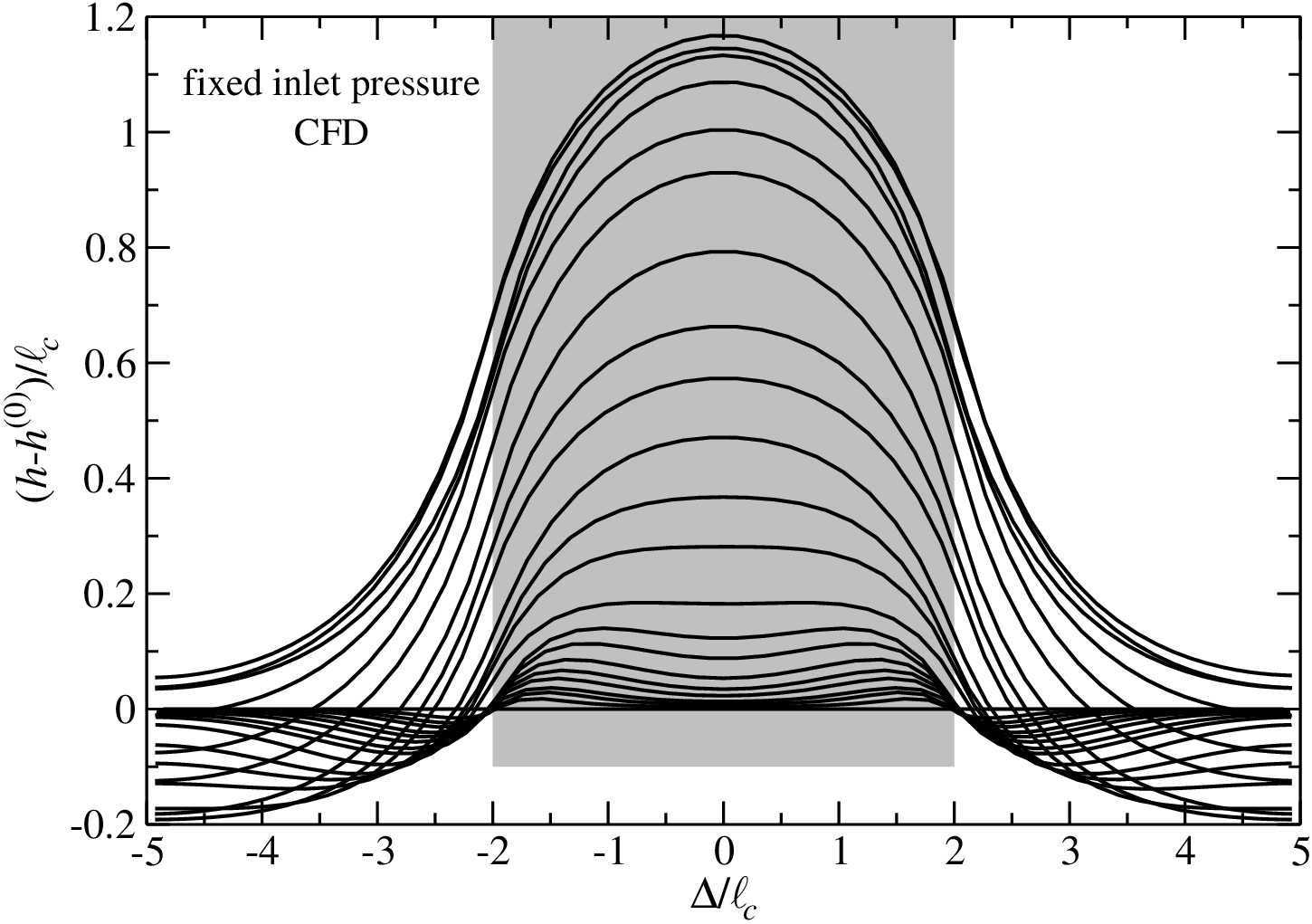}
    \includegraphics[width=0.45\linewidth]{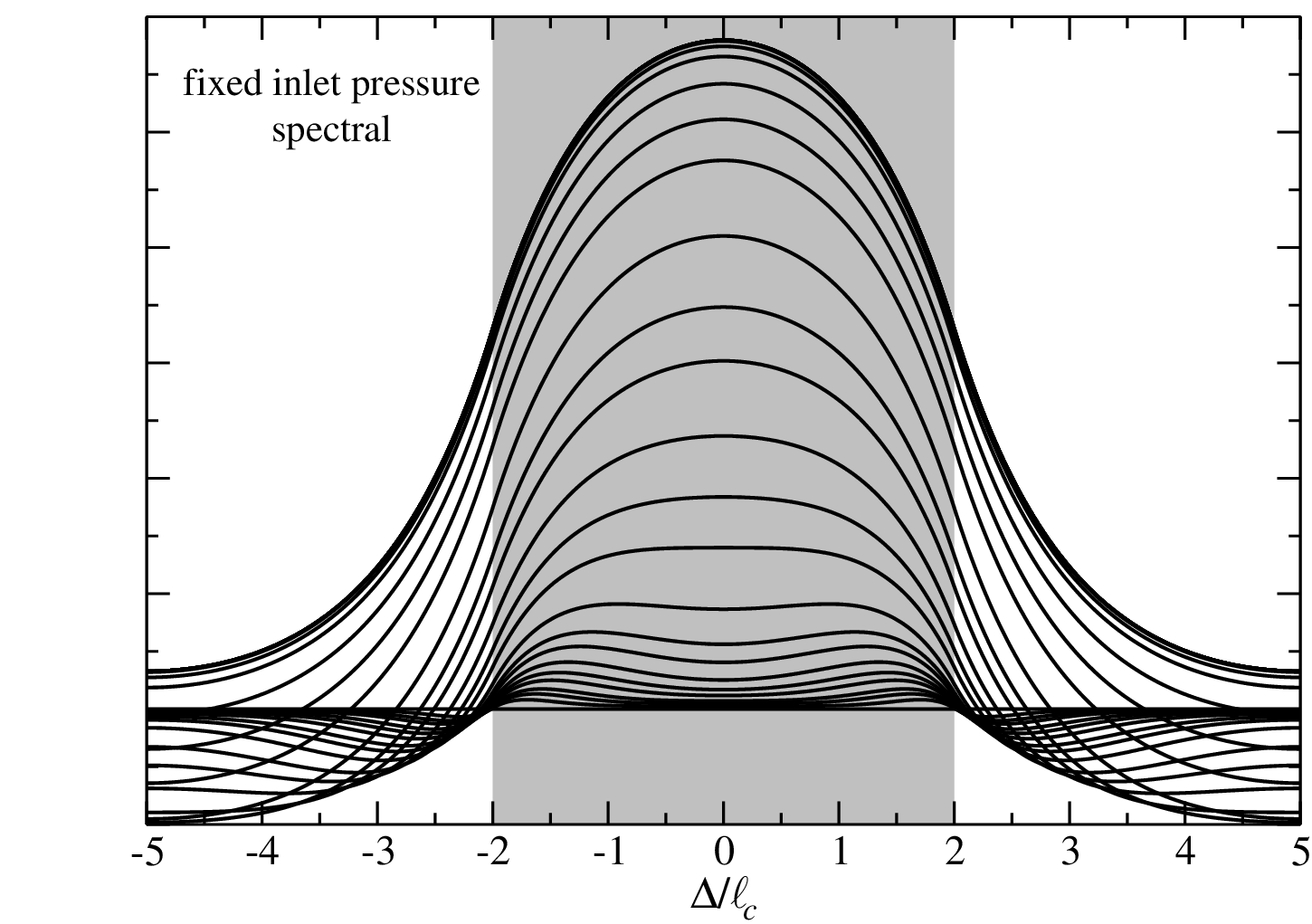}
    \caption{For a Hele-Shaw cell of width $W/\ell_c=10$ and {aperture} $b_0/\ell_c=0.1$, with a single defect of width $w/\ell_c=4$ and height $|\delta b|/\ell_c=0.005$, the evolution of the liquid-air interface shape for a closed inlet (top row) and an open inlet with a fixed pressure at the inlet such that without the defect the system would be in equilibrium initially (bottom row). 
    {
    The shading shows the position and extent of the defect (it is formally infinitely long in both directions in the spectral calculation, but the results only rely on the assumption that the part of the interface with $|\Delta|<w/2$ is within the defect at all times).
    } 
    These data are obtained using CFD simulations (left column) and the spectral approach introduced in this paper [Eq. \ref{eq:hmesa}] (right column). 
The curves correspond to dimensionless times $\hat{t}$ [Eq.~(\ref{eq:t-hat})]: 0, $\{1, 2, 3, 5, 7\}\times 10^{-2}$, $\{1, 1.5, 2, 3, 5, 7\}\times\{10^{-1},\ 10^0\}$, 10, 15, 20 and (in the bottom panels only) 30. The time increases from the bottom up at $\Delta=0$. The time step of CFD simulations is $10^{-2}$. Other parameters and details are in the text.}
    \label{fig:relax_compar}
\end{figure}

Finally, in Fig.~\ref{fig:dissip_compar} we compare the dissipation rates obtained in CFD simulations and spectral calculations for the closed-inlet case. For the CFD calculations we use the full 2D velocity field obtained in the simulations in Eq.~(\ref{eq:Rgen}) (thus, not making the $b=b_0$ assumption). For the spectral calculations, Eq.~(\ref{eq:Rmesa}) is used.
The agreement is likewise excellent, {aside from late times, where in CFD simulations weak spurious flows are present.}

\begin{figure}[t!]
    \centering
    \includegraphics[width=0.5\linewidth]{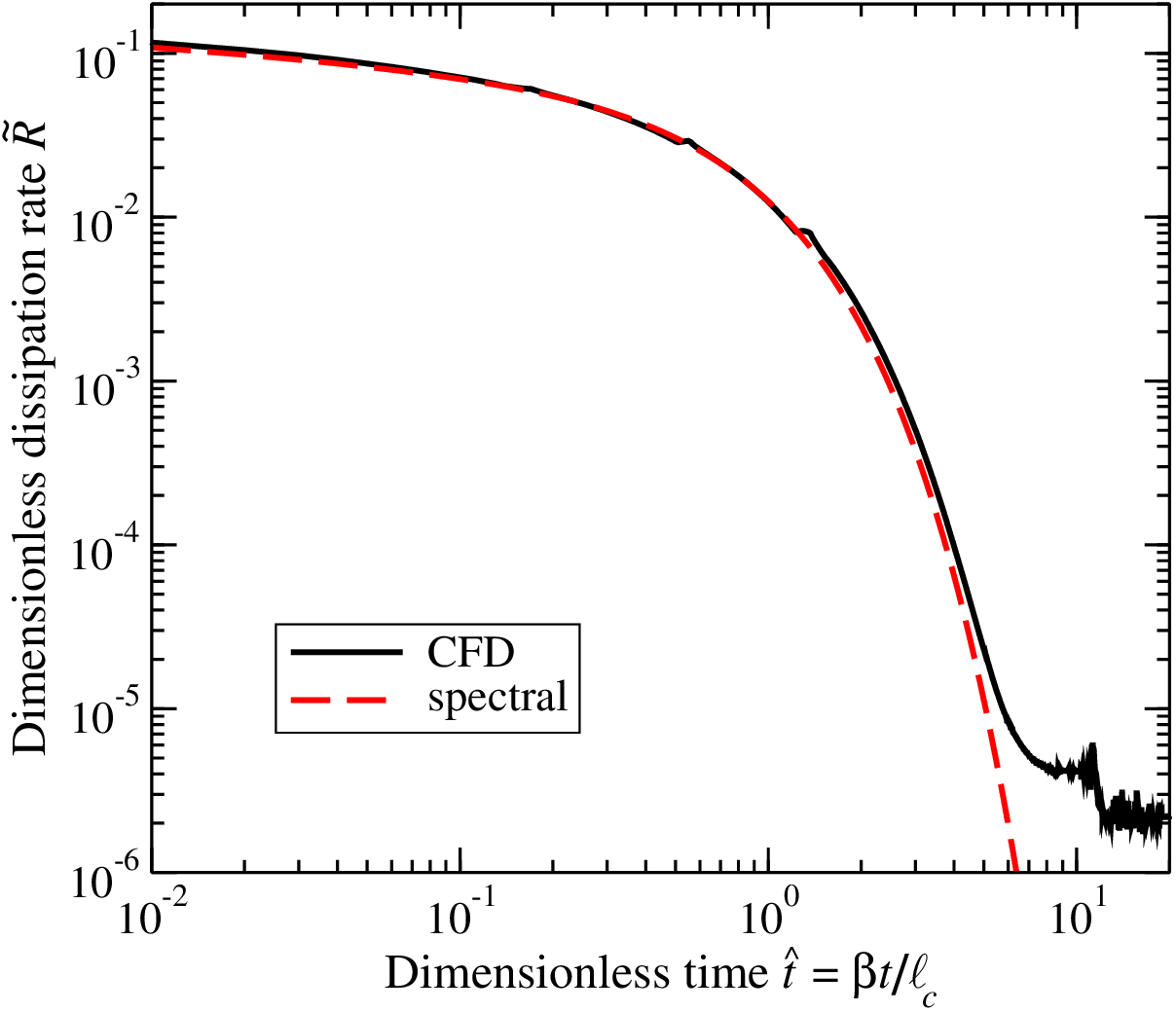}
    \caption{Comparison of the energy dissipation rates obtained in CFD simulations and spectral calculations [Eq.~(\ref{eq:Rmesa})] for a Hele-Shaw cell with a single defect and a closed inlet, with the same parameters as in Fig.~\ref{fig:relax_compar}.}
    \label{fig:dissip_compar}
\end{figure}

\subsection{Imbibition and drainage} 
\label{sec:imbdra}

Here, we test our model against CFD simulations and experiments for both imbibition and drainage, for a fixed flow rate, reported in \citet{Lavi_PRF_2023}.
Although the defect was long relative to its width (length $L=Y_2-Y_1=60$~mm, width $w=3$~mm), it 
did not extend the entire cell length. Unlike the relaxation case discussed in the previous section, here we consider an initially flat interface which approaches the defect
during the imbibition/drainage process.

\subsubsection{Spectral calculations} 
Since the defect has a finite length and
part of the interface may be on the defect,
$b=b(x)$ can no longer be assumed, and a simulation with time stepping is required, using the algorithm of Sec.~\ref{sec:alg}. Notably, in the drainage experiments the inlet speed $V$ was as high as $\approx$70\% of the Saffman-Taylor instability threshold. As a result, the effective gravity $g_{eV}$ was low and the equilibrium deformation for the interface crossing the defect was high, with $h'$ significantly exceeding unity, way beyond the linear regime. 
Since our theory does not apply to such conditions, simulating this case provides a very stringent test beyond the expected applicability limits. The only change we make in the algorithm to accommodate this is using the exact expression for $p_{eV}$, Eq.~(\ref{eq:peV}), with the full nonlinear expression for the curvature. This ensures that the equilibrium state is the same as in the CFD [neglecting the fact that Eq.~(\ref{eq:equilFR}) uses the approximation of Eqs.~(\ref{eq:uappr})--(\ref{eq:contappr})], even if the dynamics of approach to equilibrium still assumes the linear approximation (around a flat interface). We have not made explicit use of the symmetry of the problem, simulating the full width of the cell, subdividing it into $N=2^{14}$ mesh steps and using the same number of Fourier components. This corresponds to about 260 mesh points within the width of the defect. The time step is $\Delta t = 0.1$ s.


\subsubsection{Quantitative comparison}
The results of our spectral simulations agree well with the experiments and simulations of \citet{Lavi_PRF_2023}, in terms of both the interface profiles (Fig.~\ref{fig:imb_dra_comp}) and the deformation amplitude (Fig.~\ref{fig:imb_dra_ampl_comp}). 
Parameter values used here are detailed in Appendix \ref{sec:imbdra_param}.
The relaxation of the interface shape towards the new equilibrium as the defect is entered and exited match nicely the published data (Fig.~\ref{fig:imb_dra_comp}). As expected, the largest deviations are observed for drainage at the largest speed. There are two notable discrepancies with the CFD: in the CFD (i) the maximum amplitude is somewhat larger; and (ii) this maximum is reached earlier, while the interface still crosses the defect. 
This can be explained as follows: The spectral approach assumes $b=b_0$ (except when calculating the capillary pressure). 
Since at the defect the aperture $b_0+\delta b$ is smaller (as $\delta b<0$), 
the actual effective gravity $g_{eV}$ corresponds to a smaller aperture than that used in the spectral approach ($b_0$).
Therefore, the effective gravity in the spectral theory is larger than the actual one, hence the interface deformation is smaller.
It is also reasonable to assume that this effective aperture depends on the interface position and approaches $b_0$ when the interface approaches the end of the defect, resulting in the decrease of the amplitude. 
Nonetheless, the agreement is remarkable given the very large interface deformation, and the discrepancy is comparable to that between the CFD simulation and the experiment.


\begin{figure}
    \centering
    \includegraphics[width=0.45\linewidth]{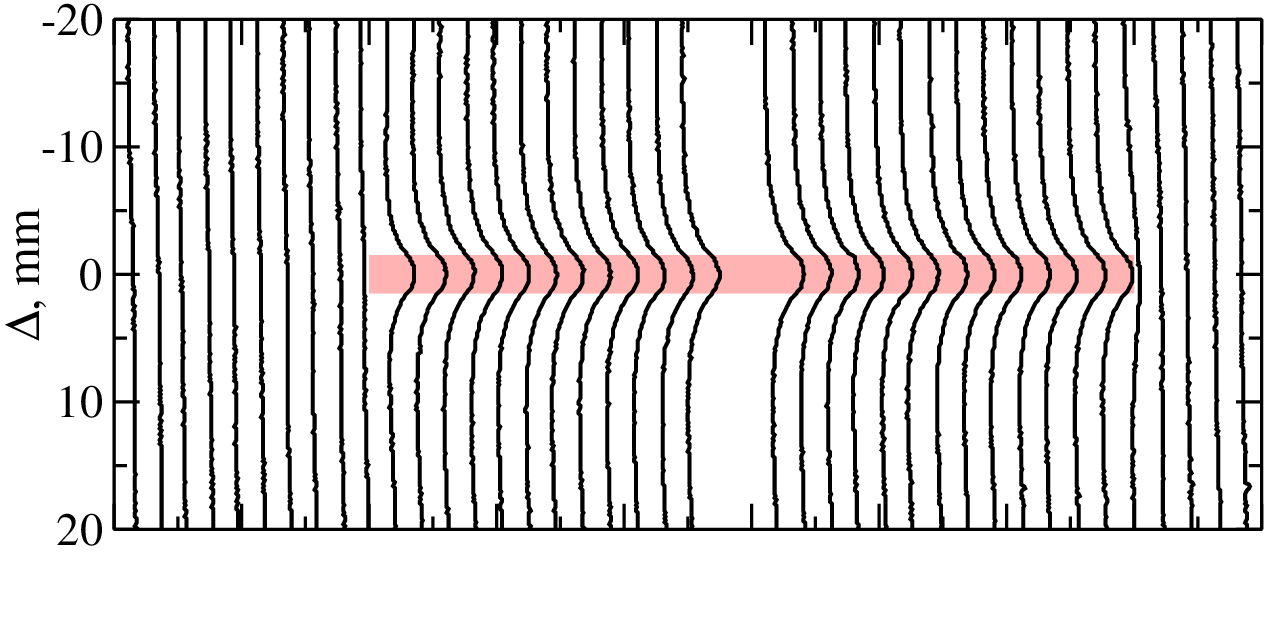}
    \includegraphics[width=0.45\linewidth]{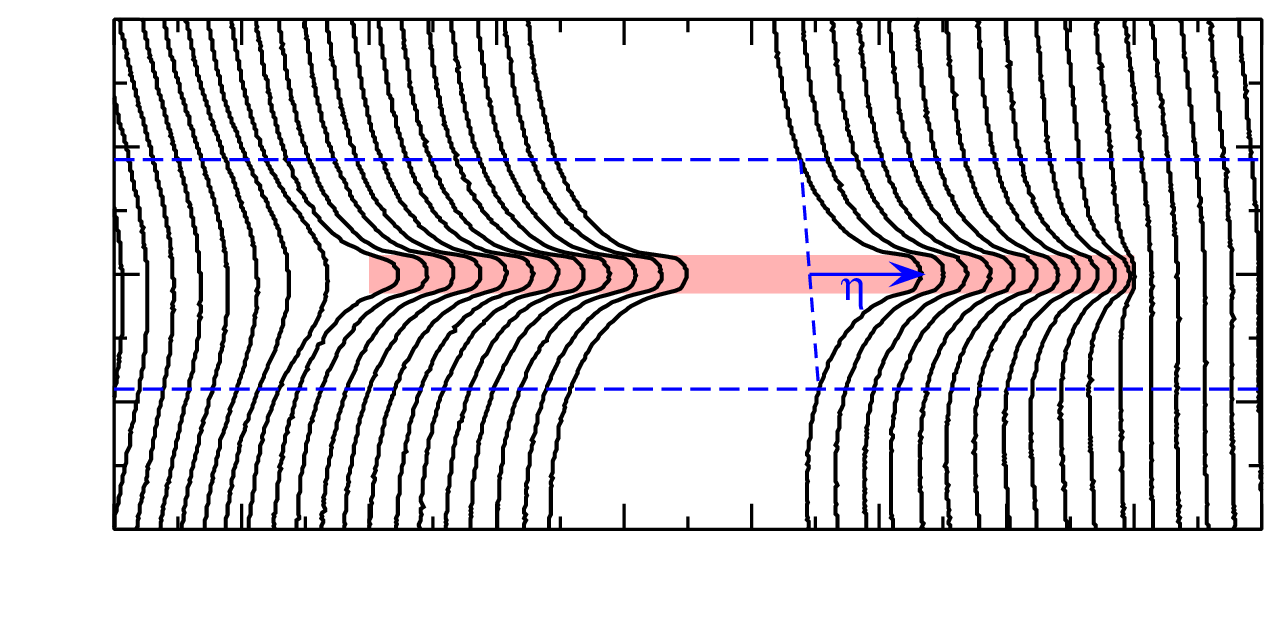}
    \includegraphics[width=0.45\linewidth]{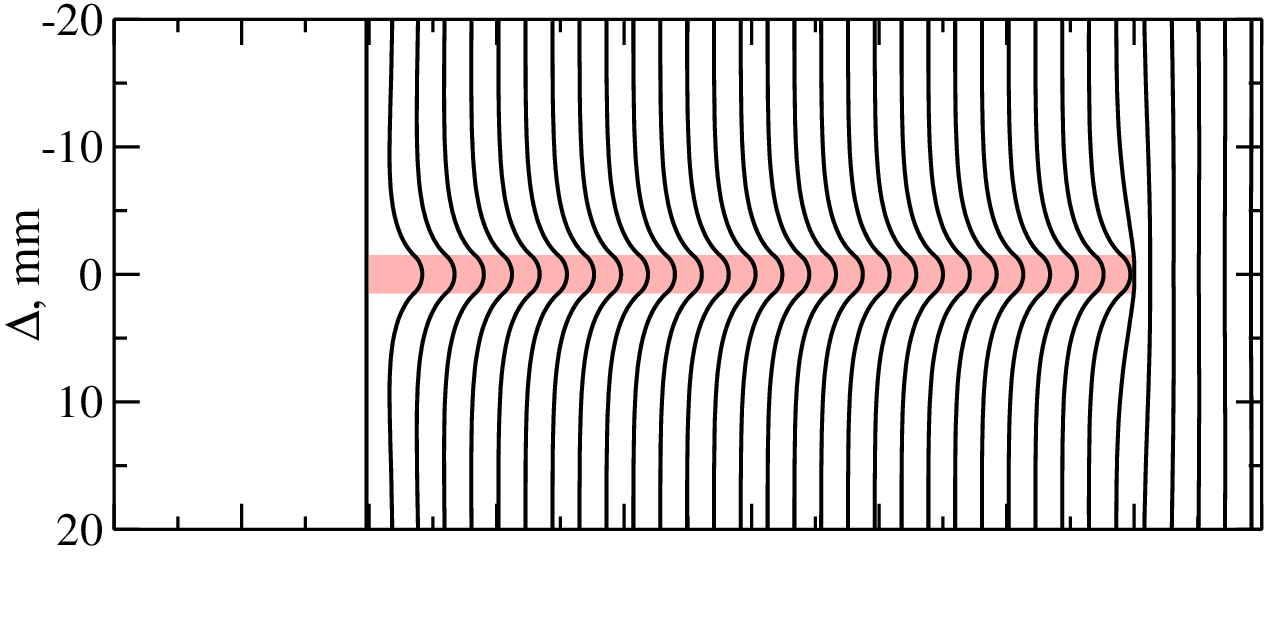}
    \includegraphics[width=0.45\linewidth]{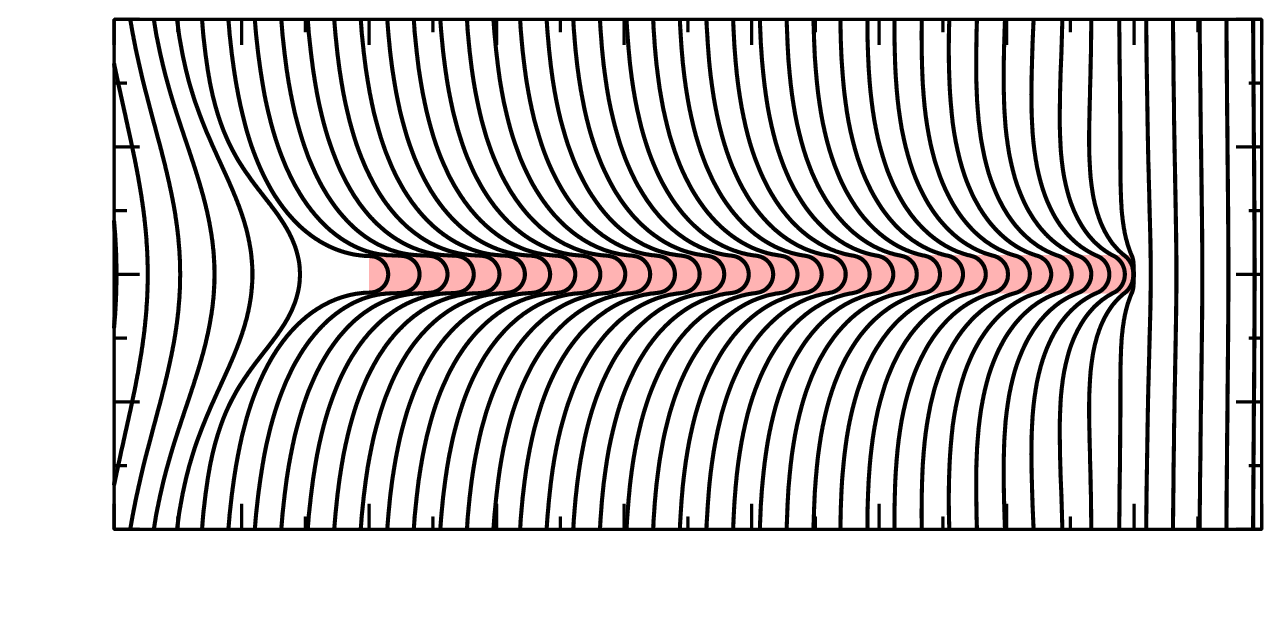}
    \includegraphics[width=0.45\linewidth]{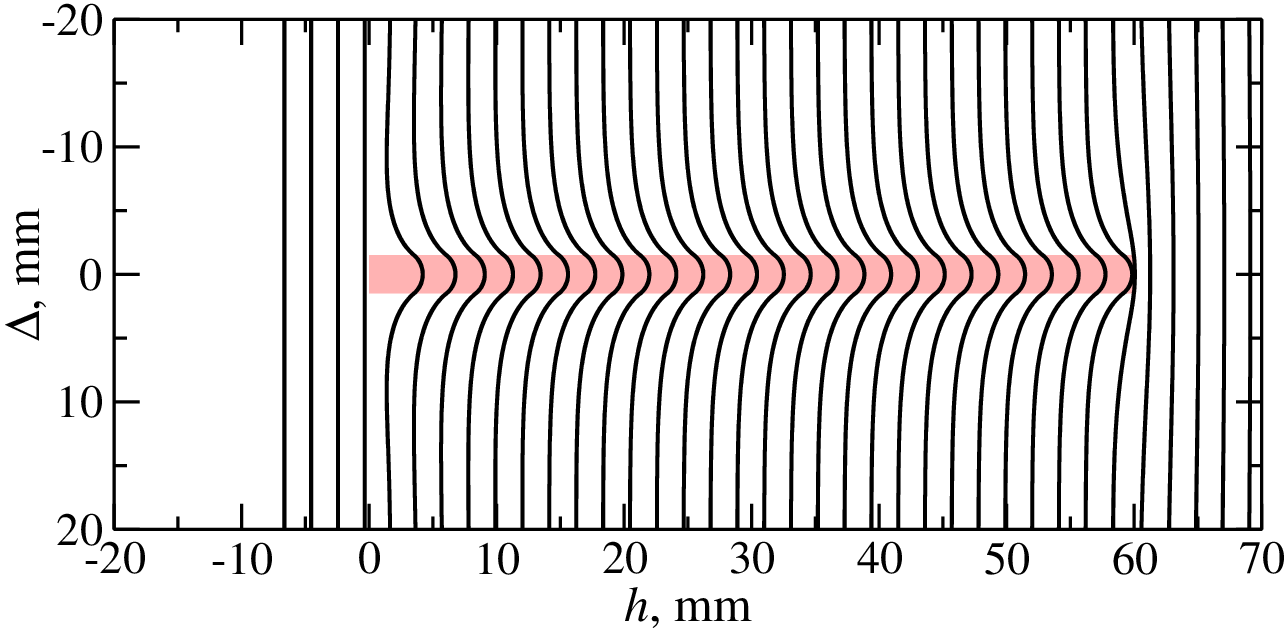}
    \includegraphics[width=0.45\linewidth]{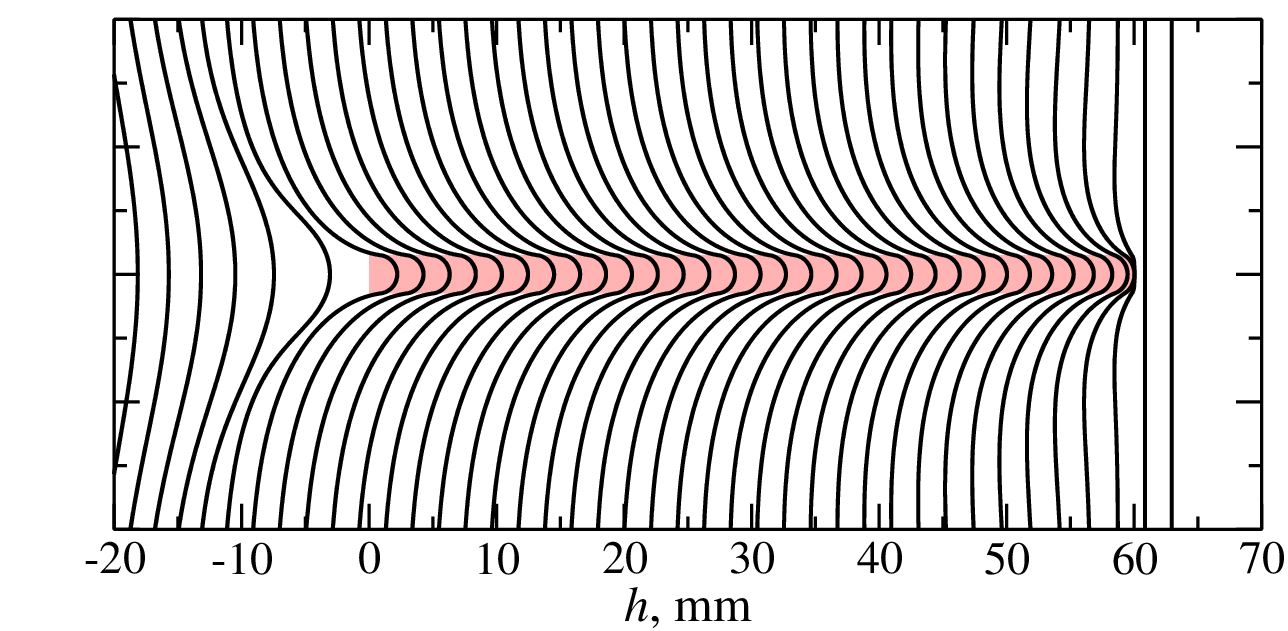}
    \caption{
    Interface evolution in experiments (top) and CFD simulations (middle) \citep{Lavi_PRF_2023} compared to our spectral simulations (bottom; see also Movies 1 and 2 in SI), for imbibition (left column) and drainage (right column). For both imbibition and drainage, for CFD and spectral simulations the inlet speed is $210~\mu$m/s; experiments were pressure-controlled with inlet pressure varying to ensure inlet speed uncertainty of less than 10\%. The time interval between the profiles is 10~s (although there are gaps in experimental data). In all plots, the inlet is far to the left of the plot. The defect is shown in red, and in the top right panel blue lines illustrate how the interface deformation (plotted in Fig.~\ref{fig:imb_dra_ampl_comp}) is calculated as the interface height on the defect axis relative to the average of the heights at $\Delta=\pm 9$~mm (horizontal dashed lines). 
    The parameters used here are listed in Appendix D.3.}
    \label{fig:imb_dra_comp}
\end{figure}

\begin{figure}
    \centering
    \includegraphics[width=0.6\linewidth]{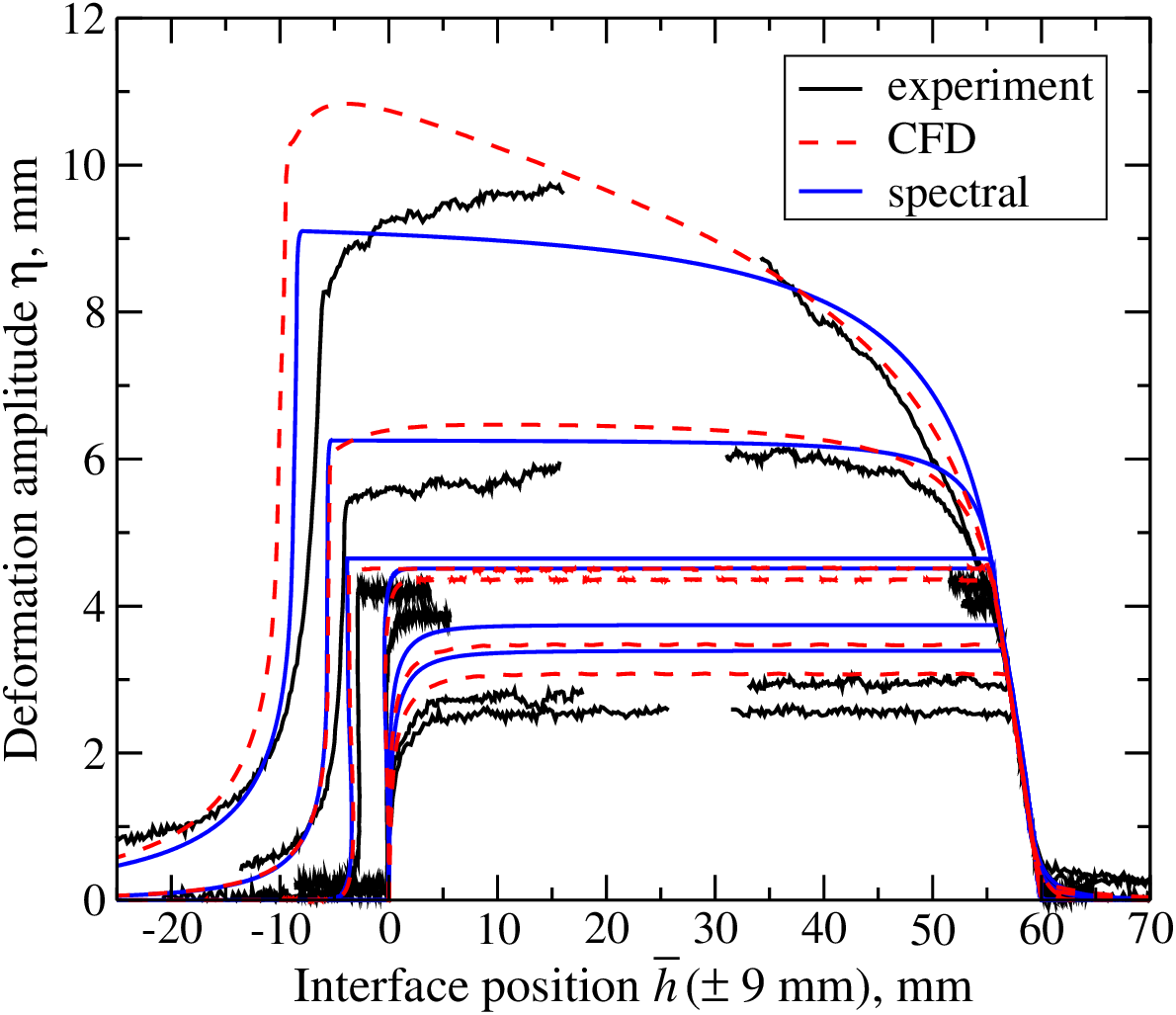}
    \caption{The interface deformation amplitude $\eta$ (the difference between the height at the defect axis and the average of the heights at distances $\pm 9$ mm from the axis; see Fig.~\ref{fig:imb_dra_comp}) as a function of the latter average, observed in experiments [black lines; from Ref.~\citet{Lavi_PRF_2023}], CFD simulations [red dashed lines; from Ref.~\citet{Lavi_PRF_2023}] and spectral simulations (blue lines). 
    Different curves correspond to different liquid speeds at the inlet: for each set of curves, top to bottom, the speeds are $-210$, $-130$, $-8$, +8, +130, +210 $\mu$m/s (``$-$'' and ``$+$'' signs correspond to drainage and imbibition, respectively; in experiments, the inlet pressure was controlled to limit uncertainty in inlet speeds to below 10\%). 
    The parameters used here are listed in Appendix D.3.
    }
    \label{fig:imb_dra_ampl_comp}
\end{figure}


\section{Dynamics of interface relaxation: Single defect}
\label{sec:relaxation}


In our validation tests (Section \ref{sec:relax_test}) we have seen that even for a single defect the evolution dynamics of the interface is rather complex {(see also Movie 3 in SI)}.
Peaks first appear near the edges and only then broaden, approaching the equilibrium shape, with appreciable (qualitative) differences between early- and late-time evolution. 
These differences are expected to be most apparent when the ratios of the relevant length scales are very large or very small, i.e., the capillary length $\ell_c$ is either much smaller or much larger than the defect width $w$, and both are much smaller than the cell width $W$. 

Here we demonstrate the power of the spectral approach by considering these cases analytically and computationally, which would be very costly with CFD. 
The single defect case is illustrative as it allows for an analytical expression (unavailable for more complicated settings). 
In Sec.\ \ref{Sec_Results_rate} we examine the dissipation rate, a useful quantity as it is a single function of time, making different temporal stages in the system behaviour particularly clear. In Sec.\ \ref{sec:Results_Interface_evolution} we show that these stages are also apparent in terms of the interface evolution.
As before, we use the simplifying assumption that the inlet is infinitely far from the interface (for pressure-control, and for the flow-rate control with $V=0$, which is equivalent); at the end of this section we comment on the more general case. 
%


\subsection{Dissipation rate}
\label{Sec_Results_rate}

Here, we show that when the ratios of different length scales are either very large or very small, the time dependence of the dissipation rate exhibits several distinct stages (for intermediate values of the ratios, these stages become less distinct). 
The dissipation rate [Eq.~(\ref{eq:Rmesa})] is a sum of terms with a time-independent prefactor $\sin^2(k_{2n}w/2)/k_{2n}$ and an exponential time-dependent factor.  
At $t=0$ all the exponential factors are unity, and the dissipation rate is proportional to the infinite sum of all the prefactors. The exponential factors decay with time, \emph{at different rates}. 

We first examine very large times, $t\gg t_e$, where only the term with the slowest decay rate ($n=1$) remains important; the relaxation rate then decays exponentially. The ``exponential time'' $t_e$ satisfies
\begin{equation}
    \beta k_2(1+\ell_c^2 k_2^2)t_e\approx \beta k_2 t_e\sim \frac{\beta t_e}{W}\sim 1,
\end{equation}
thus,
\begin{equation}
    t_e\sim\frac{W}{\beta}.
\end{equation}

For earlier times $t\ll t_e$, many terms significantly contribute to the sum in Eq.~(\ref{eq:Rmesa}), which we approximate by
\begin{equation}
\sum_{n=1}^{\infty}\ldots\approx \frac{W}{2\pi}\int_0^{\infty}dk\, \ldots
\end{equation}
Thus, 
\begin{equation}
    |R(t)|\approx\frac{4b_0\beta\rho g_e \mathcal{H}^2}{\pi}\int_0^{\infty}\frac{\sin^2(wk/2)}{k}\exp[-2\beta k(1+\ell_c^2 k^2)t]\, dk.
\end{equation}
At a given time $t$, the exponential factor is small for sufficiently large $k$ and around one for small $k$.
The cutoff value $k_{\rm cut}$ at which the exponential becomes small satisfies
\begin{equation}
    \beta k_{\rm cut}(1+\ell_c^2 k_{\rm cut}^2)t\sim 1.
\end{equation}
If $k_{\rm cut}\ll\ell_c^{-1}$, then the exponent can be approximated as $-2\beta kt$; for all other $k$ values, for which the approximation is not valid, the exponential is small anyway and thus the error in its value has minor effect. 
On the other hand, if $k_{\rm cut}\gg\ell_c^{-1}$, then we can approximate the exponent as $-2\beta\ell_c^2 k^3 t$; noting that while this is not valid for small $k$, the exponential is approximately unity regardless. The rate becomes
\begin{equation}
    |R(t)|\approx
    \begin{cases}
     \displaystyle   \frac{4b_0\beta\rho g_e \mathcal{H}^2}{\pi}\int_0^{\infty}\frac{\sin^2(wk/2)}{k}\exp[-2\beta \ell_c^2k^3t]\, dk, & t\ll t_c,\\
      \displaystyle   \frac{4b_0\beta\rho g_e \mathcal{H}^2}{\pi}\int_0^{\infty}\frac{\sin^2(wk/2)}{k}\exp[-2\beta kt]\, dk, & t\gg t_c,
    \end{cases}
\end{equation}
where we define a ``capillary crossover time'' as
\begin{equation}
    t_c\sim\frac{\ell_c}{\beta}.
    \label{eq:tc}
\end{equation}
Another crossover occurs at time $wk_{\rm cut} \sim 1$.
For $wk_{\rm cut}\gg 1$, the term $\sin^2(wk/2)$ oscillates many times 
and can be replaced with its average 1/2, except in the vicinity of zero where that would cause the integral to diverge. The latter can be resolved by replacing the lower integration limit with $\sim$$w^{-1}$.
For $wk_{\rm cut}\ll 1$, we get $\sin^2(wk/2)\approx (wk/2)^2$.
This defines a ``width crossover time''
\begin{equation}
    t_w\sim\frac{w^3}{\beta(w^2+\ell_c^2)}.
\end{equation}
For a narrow defect ($w\ll\ell_c$), 
\begin{equation}
    t_w\sim\frac{w^3}{\beta\ell_c^2},
    \label{eq:twnarrow}
\end{equation}
whereas for a wide defect ($w\gg\ell_c$),
\begin{equation}
    t_w\sim\frac{w}{\beta}.
\end{equation}

The three crossover times, $t_c$, $t_w$ and $t_e$, can be used to define five stages, I--V:
\begin{itemize}
    \item \textbf{Stage I}: $t\ll t_c,t_w\ll t_e$, and
    \begin{equation}
        |R(t)|\approx\frac{2b_0\beta\rho g_e \mathcal{H}^2}{\pi}\int_{w^{-1}}^{\infty}\frac{\exp(-2\beta \ell_c^2k^3t)}{k}\, dk\approx \frac{2b_0\beta\rho g_e \mathcal{H}^2}{3\pi}\left[\ln\frac{w^3}{\beta\ell_c^2 t}+O(1)\right],
        \label{eq:RI}
    \end{equation}
    \item \textbf{Stage II}: $t_c\ll t\ll t_w\ll t_e$, and
    \begin{equation}     |R(t)|\approx\frac{2b_0\beta\rho g_e \mathcal{H}^2}{\pi}\int_{w^{-1}}^{\infty}\frac{\exp(-2\beta kt)}{k}\, dk\approx \frac{2b_0\beta\rho g_e \mathcal{H}^2}{\pi}\left[\ln\frac{w}{\beta t}+O(1)\right],
        \label{eq:RII}
    \end{equation}
    \item \textbf{Stage III}: $t_w\ll t\ll t_c\ll t_e$,
    and
    \begin{equation}
        |R(t)|\approx \frac{b_0\beta\rho g_e \mathcal{H}^2w^2}{\pi}\int_0^{\infty} k\exp(-2\beta\ell_c^2 k^3t)\, dk=\frac{b_0(2\beta)^{1/3}\Gamma(2/3)\rho g_e \mathcal{H}^2 w^2}{6\pi\ell_c^{4/3}t^{2/3}},
        \label{eq:RIII}
    \end{equation}
\noindent    where $\Gamma$ is the gamma function; 
    \item \textbf{Stage IV}: $t_c,t_w\ll t\ll t_e$, and
    \begin{equation}
        |R(t)|\approx \frac{b_0\beta\rho g_e \mathcal{H}^2w^2}{\pi}\int_0^{\infty} k\exp(-2\beta kt)\, dk=\frac{b_0\rho g_e \mathcal{H}^2 w^2}{4\pi\beta t^2},
        \label{eq:RIV}
    \end{equation}
    \item \textbf{Stage V}: $t_c,t_w\ll t_e\ll t$, and
    \begin{equation}
        |R(t)|\approx \frac{4\pi b_0\beta\rho g_e \mathcal{H}^2 w^2}{W^2}\exp\left(-\frac{4\pi\beta}{W}t\right),
        \label{eq:RV}
    \end{equation}
\noindent    obtained by retaining only the $n=1$ term (with $k_2=2\pi/W$) in Eq.~(\ref{eq:Rmesa}).
\end{itemize}

Note that for a given system, only one of the two stages II and III is possible: for a narrow defect ($w\ll\ell_c$), $t_w\ll t_c$, and the sequence of the stages in time is $\mathrm{I}\to\mathrm{III}\to\mathrm{IV}\to\mathrm{V}$, while for a wide defect ($w\gg\ell_c$), $t_w\gg t_c$ and the sequence is $\mathrm{I}\to\mathrm{II}\to\mathrm{IV}\to\mathrm{V}$. 
In summary, the dissipation rate in stages I and II shows a logarithmic time dependence, with two different prefactors; a power law dependence in stages III and IV, with different exponents; and an exponential decay in stage V.

Our numerical calculations of the dissipation rate for different cases (in terms of defect and cell widths) and the consequent stages are shown in Fig.~\ref{fig:dissip_regimes}.
As expected from Eqs.~(\ref{eq:RIII})--(\ref{eq:RV}), the dimensionless dissipation rate
\begin{equation}
    \tilde{R}=\frac{\ell_c^2}{\beta\rho g_e b_0 w^2 \mathcal{H}^2}|R|
\end{equation}
as a function of the dimensionless time $\hat{t}$ [Eq.~(\ref{eq:t-hat})] is universal in stages III and IV, and only varies with $W/\ell_c$ in stage V.
The expected weak time-dependence for small $t$ (consistent with the predicted logarithmic dependence), and exponential cutoffs at large $t$, are also evident. 
For intermediate times, either one or two power-law stages are observed, depending on $w/\ell_c$. 
It is seen that the existence of a clear power law in stage III requires $w$ to be at least an order of magnitude smaller than $\ell_c$. 
Similarly, the power law associated with stage IV requires $W$ to exceed both $w$ and $\ell_c$ by at least two orders of magnitude. 

\begin{figure}
    \centering
    \includegraphics[width=0.6\linewidth]{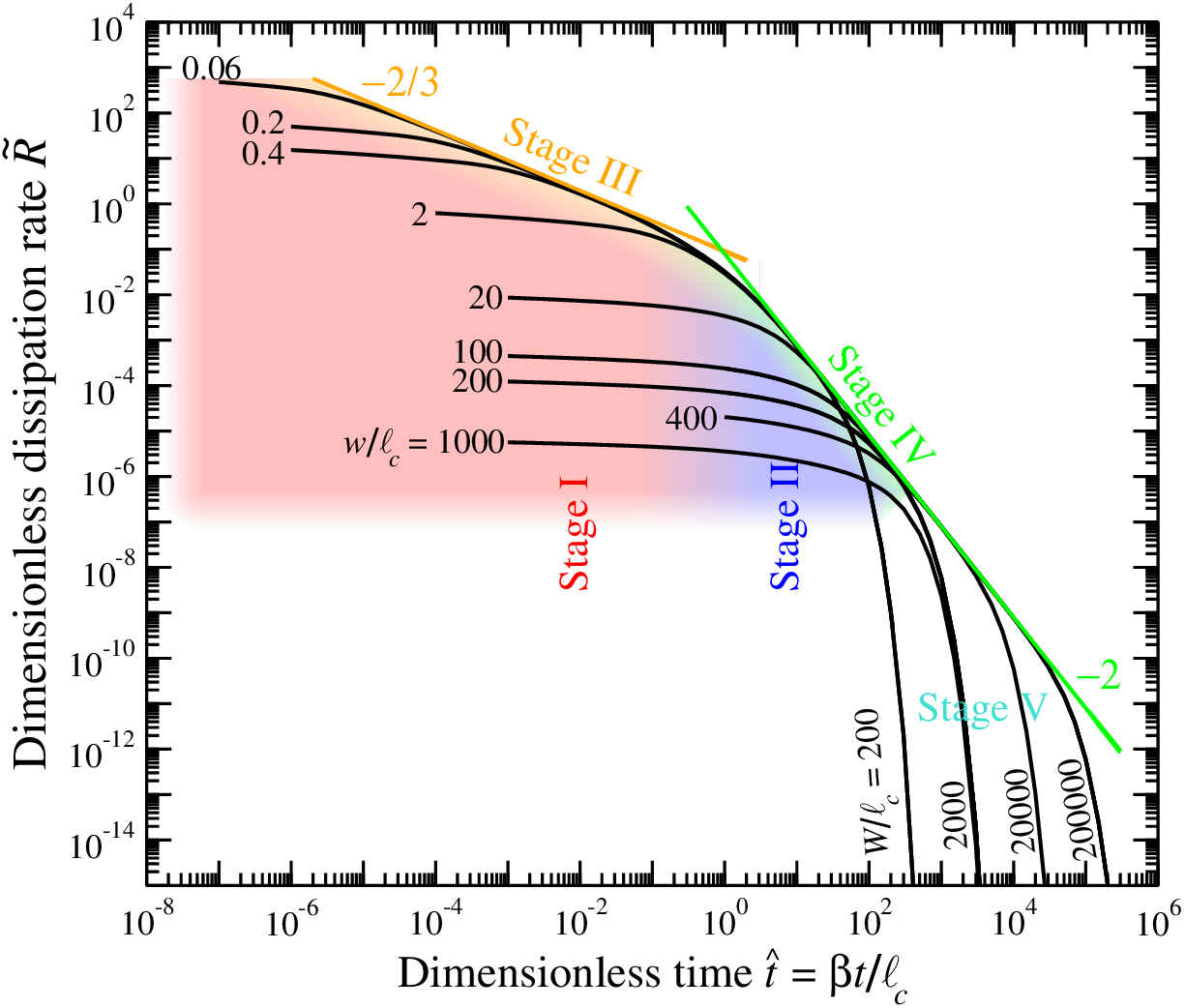}
    \caption{Dimensionless dissipation rates as a function of time for Hele-Shaw cells with a single infinitely long mesa defect, for different defect widths $w$ and cell widths $W$ [computed with the spectral approach, Eq.~(\ref{eq:Rmesa})]. The power laws correspond to stages III and IV (as described in the text), and the black lines are Eqs.~(\ref{eq:RIII}) and (\ref{eq:RIV}) (with the correct prefactor, as well as exponent). The sharp exponential drops on the right side are stage V, and the nearly flat parts of the curves on the left correspond to the logarithmic dependences of stages I and II (the crossover between these stages at $\hat{t}\sim 1$ is subtle and not immediately apparent). 
    }
    \label{fig:dissip_regimes}
\end{figure}

\subsection{Interface evolution}
\label{sec:Results_Interface_evolution}
Here, we show that the interface dynamics exhibits the same temporal stages as the dissipation rate. 
We quantify this using the interface height $h$,  
based on the explicit expression, Eq.~(\ref{eq:hmesa}). 
In what follows we will ignore the $h^{(0)}$ term in that equation, thus measuring the interface height $h$ from its initial position rather than the inlet. 

Motivated by the findings for the dissipation rate in Sec.\ \ref{Sec_Results_rate}, we first examine the cases of a wide ($w/\ell_c=200$) and a narrow defect ($w/\ell_c=0.2$), in  Fig.~\ref{fig:profiles}.
The value of $w/\ell_c$ is quite close to unity for the narrow defect, which is acceptable, since $t_c/t_w\sim (\ell_c/w)^3$ in that case [see Eqs.~(\ref{eq:tc}) and (\ref{eq:twnarrow})], thus $t_c/t_w$ is still large. 
The values of $W$ for Figs.~\ref{fig:profiles}--\ref{fig:hmaxmin} 
are chosen to be large enough such that finite cell width effects are negligible, and stage V is delayed until very large $t$ and thus can be ignored. 
It is seen clearly that the interface shape changes qualitatively (highlighted by different colours in Fig.~\ref{fig:profiles}) between the different stages, as identified for the dissipation rate.
{For the calculations we use a defect which is infinite in $y$ in both directions; as discussed in Sec.~\ref{sec:relax_test}, this is equivalent to the physical situation of an interface approaching a defect which starts at $y=0$, as in Fig.~\ref{fig:profiles}.}

At early times, the perturbation of the interface is concentrated near the defect edges and is antisymmetric with respect to the edge. 
Indeed, the rate of change of $h$ at $t=0$, which can be derived analytically (see Appendix~\ref{sec:mesa_details_general}),
\begin{equation}
    \left.\dot{h}\right|_{t\to 0}=\frac{\beta \mathcal{H}}{\pi}\left(\frac{1}{\Delta+w/2}-\frac{1}{\Delta-w/2}\right),
    \label{eq:hdottzero}
\end{equation}
diverges at the defect edges. Nevertheless, this rate is nonzero everywhere,
despite the fact that pressure imbalance is only present at the edges. This emphasises nonlocality of the process of interface deformation in response to pressure imbalance and is in contrast to solutions of any local PDE approximations, as discussed in the SI.

The initial antisymmetry of interface deformation with respect to the defect edge is later lost, as the interface relaxes towards the equilibrium shape, which for $W\to\infty$ is
\begin{equation}
    h_{\rm eq}(\Delta)=
    \begin{cases}
       \displaystyle \mathcal{H}\left[1-\exp\left(-\frac{w}{2\ell_c}\right)\cosh\left(\frac{\Delta}{\ell_c}\right)\right], & |\Delta|<\frac{w}{2},\\
       \displaystyle \mathcal{H}\sinh\left(\frac{w}{2\ell_c}\right)\exp\left(-\frac{|\Delta|}{\ell_c}\right), & |\Delta|>\frac{w}{2}.
    \end{cases}
    \label{eq:heq}
\end{equation}
The quantities that reflect this behaviour are thus 
the time dependences of the interface deformation amplitude $h_{\rm max}-h_{\rm min}$ and of a measure of antisymmetry breaking, $h_{\rm max}+h_{\rm min}$. Here $h_{\rm max}$ and $h_{\rm min}$ are the heights of the maximum and the minimum of the interface profile at a given time, respectively.

\begin{figure}
    \centering
    \includegraphics[width=0.45\linewidth]{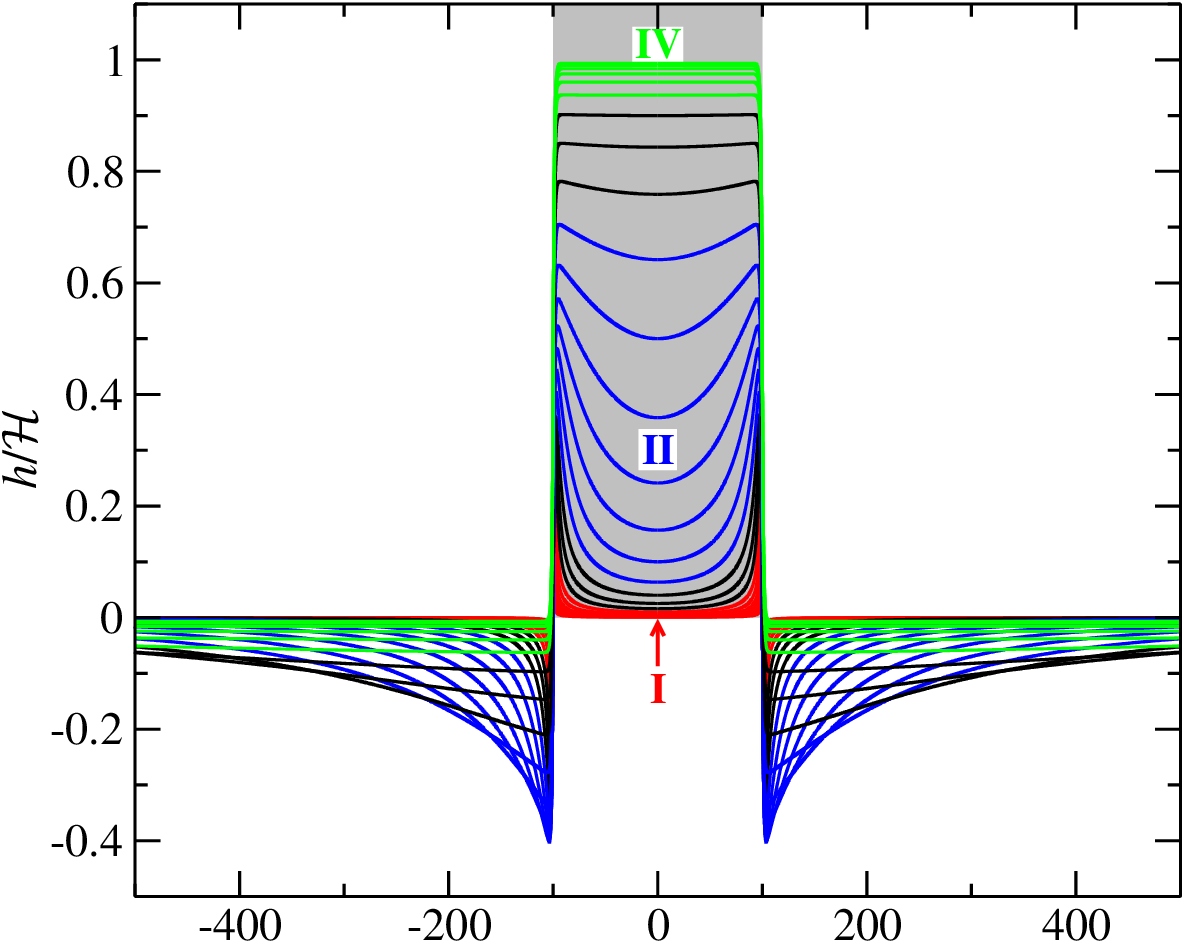}
    \includegraphics[width=0.45\linewidth]{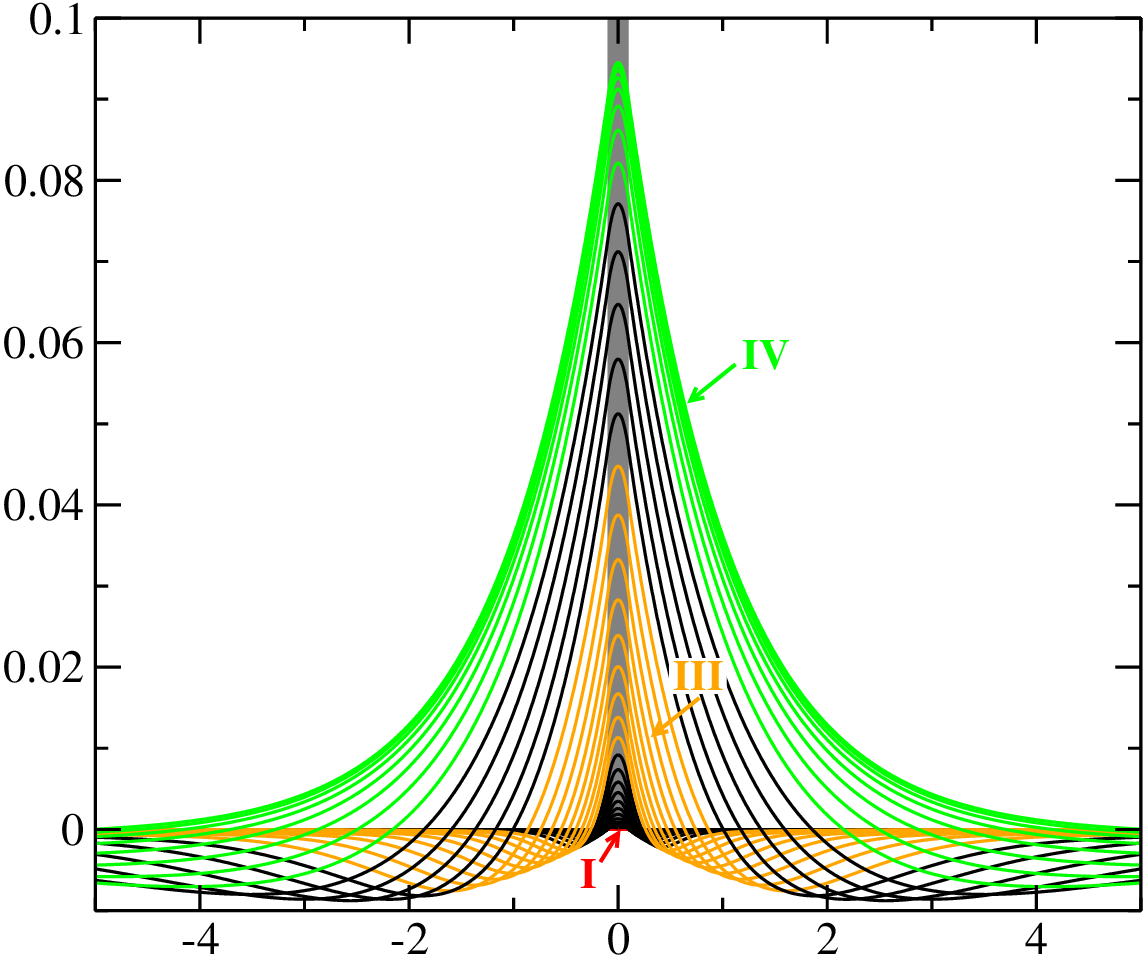}
    \includegraphics[width=0.45\linewidth]{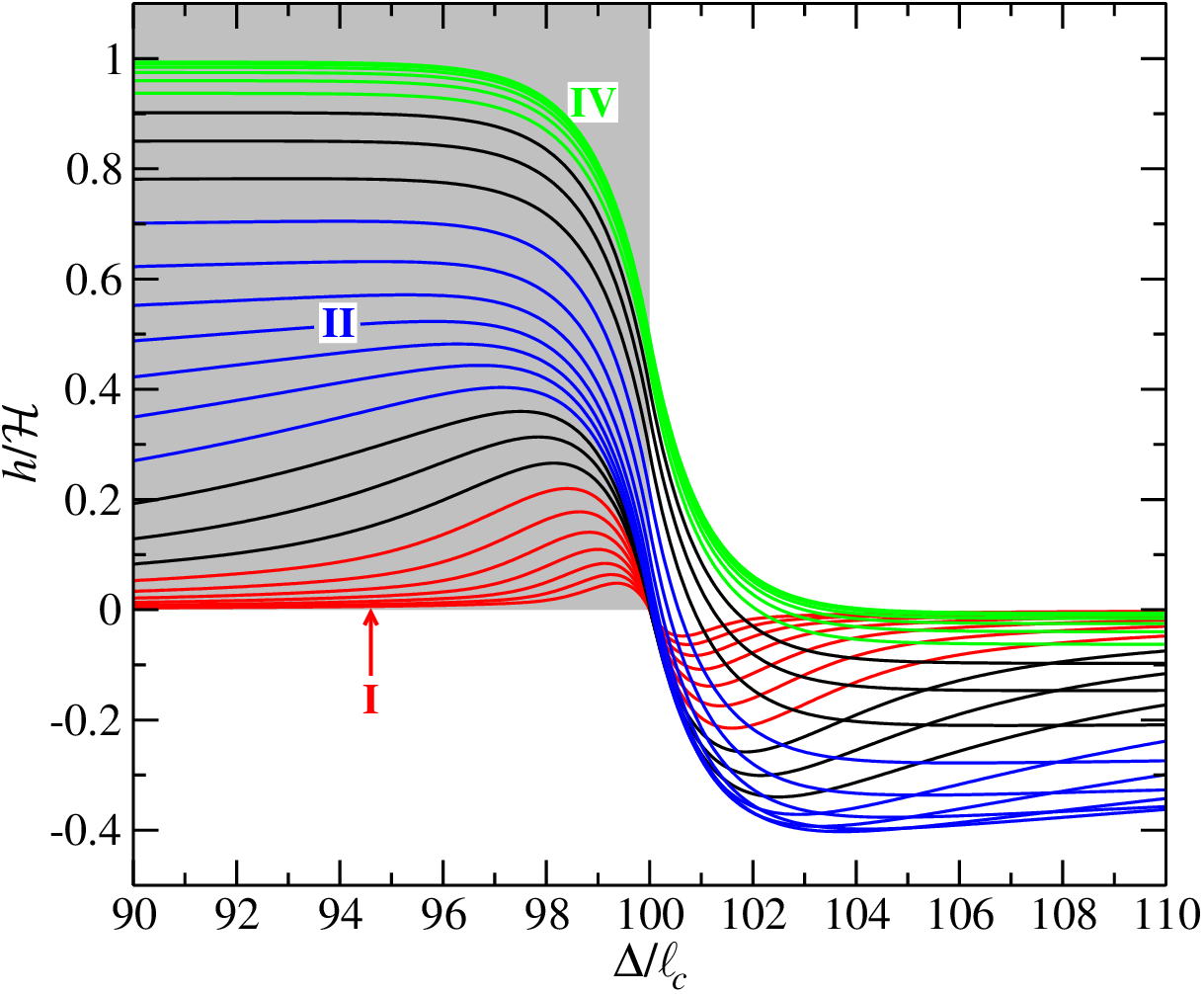}
    \includegraphics[width=0.45\linewidth]{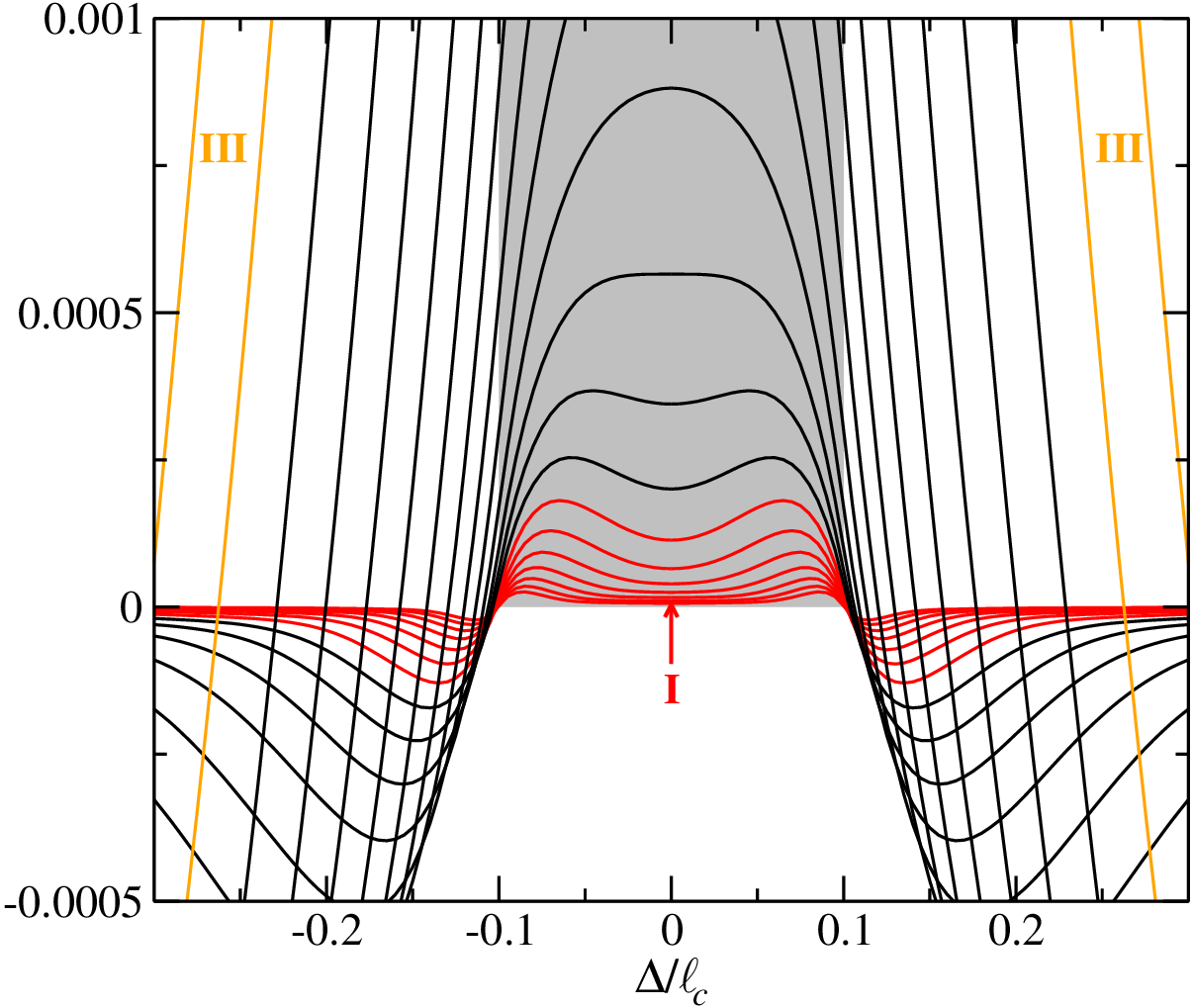}
    \caption{The evolution of the interface shape for Hele-Shaw cells with a single infinitely long mesa defect, for two defect widths, $w/\ell_c=200$ (left column) and $w/\ell_c=0.2$ (right column) [computed with the spectral approach, Eq.~(\ref{eq:hmesa})].
    The bottom row shows a zoom-in.
    The profiles are equally spaced in time on the logarithmic scale, with five profiles per decade, starting with $\hat{t}=0.1$ for the wide defect and $\hat{t}=10^{-6}$ for the narrow defect [the dimensionless time $\hat{t}$ is given by Eq.~(\ref{eq:t-hat})]. The profiles are coloured according to the stage that they belong to (stage I in red; II in blue; III in orange; IV in green), with those corresponding to the crossovers between the stages in black (noting that the \emph{exact} locations of the boundaries between stages are somewhat arbitrary, as the transitions are continuous). 
    Cell widths $W$ are chosen large enough that there is no visible deviation from the $W\to\infty$ limit.
    The defects are shown here (grey) as semi-infinite, while taken as extending infinitely in $y$ in the calculations (see text for explanation).
For visualisation of the dynamics see Movie 3 in SI (shown for an intermediate case of $W/l_c=20$).}
 \label{fig:profiles}
\end{figure}

Similarly to how different stages in the evolution of the dissipation rate were derived above, we can do the same for the evolution of the interface shape by using the appropriate approximations of Eq.~(\ref{eq:hmesa}). 
Here we provide a brief summary of the results; the derivation details are given in Appendix~\ref{sec:mesa_details}.
First, in \textbf{stages I and III} we find the rescaled height
\begin{equation} \tilde{h}=\frac{\ell_c^2 h}{\mathcal{H} w^2}=\tilde{h}(\tilde{\Delta},\tilde{t}),
    \label{eq:htilde}
\end{equation}
where
\begin{equation}
\tilde{t}=\frac{\beta\ell_c^2 t}{w^3}\label{eq:ttilde}
\end{equation}
[note that this is different from the dimensionless time $\hat{t}$ given by Eq.~(\ref{eq:t-hat}) and used in Figs.~\ref{fig:relax_compar}, \ref{fig:dissip_regimes} and \ref{fig:profiles}] and
\begin{equation}
\tilde{\Delta}=\frac{\Delta}{w}.
\label{eq:Deltatilde}
\end{equation}
Based on the above, we plot $\tilde{h}_{\rm max}-\tilde{h}_{\rm min}$ and $\tilde{h}_{\rm max}+\tilde{h}_{\rm min}$ vs. $\tilde{t}$ for different $w/\ell_c$ in Fig.~\ref{fig:hmaxmin}, showing the collapse of the data onto master curves (stages I and III) and the eventual deviations from these curves.


\begin{figure}
    \centering
    \includegraphics[width=0.6\linewidth]{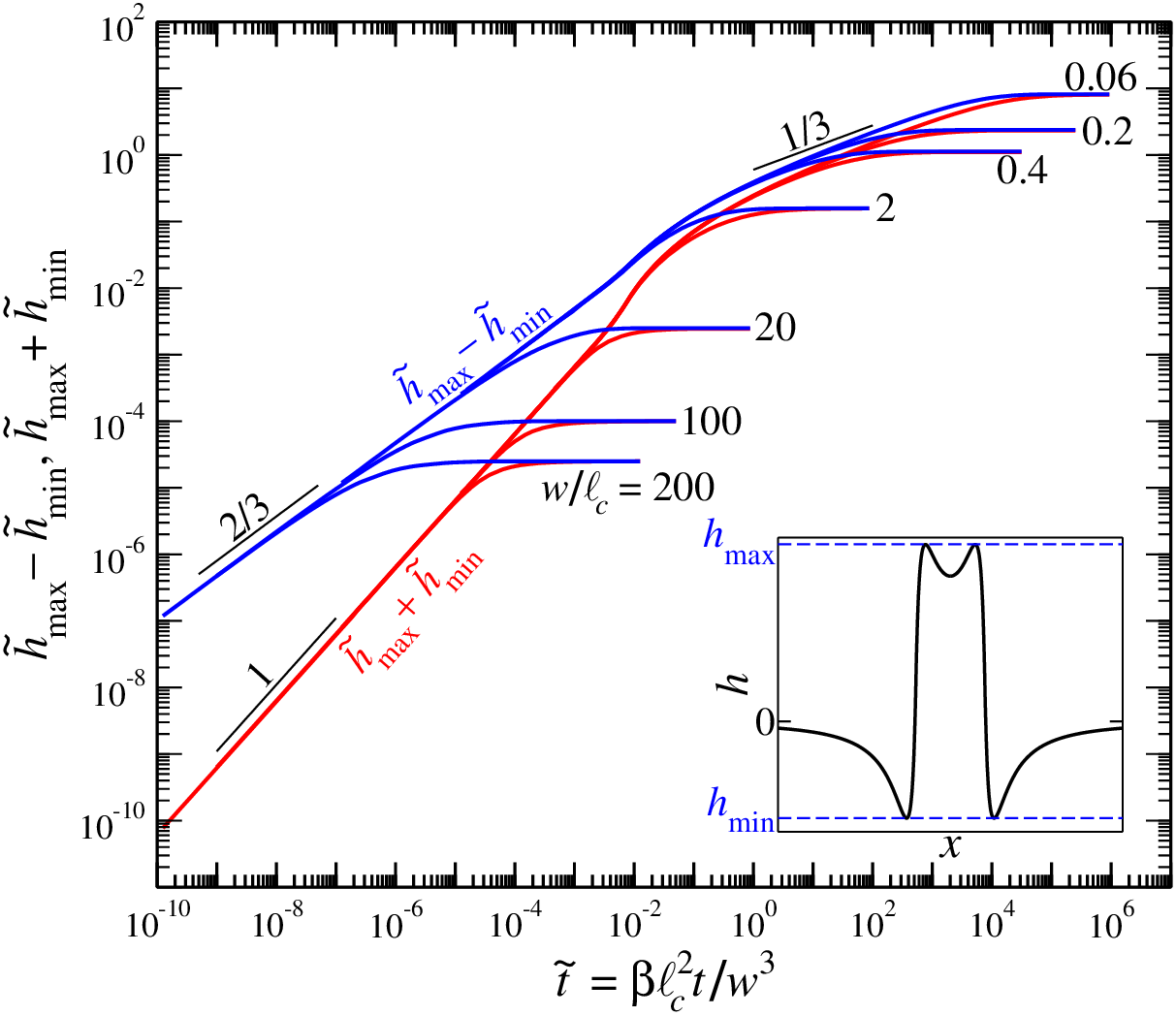}
    \caption{Functionals of the interface deformation, (i) amplitude $h_{\rm max}-h_{\rm min}$ and (ii) mean $h_{\rm max}+h_{\rm min}$, both rescaled [see Eq.~(\ref{eq:htilde})], as a function of the rescaled time $\tilde{t}$ [Eq.~(\ref{eq:ttilde})] for cells with an infinitely long mesa defect 
    [computed with the spectral approach, Eq.~(\ref{eq:hmesa})].
    The plot is for different defect widths $w$ and very large cell widths $W$. In stages I and III all the data collapse onto master curves. The stages are: I (exponents 2/3 and 1), III (both exponents are 1/3), II (the amplitude approaches saturation while the mean remains linear in $t$), IV and V (both quantities approach saturation).
    }
    \label{fig:hmaxmin}
\end{figure}

Further, we find that in \textbf{stage I} near the edge of the defect at $\Delta=w/2$
\begin{equation}
    h\sim \frac{\mathcal{H}\beta^{2/3}t^{2/3}}{\ell_c^{2/3}}f_1\left(\frac{w/2-\Delta}{(\beta\ell_c^2 t)^{1/3}}\right),
    \label{eq:hscalI}
\end{equation}
where $f_1$ is an odd function of its argument. This shows that there is an invariant shape that gets broader with time as $t^{1/3}$ and higher as $t^{2/3}$. The positions of the maximum and the minimum near the edge at $\Delta=w/2$ are given by
\begin{equation}
    w/2-\Delta_{\rm max}\approx\Delta_{\rm min}-w/2\sim(\beta\ell_c^2t)^{1/3}
    \label{eq:maxI}
\end{equation}
and the heights
\begin{equation}
    h_{\rm max}\approx -h_{\rm min}\sim\frac{\beta^{2/3}t^{2/3}}{\ell_c^{2/3}}.\label{eq:hmaxI}
\end{equation}
A small amount of asymmetry (not accounted for in the equations above) grows linearly in time. These findings are corroborated by numerical results in Fig.~\ref{fig:hmaxmin}, where at early times the perturbation amplitude $h_{\rm max}-h_{\rm min}$ indeed grows as $t^{2/3}$ and $h_{\rm max}+h_{\rm min}$ is linear in $t$, and by the shapes of the profiles (red curves) in Fig.~\ref{fig:profiles}. For narrow defects, stage I terminates at $t \sim t_w$, at which point the maximum reaches the centre of the defect; antisymmetry is, obviously, lost. This time is very short and by its end the maximum interface height is only around $(w^2/\ell_c^2)\mathcal{H}$, considerably below the equilibrium value [of the order of $(w/\ell_c)\mathcal{H}$].  On the other hand, for wide defects this stage terminates at $t\sim t_c$, at which point the perturbation amplitude $h_{\rm max}-h_{\rm min}$ is already on the order of $\mathcal{H}$ (the equilibrium value), but the perturbation is still approximately antisymmetric and confined to within $\ell_c$ ($\ll w$) of the edge.

In \textbf{stage III} (only observed for narrow defects), we get
\begin{equation}
    h\sim \frac{\mathcal{H} w\beta^{1/3}t^{1/3}}{\ell_c^{4/3}}f_2\left(\frac{\Delta}{(\beta\ell_c^2 t)^{1/3}}\right),
    \label{eq:hIII}
\end{equation}
{where $f_2$ is an even function.} We again get an invariant shape that stretches horizontally and vertically with time, but the axis around which horizontal stretching occurs is now at the centre of the defect, rather than its edge, and the vertical stretching has a different power law associated with it (power 1/3 instead of 2/3). This shape invariance and stretching are seen in Fig.~\ref{fig:profiles} (orange lines). Since the perturbation loses its antisymmetry by the end of stage I, both the sum and the difference of $h_{\rm max}$ and $h_{\rm min}$ should evolve with the same power law exponent 1/3, as is indeed observed in Fig.~\ref{fig:hmaxmin}. By the end of stage III at time $t\sim t_c\sim\ell_c/\beta$, $h_{\rm max}\sim \mathcal{H} w/\ell_c$ is of the same order as the final equilibrium height; only relatively minor final adjustment occurs in stages IV and V.
We also note the very broad transition region between stages I and III for narrow defects in Fig.~\ref{fig:profiles}. The deviations from the power-law behaviours in $h_{\rm max}\pm h_{\rm min}$ start to be significant more than two orders of magnitude in time before the already very short estimate of Eq.~(\ref{eq:twnarrow}) for $t_w$ with prefactor 1. This makes stage I difficult to observe for narrow defects.

For \textbf{stages II and IV}, we get a closed-form expression,
\begin{equation}
    h(\Delta,t)
    =
    \begin{cases}
      \displaystyle  \frac{\mathcal{H}}{\pi}\left(\arctan\frac{\beta t}{w/2+\Delta}+\arctan\frac{\beta t}{w/2-\Delta}\right)-\mathcal{H}\exp\left(-\frac{w}{2\ell_c}\right)\cosh\left(\frac{\Delta}{\ell_c}\right), & |\Delta|<\frac{w}{2},\\
       \displaystyle  \frac{\mathcal{H}}{\pi}\left(\arctan\frac{\beta t}{w/2+\Delta}+\arctan\frac{\beta t}{w/2-\Delta}\right)+\mathcal{H}\sinh\left(\frac{w}{2\ell_c}\right)\exp\left(-\frac{|\Delta|}{\ell_c}\right), & |\Delta|>\frac{w}{2}.
    \end{cases}
    \label{eq:hII_IV}
\end{equation}
which provides excellent fits ({Fig.~\ref{fig:profiles_vs_theory},} Appendix~\ref{sec:mesa_details}).
It is particularly remarkable that this expression is valid in stage II, where the shape is still far from equilibrium. Moreover, for wide defects this expression is also valid in stage I far enough from the defect edges and, in particular,
\begin{equation}
    h(\Delta=0,t)=\frac{2 \mathcal{H}}{\pi}\arctan\frac{2\beta t}{w},\ w\gg\ell_c,
    \label{eq:h0wide}
\end{equation}
is accurate at all times (it misses the crossover to exponential relaxation in stage V, but by that time $h(0,t)$ is already extremely close to its equilibrium value, $\mathcal{H}$). In {stage II}
\begin{equation}
    \Delta_{\rm min,max}\approx\frac{w}{2}\pm \ell_c\ln\frac{\pi\beta t}{\ell_c},
    \label{eq:maxII}
\end{equation}
a logarithmic time dependence even weaker than the $t^{1/3}$ dependence in stage I (although the two may be hard to distinguish numerically except for very large $w/\ell_c$), and
\begin{eqnarray}
    h_{\rm max}-h_{\rm min}&\approx&\mathcal{H}\left[1-\frac{\ell_c}{\pi\beta t}\left(1+2\ln\frac{\pi\beta t}{\ell_c}\right)\right],\label{eq:hamplII}\\
    h_{\rm max}+h_{\rm min}&\approx&\frac{2\beta \mathcal{H}}{\pi w}t.
    \label{eq:hsumII}
\end{eqnarray}
Thus, the interface perturbation amplitude, $h_{\rm max}-h_{\rm min}$, which was already $O(\mathcal{H})$ by the start of stage II, is now relaxing towards its equilibrium value $\mathcal{H}$, approaching it roughly as $1/t$ (aside from a logarithmic factor). At the same time, the mean of the maximum and minimum heights is small at the start of the stage and increases linearly, becoming $O(\mathcal{H})$ towards the end of the stage. Both the saturation of $h_{\rm max}-h_{\rm  min}$ and the linear dependence of $h_{\rm max}+h_{\rm min}$ are confirmed numerically (see Fig.~\ref{fig:hmaxmin}); in fact, there is no change in slope or any other visible peculiarity in the latter quantity at the boundary of stages I and II, similar to the behaviour of $h(\Delta=0)$. 
Note that while the interface peak and dip remain close to the defect edge, the width of the interface's tails grows linearly in time, with the inner tail reaching the centre of the defect by the end of the stage. 
Note also that by subtracting Eq.~(\ref{eq:hsumII}) from Eq.~(\ref{eq:hamplII}) we can see that $|h_{\rm min}|$ exhibits a maximum, i.e., the dip outside the defect first gets deeper and then shallower. The maximum is at $t\sim(w\ell_c)^{1/2}/\beta$ (aside from a logarithmic factor), i.e., in the middle of stage II on the logarithmic scale. All these features of the profiles are observed in Fig.~\ref{fig:profiles} (blue curves).

For \textbf{stage IV}, in the vicinity of the defect, $|\Delta|\lesssim {\rm max}(w,\ell_c)$, we get
\begin{equation}
    h_{\rm eq}(\Delta)-h(\Delta,t)\approx\mathcal{H}\frac{w}{\pi\beta t}.
    \label{eq:hIV}
\end{equation}
That is, in the lowest order in $1/t$ the correction to the eventual equilibrium shape is independent of $\Delta$, i.e. the interface preserves its shape, merely translating vertically (see green curves in Fig.~\ref{fig:profiles}, for both the narrow and the wide defect). Functionals of $h(\Delta,t)$ describing this shifting, such as $h(\Delta=0)$, $h_{\rm min}+h_{\rm max}$, or $h_{\rm max}$ and $|h_{\rm min}|$ individually, approach their equilibrium values as $1/t$ (aside from possible logarithmic corrections), while those related to the interface shape change, such as $h_{\rm max}-h_{\rm min}$, decay faster.
Since for narrow defects $|h_{\rm min}|$ grows (as $t^{1/3}$) in stage III and decays (as $1/t$) in stage IV, the maximum of $|h_{\rm min}|$ is expected on the boundary of stages III and IV, i.e., for $t\sim\ell_c/\beta$, as confirmed numerically (see Fig.~\ref{fig:profiles}). 
The shape of the interface tails, however, does continue to evolve in stage IV. For $|\Delta|\gg{\rm max}(w,\ell_c)$,
\begin{equation}
    h(\Delta,t)\approx -\frac{\beta w\mathcal{H} t}{\pi[(\beta t)^2+\Delta^2]},
    \label{eq:tail}
\end{equation}
namely, the tails become broader ($\propto t$) and lower in height ($\propto 1/t$), until they span the whole width of the system, at which point \textbf{stage V} kicks in, where the final exponential relaxation towards equilibrium, 
\begin{equation}
    h(\Delta,t)\approx h_{\rm eq}(\Delta)-\frac{2 w \mathcal{H}}{W}\exp\left(-\frac{2\pi\beta}{W}t\right)\cos\left(\frac{2\pi}{W}\Delta\right),
    \label{eq:hV}
\end{equation}
is observed for finite cell widths. {Equation~(\ref{eq:tail}) describes propagation of a perturbation pointlike in space ($\Delta=0$) and instantaneous in time ($t=0$) for $W\to\infty$, with a similar spatiotemporal dependence as Eq.~(2.34) in \cite{Reis_JFM_2026}.}

The above analysis assumes that the mean distance between the interface and the inlet (or the bottom boundary of the cell) is infinite, justified when 
this distance is substantially larger than the cell width. This removes the distinction between the pressure control and flow-rate control (with $V=0$) cases, and means that the liquid volume [the area under the interface in our 2D approximation (with $b=b_0$)] is preserved. 
If this assumption about the inlet-interface distance cannot be made, then (i) the distinction between the two processes remains, and (ii) for both there is yet another governing length scale, which introduces new stages depending on the relation between that length and others [see Appendix~\ref{sec:mesa_details_finite_inlet_distance} for more details].

\section{Dynamics of interface relaxation: Disordered medium}
\label{sec:disorder}

Finally, we demonstrate the utility and efficiency of the spectral approach by considering an IHSC with random roughness, a proxy for a rough fracture (Fig.~\ref{fig:fig1_schematic}). 
There are various procedures to generate rough surfaces, e.g. see review in~\citet{borodich2024multiscale}. 
Here, we generate a random two-dimensional aperture field by summing up Fourier components with random phases and deterministic amplitudes chosen so that the correlation function decays exponentially:
\begin{equation}
    \langle [b(x,y)-b_0][b(x+\Delta x,y+\Delta y)-b_0]\rangle=B^2\exp\left[-\frac{(\Delta x^2+\Delta y^2)^{1/2}}{L}\right].
    \label{eq:apercorr}
\end{equation}
The $x$ and $y$ components of the wave vector are multiples of the same wavenumber, thus, the random field is periodic in $x$ and $y$ directions with the same period, and the width of the cell $W$ is chosen equal to the period. The resulting aperture field is characterised by two parameters---variance $B^2$ and correlation length $L$---chosen such that $L\ll W$. The distribution of apertures is normal.
Details of the surface generation procedure are provided in Appendix~\ref{sec:generating_rough_FR}. 
We use here an exponential correlation structure to represent a rough fracture.
While many models of two-phase flow in rough fractures 
use self-affine structure (characterised by power-law spatial correlations), e.g., see \citet{Candela2010_chapter}, there are other types of fracture surface models~\citep{cvetkovic2014upscaling}.


We use the spectral approach to simulate relaxation of the liquid-air interface under pressure control, keeping the inlet pressure $P$ constant. The generated random aperture field is used in the expression for pressure imbalance, Eq.~(\ref{eq:pe}), which in the linear approximation in $h-h_0$ and $b-b_0$ gives
\begin{eqnarray}
    p_e\approx\rho g_e\left[h-\ell_c^2\left(h''+\frac{8}{\pi b}\right)\right]-P
    &\approx&\rho g_e\left[h-\ell_c^2\left(h''+\frac{8}{\pi b_0}-\frac{8}{\pi b_0^2}\delta b\right)\right]-P\nonumber\\
    &=&\rho g_e(-\ell_c^2 h''+h-{H_\mathrm{eq}}+\tilde{\delta b}),
\end{eqnarray}
where 
\begin{equation}
    {H_\mathrm{eq}}=\frac{P}{\rho g_e}+\frac{8\ell_c^2}{\pi b_0},
\end{equation}
is the equilibrium height of the interface without the disorder ($b=b_0$), and
\begin{equation}
    \tilde{\delta b}=\frac{8\ell_c^2}{\pi b_0^2}(b-b_0).
\end{equation}
For our test, we choose
\begin{equation}
    \left\langle\tilde{\delta b}^2\right\rangle=\ell_c^2,
\end{equation}
corresponding to $B=\pi b_0^2/(8\ell_c)$ in Eq.~(\ref{eq:apercorr}). Further, we use $W/\ell_c={H_\mathrm{eq}}/\ell_c=20$. The interface is initially flat, and $h(t=0)={H_\mathrm{eq}}$, thus, the interface is expected to relax ``in place'', with little motion as a whole. The disorder correlation length is $L/\ell_c=0.1$. The aperture field $\tilde{\delta b}$ is generated on a $2^{14}\times 2^{14}$ grid, corresponding to about 82 grid steps per correlation length. The computational procedure is specified in Sec.~\ref{sec:alg}, with the grid in the $x$ direction matching the grid of the aperture field, while in the $y$ direction linear interpolation between grid points is used when calculating $\tilde{\delta b}$ at the current position of the interface.

The simulated interface profile is shown at several different times for two different time steps $\Delta t$ in Fig.~\ref{fig:disordered}.
The fact that differences between simulations with different $\Delta t$ are very minor indicates convergence of the numerical procedure. 
There is no visible difference between the profiles at the two latest times ($\beta t/\ell_c=20$ and 100), suggesting that the final relaxed shape is reached by the end of the simulation. As in the single-defect case, disparate length scales ($L\ll\ell_c\ll W$) result in nontrivial evolution over several orders of magnitude in time, with the interface shape generally coarsening with time. It is also notable that different parts of the interface reach the final shape at very different times. The computational efficiency is noteworthy: about 10 minutes on an ordinary laptop even for the finer $\Delta t$.

\begin{figure}
    \centering
\includegraphics[width=0.45\linewidth]{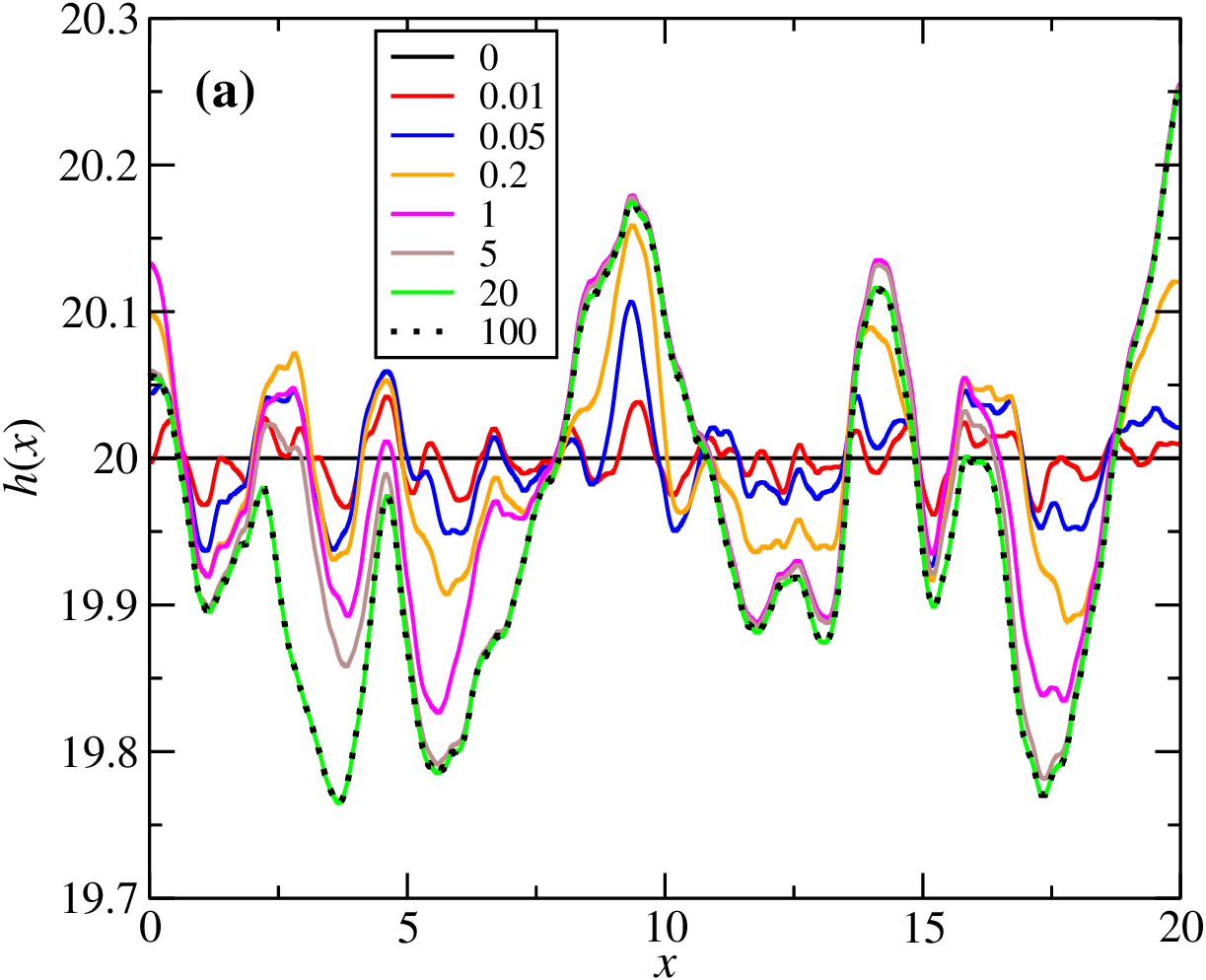}
\hspace{0.5cm}
\includegraphics[width=0.45\linewidth]{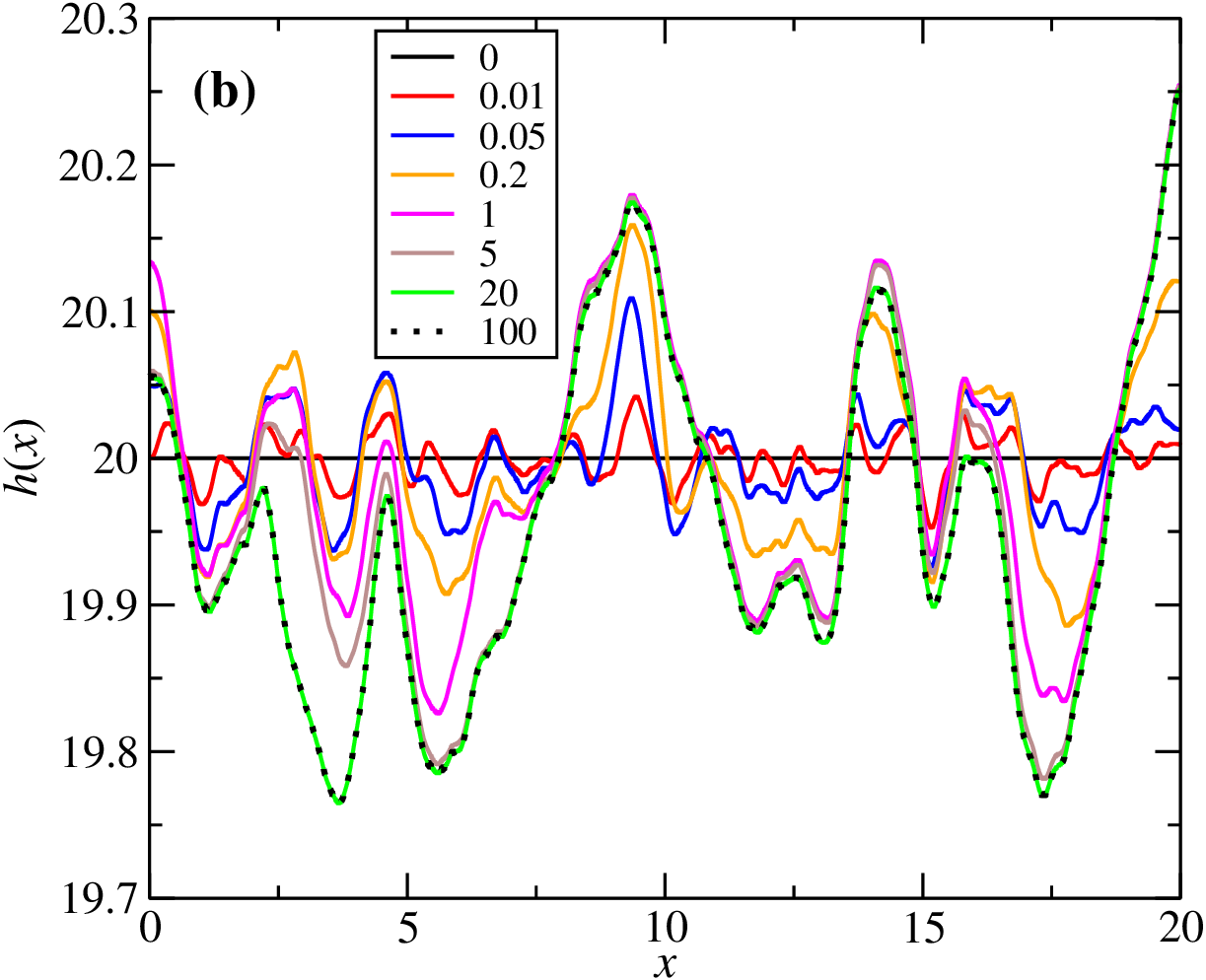}
    \caption{Evolution of the interface profile from the initial flat shape to the final relaxed shape in a cell with disorder, obtained using spectral simulations [Sec.~\ref{sec:alg}] with the time step [given in units of dimensionless time $\hat{t}$, Eq.~(\ref{eq:t-hat})] (a) 0.01 and (b) 0.001. Different curves correspond to times (in the same dimensionless units) specified in the legends. All other parameters of the simulation and the cell are given in the text.
    {For visualisation of the dynamics see Movie 4 in SI.}
    }
    \label{fig:disordered}
\end{figure}


\section{Discussion and conclusions}
\label{sec:conclusions}


In this paper, we present a novel approach to compute interface evolution and energy dissipation in two-phase flows through confined disordered media. 
We consider an imperfect Hele-Shaw cell (IHSC), where variations in the aperture give rise to varying capillary forces on the fluid-fluid interface.
%
%
We consider two fluids:
a viscous, denser wetting fluid and an inviscid, much lighter nonwetting fluid (e.g. liquid and gas), in the Stokes regime.
%

Our spectral approach, based on a Fourier expansion (using Fast Fourier Transform) of the interfacial configurations, provides a much higher computational efficiency compared to CFD methods (Sec. \ref{sec:spectral}). 
This allows us to follow the interfacial evolution over many orders of magnitude
in time and space, which is essential for the resolution of even simpler problems such as a single defect. 
%
Our approach, which we validate against CFD and experiments (Sec.\ \ref{sec:validation}), goes beyond the quasistatic limit and provides means to quantify the impact of the driving rate, for both pressure- and flow-rate-control driving. 
The combined accuracy and efficiency of the presented approach allows detailed analysis of large disordered systems, that (to the best of our knowledge) is not feasible with existing methods (CFD), or differential equations for the interface (SI), opening up new possibilities for generation of novel data, new analyses, and upscaling.

Analysis of the equations for IHSC (Sec.\ \ref{sec:HS}) has enabled us to obtain analytical expressions of the energy dissipation rate for different modes of driving (both rate- and pressure-controlled). 
In particular, we have shown that in quasistatically-driven
displacements, the energy dissipated (in Haines jumps) is of viscous origin. 
We prove that the total amount of energy dissipated in a Haines jump is independent of the actual fluid viscosity, which only controls its rate. 
This explains why the total energy dissipated could be computed from the stationary interfacial configurations alone, without details of the dynamics, e.g. intermediate configurations or velocities [as done in \citet{Holtzman2023}].
Detailed analysis of the apparently simple case of a single mesa defect reveals the complexity of displacement dynamics (Sec.\ \ref{sec:relaxation}). 
We identify different temporal stages in the system behaviour, in terms of both the dissipation rate and the interface evolution. 
We end with a demonstration of the ability of our spectral approach to handle large disordered systems (out of reach for other methods). 


The high efficiency of our approach is gained by making simplifying assumptions which can be divided 
into two categories: (a) general, which are inherent in the basic 2D Hele-Shaw equations (Sec.~\ref{subsec:HS-equations}); and (b) specific to the spectral approach (Sec.\ \ref{sec:spectral}).
The first includes simplified representation of the interface curvature, boundary conditions (slip and contact angle with side and top walls), and exclusion of inertial effects as well as interface disconnections (snapoff) and fluid trapping. 
The second category includes small deformation (linear approximation) and consideration of the effect of the gap variations only in the capillary pressure. 
The latter implies that the interface motion is affected at a given time only by the features the interface is in contact with then, with no influence of a defect entirely inside the liquid phase, or of a defect end during drainage until the interface reaches it. 
The good agreement with experiments using mesa defects indicates that the impact of at least some of these is small (for the conditions tested here). 
More detailed validation, in particular through experiments with disordered media, is left for future work.

Several physical mechanisms, which are not included in the current approach, are of interest for future work. These include the aforementioned interface disconnections and fluid trapping \citep{Bonn2009}, 
and inertial effects which may be important at early times during fast Haines jumps \citep{Berg2013, Moebius2012}. 
The impact of non-negligible viscosity and density of the nonwetting fluid \citep{Paune2002, Paune2003} is also of interest. 
It is also interesting to consider wetting effects, such as contact angle dynamics and hysteresis \citep{PahlavanPRL2015}, and 3D wetting effects; these include thin residual coating layers of the wetting liquid along the solid walls left in the drained region during imbibition, which modify the kinematic and dynamic boundary conditions to account for volume conservation and enhanced curvature and viscous stresses at the interface \citep{Park/Homsy:1984, Reinelt1985, Morrow2023}, as well as instabilities caused by entrainment of the less-wetting fluid \citep{Levache2014,Odier2017} and corner films \citep{Chauvet2009} which may appear along mesa defects and possibly other locations where there are abrupt changes of aperture.

In conclusion, 
our approach tackles successfully a highly challenging problem---the nonlocality of the response of the flow to interface deformation, which was out of reach of existing methods.
This provides a quantitative tool to resolve fundamental questions such as the individual contribution of viscous friction during continuous flow (which is the only source of energy dissipation in simple homogeneous systems) and of Haines jumps, and the impact of the mode of driving (pressure- vs. rate-control, and the rate they are changed), on hysteresis and dissipation.
The ability to handle large disordered systems opens up interesting possibilities for upscaling the flow behaviour from the pore- to the meso-scale, deriving novel constitutive relations in terms of disorder and fluid properties.
%
%
Furthermore, the efficiency of our spectral approach makes it an attractive alternative to existing methods used to simulate problems in which fluid-fluid interface dynamics is strongly coupled with other processes such as transport of solutes \citep{SAEIBEHROUZI_ESR_2024} or alteration of the porous medium properties by capillary forces \citep{juanes_multiphase_2020} or chemical reactions \citep{Ladd_annurev2021}.


\begin{bmhead}[Acknowledgments]
We thank Ido Lavi for help with the CFD code and simulations. \end{bmhead}

\begin{bmhead}[Funding]
RH acknowledges support from the Ministry of Science and Innovation, Spain (project ATR2024-154752 funded by MICIU/AEI/10.13039/501100011033), from the Engineering and Physical Sciences Research Council, UK (EP/V050613/1), and from the Royal Society, UK (IES/R2/232054). JO and MD acknowledge the support from the Spanish research agency (10.13039/501100011033) and
Spanish Ministry of Science and Innovation through project HydroPore-II (PID2022-137652NB-C41 and C42). \end{bmhead}

\begin{bmhead}[Declaration of interests]
The authors report no conflict of interest.\end{bmhead}

\begin{bmhead}[Data availability statement]
The data that support the findings of this study can be made available upon
request.\end{bmhead}

\begin{bmhead}[Author contributions]
M.V. Chubynsky: Conceptualization, Methodology, Software, Validation, Formal Analysis, Investigation, Writing. J. Ort{\'\i}n: Conceptualization, Methodology, Formal Analysis, Investigation, Writing. M. Dentz: Conceptualization, Methodology, Formal Analysis, Investigation, Writing. R. Holtzman: Conceptualization, Methodology, Formal Analysis, Investigation, Writing, Supervision, Funding Acquisition.
\end{bmhead}




\begin{appen}

\section{Direct derivation of the dissipation rate}
\label{sec:dissip_direct}

The derivation of the dissipation rate in Sec.~\ref{subsec:viscosity-independence} is based on the fact that work of non-viscous forces equals viscous dissipation, as these forces pump energy into the system that is dissipated via viscous forces. This has the advantage that since these forces do not depend on the transverse coordinate $z$ in the approximation we are considering, the $z$ integral is trivial and only a 2D integral remains. Work of viscous forces, and thus the viscous dissipation rate, can also be calculated directly, although this requires explicit $z$ integration.

Equation~(\ref{eq:u}) is derived based on the assumption that the transverse velocity profile is parabolic \citep{Krishna_JFM_2025}, thus, the full 3D velocity field is given by
\begin{equation}
    \mathbf{v}(x,y,z)=6\mathbf{u}(x,y)z[b(x,y)-z]/b(x,y)^2,
\end{equation}
where the prefactor of the parabolic dependence is chosen so that the velocity averaged over the $z$ coordinate gives the 2D velocity field $\mathbf{u}$:
\begin{equation}
    \frac{1}{b(x,y)}\int_0^{b(x,y)} \mathbf{v}(x,y,z)\, dz=\mathbf{u}(x,y).
\end{equation}
Vectors $\mathbf{v}$ and $\mathbf{u}$ lie in the $x-y$ plane. Further, this parabolic approximation is valid when $x$ and $y$ derivatives of the velocity are much smaller than the $z$ derivative. Then the volume density of the viscous force is
\begin{equation}
    \mathbf{f}=\mu\frac{\partial^2 \mathbf{v}}{\partial z^2}=-12\mu\mathbf{u}(x,y)/b(x,y)^2.
\end{equation}
The work per unit time of this viscous force is equal to the dissipation rate, thus,
\begin{equation}
    |R|=-\int_V \mathbf{v}\cdot\mathbf{f}\, dV=72\mu\int_A \frac{u(x,y)^2}{b(x,y)^4}\left(\int_0^{b(x,y)} z[b(x,y)-z]\, dz\right)\, dA,
\end{equation}
where the first integral is over the liquid volume $V$. This gives Eq.~(\ref{eq:Rgen}).

\section{Deriving the expression for the dissipation rate $R$ based on interface dynamics and relating $R$ to the work of external forces (Sec. \ref{subsec:dissipation_rate})}
\label{sec:demo_energy_diss_visc}

For the derivation below, a useful equality is
\begin{equation}
\nabla\cdot[(p+\rho g_e y)\mathbf{u}]=\mathbf{u}\cdot\nabla(p+\rho g_e y)+(p+\rho g_e y)
(\nabla\cdot\mathbf{u})
=\mathbf{u}\cdot[\nabla p+\rho g_e\hat{y}],
\end{equation}
where we have used Eq.~(\ref{eq:contappr}). Then, from Eq.~(\ref{eq:Rgen}) the dissipation rate is
\begin{eqnarray}
|R|&=&-\int_A b(\mathbf{u}\cdot[\nabla p+\rho g_e\hat{y}])\, dA\approx -b_0\int_A (\mathbf{u}\cdot[\nabla p+\rho g_e\hat{y}])\, dA\nonumber\\
&=&-b_0\int_A \nabla\cdot[(p+\rho g_e y)\mathbf{u}]\, dA.
\end{eqnarray}
By Gauss theorem, this is
\begin{equation}
-b_0\oint (p+\rho g_e y)u_n\, ds.
\end{equation}
The integral is taken along the contour surrounding the liquid, with $ds$ the length element of the contour. The normal velocity $u_n$ is the projection onto the normal pointing outwards. The parts of the integral along the side walls are zero, because $u_n$ is zero there. The contribution along the inlet gives
\begin{equation}
+b_0\int_{x=0}^W p(x,y=0)u_y(x,y=0)\, dx
\end{equation}
(the plus sign is because the normal points downwards). In \textit{\textbf{pressure-controlled displacements}}, $p(x,y=0)=P$, this gives
\begin{equation}
b_0 P\int_{x=0}^W u_y(x,y=0)\, dx
=b_0 P\int_{x=0}^W u_n(x,y=h(x))\, ds,
\end{equation}
where the second expression follows from mass conservation.
The normal velocity component at the interface is
\begin{equation}
u_n(x,y=h(x))=\frac{u_y-u_x(\partial h/\partial x)}{\sqrt{1+(\partial h/\partial x)^2}}=\frac{\partial h/\partial t}{\sqrt{1+(\partial h/\partial x)^2}},
\end{equation}
and the length element on the interface is
\begin{equation}
ds=\sqrt{1+(\partial h/\partial x)^2}\, dx.
\end{equation}
Then
\begin{eqnarray}
|R(t)|&\approx& -b_0\oint (p+\rho g_e y)u_n\, ds\nonumber\\
&=&-b_0\int_{x=0}^W [p(x,h(x),t)+\rho g_eh(x,t)-P]u_n(x,h(x),t)\, ds\nonumber\\
&=&-b_0\int_{x=0}^W p_e(x,t)\frac{\partial h(x,t)}{\partial t}\, dx,
\label{eq:Rfinalapp}
\end{eqnarray}
where
$p_e(x,t)=p(x,h(x),t)+\rho g_eh(x,t)-P$ and thus coincides with Eq.~(\ref{eq:pe}). Equation~(\ref{eq:Rfinalapp}) is the same as Eq.~(\ref{eq:Rfinal}) in the main text.

In \textit{\textbf{flow-rate-controlled displacements}}, $u_y(x,y=0)=V$, from mass conservation
\begin{equation}
\int_{x=0}^W u_n(x,y=h(x))\, ds=WV.
\end{equation}
Denoting
\begin{equation}
\frac{1}{W}\int_{x=0}^W p(x,y=0,t)\, dx=P(t),
\label{eq:PV}
\end{equation}
we get
\begin{equation}
b_0\int_{x=0}^W p(x,y=0,t)u_y(x,y=0,t)\, dx=b_0 P(t)WV=b_0 P(t)\int_{x=0}^W u_n(x,y=h(x),t)\, ds,
\end{equation}
and Eq.~(\ref{eq:Rfinalapp}) is obtained again, but here $P$ is the inlet pressure $p(x,y=0,t)$ \textit{averaged along the inlet}, which is time-dependent but $x$-independent.

Finally, we show that Eq.~(\ref{eq:Rfinalapp}) has a straightforward interpretation as the rate of work (i.e. power) of external forces acting on the liquid volume. We do this
by decomposing Eq.~(\ref{eq:Rfinalapp}) into three parts, corresponding to the power of different external forces. 
We still use the approximation of Sec.~\ref{subsec:b_0}, i.e, we assume $b=b_0$ everywhere, except in the capillary pressure. The power of the capillary force is
\begin{equation}
    -b_0\int_0^W p(x,h(x),t)\frac{\partial h(x,t)}{\partial t} \; dx
    =b_0\int_0^W\gamma\left(\frac{\pi}{4}\kappa(x,t)+\frac{2}{b(x,h(x,t))}\right)\frac{\partial h(x,t)}{\partial t} \; dx.
\end{equation}
The power of the gravity force is
\begin{equation}
    -\rho g_e b_0
 \int_0^W h(x,t)\frac{\partial h(x,t)}{\partial t} \; dx.
\end{equation}
Finally, the power of the pressure force at the inlet is
\begin{equation}
    Pb_0
\int_0^W\frac{\partial h(x,t)}{\partial t} \; dx
\end{equation}
in the \textit{\textbf{pressure-controlled}} case and
\begin{equation}
    b_0V\int_0^W p(x,0,t) \;dx=b_0P(t)\int_0^W\frac{\partial h(x,t)}{\partial t} \; dx
\end{equation}
in the \textit{\textbf{flow-rate-controlled}} case, where in both cases we have used mass conservation to relate the rate of flow through the inlet to the rate of interface advancement.
As expected, the sum of these three contributions gives Eq.~(\ref{eq:Rfinalapp}). 

%

\section{Dissipation rate in the spectral approach (Sec. \ref{subsec:dissip_spectral}): Derivation} 
\label{APPENDIX_dissip_spectral}


We use Eq.~(\ref{eq:Rfinal}) for the dissipation rate expressed in terms of the interface evolution rate and the pressure imbalance.
In the \textit{\textbf{pressure-controlled}} case, we start with Eq.~(\ref{eq:Rfinal}) and first use Eqs.~(\ref{eq:dpi}) and (\ref{eq:hfour}) giving Fourier decompositions of these quantities and the orthogonality relation for the cosines, to obtain
\begin{eqnarray}
    |R(t)|&=&-b_0\int_{x=0}^W p_e(x,t)\frac{\partial h(x,t)}{\partial t}\, dx\nonumber\\
    &\approx& -b_0\int_{x=0}^W \sum_{m,n=0}^{\infty}p_m(t)\dot{h}_n(t)\cos(k_m x)\cos(k_n x)\, dx\nonumber\\
    &=&-b_0 W\left(p_0(t)\dot{h}_0(t)+\sum_{n=1}^{\infty}\frac{p_n(t)\dot{h}_n(t)}{2}\right).\\
\end{eqnarray}
The spectral approach expresses $\dot{h}_n$ in terms of $p_n$ via Eqs.~(\ref{eq:dh_vs_p})--(\ref{eq:dh0_vs_p0}), which gives
\begin{equation}
    |R(t)|=\frac{b_0^3W}{12\mu}\left(\frac{p_0^2(t)}{h_0}+\frac{1}{2}\sum_{n=1}^{\infty}k_n\coth(k_n h_0)p_n^2(t)\right).
\end{equation}
We can also use Eqs.~(\ref{eq:omp})--(\ref{eq:omp0}) for the relaxation rates to obtain an alternative form, Eq.~(\ref{eq:Rspec}).
%
In the \textit{\textbf{flow-rate-controlled}} case, we start with 
$P(t)$ from Eq.~(\ref{eq:PV}), together with Eqs.~(\ref{eq:deltapV}) and (\ref{eq:dpV}) with $p_0=0$, to obtain 
\begin{equation}
    P(t)=\frac{1}{W}\int_{x=0}^W p(x,y=0,t)\, dx=\Pi(t).\label{eq:PPi}
\end{equation}
From the expressions for $p_e$ [Eq.~(\ref{eq:pe})] and $p_{eV}$ [Eq.~(\ref{eq:peV})], as well as the definition of $g_{eV}$ [Eq.~(\ref{eq:geV})], we get
\begin{equation}
    p_e(x,t)=p_{eV}(x,t)-\frac{12\mu V}{b_0^2}h(x,t),
\end{equation}
and Eq.~(\ref{eq:Rfinal}) gives
\begin{equation}
    |R(t)|=-b_0\int_{x=0}^W p_{eV}\frac{\partial h(x,t)}{\partial t}\, dx+\frac{12\mu V}{b_0}\int_{x=0}^W h(x,t)\frac{\partial h(x,t)}{\partial t}\, dx.
\end{equation}
Finally, we use Eqs.~(\ref{eq:dpiV}), (\ref{eq:hfour}) and the orthogonality relations between the cosine functions to get
\begin{equation}
    |R(t)|=-b_0W\sum_{n=1}^{\infty}\frac{p_n\dot{h}_n}{2}+\frac{12\mu WV}{b_0}\left(h_0 \dot{h}_0+\sum_{n=1}^{\infty}\frac{h_n\dot{h}_n}{2}\right),
\end{equation}
and then applying Eqs.~(\ref{eq:dh_vs_pV})--(\ref{eq:dh0_vs_p0V}), (\ref{eq:omV}) and (\ref{eq:betaV}) gives Eq.~(\ref{eq:RspecV}).
\section{Validating the spectral approach (Sec. \ref{sec:validation}): Details}

\subsection{Spectral calculations for a mesa defect:
Derivation}
\label{sec:Spectral_h_R_details}

Here we provide the derivation for the expressions of interface evolution and dissipation rate in Sec.~\ref{sec:mesa_spectral}.
Consider first the case of a \textbf{closed inlet} (i.e., flow-rate control with $V=0$). 
From Eq.~(\ref{eq:peV}) we get
\begin{eqnarray}
    p_{eV}(x,t)&=&-\frac{\pi}{4}\gamma h''(x,t)+\rho g_e [h(x,t)-h^{(0)}]+2\gamma\left(\frac{1}{b_0}-\frac{1}{b}\right)+C\nonumber\\
    &=&-\frac{\pi}{4}\gamma h''(x,t)+\rho g_e [h(x,t)-h^{(0)}]-\rho g_e \mathcal{H}{\rm Rect}\left(\frac{x-W/2}{w}\right)+C.
    \label{eq:pemesa}
\end{eqnarray}
Here $C$ is a constant the value of which is irrelevant; we have used the linear approximation $\kappa\approx h''$; {$\mathcal{H}$ is given by Eq.~(\ref{eq:H})}; $g_{eV}=g_e$ for $V=0$; and the ``rectangular function''
\begin{equation}
    {\rm Rect}(s)=\begin{cases}
        1, & |s|<1/2,\\
        0, & |s|>1/2.
    \end{cases}
\end{equation}
Since here $b=b(x)$, the entire interface evolution in the spectral approach is given directly by Eq.~(\ref{eq:DeltahnV}), where for $V=0$ $g_{eV}=g_e$,
$\ell_{cV}=\ell_c$, $\beta_V=\beta$, and we also assume $kh_0\gg 1$ (valid for all $k$ if $h_0\gg W$, but in practice $h_0\gg w,\ell_c$ is sufficient), thus,
\begin{equation}
    \omega_V(k)=\beta k(1+\ell_c^2 k^2).
    \label{eq:omegaV0_SI}
\end{equation}
Fourier transforming the initial pressure imbalance to obtain $p_n(0)$ is done analytically. From Eq.~(\ref{eq:pemesa}),
\begin{equation}
    p_{eV}(x,t=0)=-\rho g_e \mathcal{H}{\rm Rect}\left(\frac{x-W/2}{w}\right).
    \label{eq:pemesa1}
\end{equation}
Given that, based on Eq.~(\ref{eq:kn}) for $k_n$,
\begin{equation}
    \cos(k_n x)=
    \begin{cases}
        (-1)^{n/2}\cos[k_n(x-W/2)], & n\ \mathrm{even},\\
        (-1)^{(n+1)/2}\sin[k_n(x-W/2)], & n\ \mathrm{odd},
    \end{cases}
\end{equation}
we get
\begin{eqnarray}
    p_n(0)&=&\frac{2}{W}\int_0^W p_e(x,t=0)\cos(k_nx)\; dx=-\frac{2\rho g_e \mathcal{H}}{W}\int_{(W-w)/2}^{(W+w)/2}\cos(k_nx)\; dx\nonumber\\
    &=&
    \begin{cases}
    -(4\rho g_e \mathcal{H})/(Wk_n)(-1)^{n/2}\sin(k_n w/2), & n\ \mathrm{even},\\
    0, & n\ \mathrm{odd}.
    \end{cases}
    \label{eq:pn0mesa}
\end{eqnarray}
Then, finally, from Eq.~(\ref{eq:DeltahnV}), where we put $t=0$ and then replace $\Delta t$ with $t$,
\begin{equation}
    h(x,t)=h^{(0)}-\sum_{n=1}^{\infty}\frac{p_n(0)}{\rho g_e (1+\ell_c^2 k_n^2)}[1-\exp(-\omega_V(k_n) t)]\cos(k_n x),
    \label{eq:hmesa_SI}
\end{equation}
{which then gives Eq.~(\ref{eq:hmesa})}. 
For the dissipation rate, from Eq.~(\ref{eq:RspecV}) and given that
\begin{equation}
    p_n(t)=p_n(0)\exp[-\omega_V(k_n)t],
\end{equation}
we obtain
\begin{equation}
    |R(t)|=\frac{b_0W}{2\rho g_e}\sum_{n=1}^{\infty}\frac{\omega_V(k_{2n})p_{2n}^2(0)}{1+\ell_c^2 k_{2n}^2}\exp[-2\omega_V(k_{2n})t],
    \label{eq:Rmesa_SI}
\end{equation}
{from which Eq.~(\ref{eq:Rmesa}) follows immediately}.

For the \textbf{pressure-controlled} case, recall that we have chosen the inlet pressure such that the initial position of the interface, $h(x,0)=h^{(0)}$, is the equilibrium state when there is no defect ($\mathcal{H}=0$). Then Eq.~(\ref{eq:pemesa}) is still correct for $p_e$ when $C=0$ [the value of $C$ is now relevant, but $C=0$ gives $p_e(x,t=0)=0$ for $\mathcal{H}=0$, as it should be], and so is Eq.~(\ref{eq:pemesa1}).
From these we deduce that Eq.~(\ref{eq:pn0mesa}) is valid for $n\ne 0$, but we now also need to consider the zeroth component, which is the average of $p_e(x,t=0)$ over the width of the cell, or
\begin{equation}
    p_0(0)=-\frac{\rho g_e \mathcal{H} w}{W}.
\end{equation}
{As stated in Sec.~\ref{sec:mesa_spectral}, we take $h_0(t)\equiv h^{(0)}$. Then}
the expression for $h(x,t)$ [Eq.~(\ref{eq:hmesa_SI})] 
remains valid, except (i) the sum now starts from $n=0$ (that is, including the area-nonconserving mode); and (ii) $\omega_V$ of Eq.~(\ref{eq:omegaV0_SI}) is replaced by $\omega_p$ given by (a) Eq.~(\ref{eq:omp}) for $n\ne 0$ without any assumptions about $kh_0$ (i.e., the coth factor is retained) and (b) Eq.~(\ref{eq:omp0}) for $n=0$. {Equation~(\ref{eq:hmesa_P_control}) then follows.}

\subsection{Parameter values: Relaxation of an interface perturbed from equilibrium (Sec. \ref{sec:relax_test})}
\label{sec:relax_test_param}

Here we explain the rationale of chosen parameter values for validation. 
First, we chose the values of the three characteristic length scales---the capillary length $\ell_c$, the defect width $w$, and the cell width $W$---to give $w/\ell_c=4$ and $W/\ell_c=10$.
This is because
if two of them are very different (more than an order of magnitude), there may be accidental agreement if the approximate approach works well in that particular limit, whereas for other parameter values it may fail. 
Second, we use 
$|\delta b|\ll b_0$ so that the interface deformation amplitude is sufficiently small for the linear approximation to be valid. Our choice, $b_0/\ell_c=0.1$ and $|\delta b|/\ell_c=0.005$, gives $\mathcal{H}/\ell_c=1.34025$ [using the exact result in Eq.~(\ref{eq:H}), not its linear approximation], 
with a deformation amplitude of $1.1\ell_c$. 
While this is smaller than $w$, it is comparable to $\ell_c$ and thus allows testing the spectral approach under the conditions where the nonlinear effects are non-negligible albeit small. 

In the spectral calculations, $b=b(x)$, and the defect length is irrelevant, whereas in the CFD simulations, the lower end of the defect is $0.1\ell_c$ below the initial position of the interface. 
Good agreement between the two implies that the presence of the defect's end, even as close to the interface as in these simulations, does not influence the results. Finally, the bottom boundary of the cell in the closed inlet case is $20\ell_c$ below the defect's lower end. This is larger than $W$ and is treated as infinite in spectral calculations. In the pressure-controlled case, this boundary (which now serves as the inlet) is only $5\ell_c$ from the lower end of the defect, and this distance is explicitly included in spectral calculations, where $h_0=5.1\ell_c$. 


\subsection{Parameter values: Imbibition and drainage (Sec. \ref{sec:imbdra})}
\label{sec:imbdra_param}

%
{In the experiments of \citet{Lavi_PRF_2023}, the following system was used.}
Wetting fluid {(silicone oil) had the} viscosity $\mu=51.6\pm 0.6$~cP, density $\rho=972\pm 4$~kg/m$^3$, and surface tension $\gamma=21.0\pm 1.3$~mN/m. 
A cell of width $W=190$~mm with aperture $b_0=0.46$~mm was tilted at an angle $\alpha=5^{\circ}3'\pm 2'$. 
The defect dimensions were: length $L=Y_2-Y_1=60$~mm, width $w=3$~mm, and height $-\delta b=0.06$~mm. {The same parameters were used in CFD simulations in that reference and in our spectral simulations.}


\section{Interface evolution for a single mesa defect (Sec.~\ref{sec:Results_Interface_evolution}): Derivation}
\label{sec:mesa_details}

\subsection{General derivations}
\label{sec:mesa_details_general}

As mentioned in Sec.~\ref{sec:Results_Interface_evolution}, we ignore the $h^{(0)}$ term in Eq.~(\ref{eq:hmesa}), measuring the interface height $h$ from its initial position, rather than the inlet.

First, we note that Eq.~(\ref{eq:hmesa}) can be transformed as
\begin{equation}
    h(\Delta,t)=\frac{2 \mathcal{H}}{W}\sum_{n=1}^{\infty}\frac{\sin[k_{2n}(\Delta+w/2)]-\sin[k_{2n}(\Delta-w/2)]}{k_{2n}(1+\ell_c^2 k_{2n}^2)}\left\{1-\exp[-\beta k_{2n}(1+\ell_c^2 k_{2n}^2)t]\right\},
\end{equation}
where the arguments of the sines are explicitly the distances from the defect edges, reflecting their role as ``sources''. Of course, at $t\to\infty$ this should approach the known equilibrium profile, which, in particular, for $W=\infty$ is given by Eq:~(\ref{eq:heq}).
Then
\begin{eqnarray}
    h(\Delta,t)&=&h_{\rm eq}(\Delta)-\frac{4 \mathcal{H}}{W}\sum_{n=1}^{\infty}\frac{\sin(k_{2n}w/2)}{k_{2n}(1+\ell_c^2 k_{2n}^2)}\exp[-\beta k_{2n}(1+\ell_c^2 k_{2n}^2)t]\cos(k_{2n}\Delta)\nonumber\\
    &=&h_{\rm eq}(\Delta)-\frac{2 \mathcal{H}}{W}\sum_{n=1}^{\infty}\frac{\sin[k_{2n}(\Delta+w/2)]-\sin[k_{2n}(\Delta-w/2)]}{k_{2n}(1+\ell_c^2 k_{2n}^2)}\exp[-\beta k_{2n}(1+\ell_c^2 k_{2n}^2)t].
    \label{eq:hgen}
\end{eqnarray}
We also note that taking the time derivative eliminates the denominator,
\begin{eqnarray}
    \dot{h}(\Delta,t)&=&\frac{4\beta \mathcal{H}}{W}\sum_{n=1}^{\infty}\sin(k_{2n}w/2)\exp[-\beta k_{2n}(1+\ell_c^2 k_{2n}^2)t]\cos(k_{2n}\Delta)\nonumber\\
    &=&\frac{2\beta \mathcal{H}}{W}\sum_{n=1}^{\infty}\left\{\sin[k_{2n}(\Delta+w/2)]-\sin[k_{2n}(\Delta-w/2)]\right\}\exp[-\beta k_{2n}(1+\ell_c^2 k_{2n}^2)t],
\end{eqnarray}
making this convenient for some purposes.
For {stage V} ($t\gg t_e$), in Eq.~(\ref{eq:hgen}) the $n=1$ term dominates in the deviation from equilibrium; neglecting all other terms and using $k_2=2\pi/W\ll w^{-1},\ell_c^{-1}$, we get {Eq.~(\ref{eq:hV})}.
In all other stages, the sum can be replaced by the integral, as for dissipation,
giving
\begin{eqnarray}
    h(\Delta,t)&=&\frac{2 \mathcal{H}}{\pi}\int_0^{\infty}\frac{\sin(kw/2)}{k(1+\ell_c^2 k^2)}\left\{1-\exp[-\beta k(1+\ell_c^2 k^2)t]\right\}\cos(k\Delta)\, dk\nonumber\\
    &=&\frac{\mathcal{H}}{\pi}\int_0^{\infty}\frac{\sin[k(\Delta+w/2)]-\sin[k(\Delta-w/2)]}{k(1+\ell_c^2 k^2)}\left\{1-\exp[-\beta k(1+\ell_c^2 k^2)t]\right\}\, dk\nonumber\\
    &=&h_{\rm eq}(\Delta)-\frac{2 \mathcal{H}}{\pi}\int_0^{\infty}\frac{\sin(kw/2)}{k(1+\ell_c^2 k^2)}\exp[-\beta k(1+\ell_c^2 k^2)t]\cos(k\Delta)\, dk\nonumber\\
    &=&h_{\rm eq}(\Delta)-\frac{\mathcal{H}}{\pi}\int_0^{\infty}\frac{\sin[k(\Delta+w/2)]-\sin[k(\Delta-w/2)]}{k(1+\ell_c^2 k^2)}\exp[-\beta k(1+\ell_c^2 k^2)t]\, dk
    \label{eq:hmesaint}
\end{eqnarray}
and
\begin{eqnarray}
    \dot{h}(\Delta,t)&=&\frac{2\beta \mathcal{H}}{\pi}\int_0^{\infty}\sin(kw/2)\exp[-\beta k(1+\ell_c^2 k^2)t]\cos(k\Delta)\, dk\nonumber\\
    &=&\frac{\beta \mathcal{H}}{\pi}\int_0^{\infty}\left\{\sin[k(\Delta+w/2)]-\sin[k(\Delta-w/2)]\right\}\exp[-\beta k(1+\ell_c^2 k^2)t]\, dk.\label{eq:hdotmesa}
\end{eqnarray}
In Eq.~(\ref{eq:hdotmesa}), for $t=0$ the exponential is unity, and the last integral oscillates as a function of the upper limit without converging. However, this integral does converge for any finite $t$ and has a well-defined limit when $t$ approaches zero from above, Eq. ~(\ref{eq:hdottzero}).

\subsection{Derivations used in the analyses of specific stages}
\label{sec:mesa_details_stages}

We now consider different stages separately, making appropriate approximations. In \textbf{stages I and III}, similarly to our consideration of dissipation, we can keep only the cubic term in $k$ in the exponential in Eq.~(\ref{eq:hdotmesa}), which gives
\begin{eqnarray}
    \dot{h}(\Delta,t)&=&\frac{2\beta \mathcal{H}}{\pi}\int_0^{\infty}\sin(kw/2)\exp[-\beta \ell_c^2 k^3 t]\cos(k\Delta)\, dk\nonumber\\
    &=&\frac{\beta \mathcal{H}}{\pi}\int_0^{\infty}\left\{\sin[k(\Delta+w/2)]-\sin[k(\Delta-w/2)]\right\}\exp[-\beta \ell_c^2 k^3 t]\, dk.\label{eq:hdotcubic}
\end{eqnarray}
Making a change of variables
\begin{equation}
    \kappa=\beta \ell_c^2 k^3 t
    \label{eq:kappa}
\end{equation}
in the first of these expressions, we get after some transformations
\begin{equation}
    \dot{h}(\Delta,t)=\frac{2\beta \mathcal{H}}{3\pi w\tilde{t}^{1/3}}\int_0^{\infty}\sin\left(\frac{\kappa}{8\tilde{t}}\right)^{1/3}\exp(-\kappa)\cos\left(\frac{\tilde{\Delta}}{\tilde{t}}\right)^{1/3}\frac{d\kappa}{\kappa^{2/3}},
\end{equation}
where $\tilde{t}$ and $\tilde{\Delta}$ are given by Eqs.~(\ref{eq:ttilde}) and (\ref{eq:Deltatilde}), respectively.
Then
\begin{equation}
    \dot{h}=\frac{\beta \mathcal{H}}{w}f_3(\tilde{\Delta},\tilde{t}),
\end{equation}
where $f_3$ is some function of its arguments, and integrating this over time gives
\begin{equation}
    h=\frac{w^2 \mathcal{H}}{\ell_c^2}f_4(\tilde{\Delta},\tilde{t})
\end{equation}
where $f_4$ is a different function of the same arguments. From this, Eq.~(\ref{eq:htilde}), which introduces a rescaled interface height $\tilde{h}(\tilde{\Delta},\tilde{t})$, is obtained.
Thus, if we plot a linear functional of $\tilde{h}(\tilde{\Delta},\tilde{t})$ as a function of $\tilde{t}$, then such curves for different values of parameters such as the defect width should fall onto a single curve within stages I and III, though they may deviate in other stages. Examples of such functionals are the height at the centre of the defect or at its edge, or the maximum height $h_{\rm max}$, rescaled as in Eq.~(\ref{eq:htilde}). Such a plot is presented in Fig.~\ref{fig:hmaxmin}.

Looking now at \textbf{stage I} in more detail, we return to Eq.~(\ref{eq:hdotcubic}) and consider it for small but finite $t$. Far from the defect edges, the sines are still highly oscillatory on the scale of the decay of the exponential and Eq.~(\ref{eq:hdottzero}) describing linear growth is still valid. However, near one of the edges, the oscillation frequency of one of the sine terms approaches zero [we will call this sine term ``singular'' for brevity, because it gives rise to the singularity in Eq.~(\ref{eq:hdottzero}) at that edge], and for a given distance from the edge, there comes a time where there is less than one oscillation before the exponential decays, in which case the contribution of that term will get modified, while the other, ``non-singular'' term still contributes linear growth. More specifically, considering the edge at $\Delta=w/2$, for a given $\Delta\approx w/2$ the ``non-oscillatory'' stage is entered when $k_{\rm cut}|\Delta-w/2|\sim 1$, where $\beta\ell_c k_{\rm cut}^3 t\sim 1$. Thus, the corresponding time is
\begin{equation}
    t_{\rm cut}\sim\frac{|\Delta-w/2|^3}{\beta\ell_c^2}.
\end{equation}
From that time on,
\begin{equation}
    \dot{h}\sim\beta \mathcal{H}(w/2-\Delta)\int_0^{\infty}k\exp(-\beta\ell_c^2 k^3 t)dk\sim\frac{\beta^{1/3}\mathcal{H}(w/2-\Delta)}{\ell_c^{4/3}t^{1/3}}.
\end{equation}
The ``non-singular'' sine term still gives a linear contribution, but is small and is ignored here (although it contributes to the asymmetry of the perturbation, as we discuss below). After integration, we get
\begin{equation}
    h\sim
    \begin{cases}
        \frac{\beta \mathcal{H} t}{w/2-\Delta}, & |w/2-\Delta|>(\beta\ell_c^2 t)^{1/3},\\
        \frac{\beta^{1/3}\mathcal{H}(w/2-\Delta)t^{1/3}}{\ell_c^{4/3}}, & |w/2-\Delta|<(\beta\ell_c^2 t)^{1/3}.
    \end{cases}
\end{equation}
This result gives a picture of a perturbation near the edge of the defect antisymmetric with respect to the edge, with a maximum moving inwards from the edge and a minimum moving outwards, with their positions and heights given by Eqs.~(\ref{eq:maxI}) and (\ref{eq:hmaxI}), respectively.
A small amount of asymmetry is produced by the ``non-singular'' term that still grows linearly in time.

It is also possible to show explicitly that the perturbation preserves its shape as it evolves. Considering only the ``singular'' sine term in the second line of Eq.~(\ref{eq:hmesaint}) and keeping only the cubic terms in both the denominator and the exponent, and then making the change of variables of Eq.~(\ref{eq:kappa}) gives
\begin{equation}
    h\sim \mathcal{H}\int_0^{\infty}\frac{\sin[\kappa^{1/3}(w/2-\Delta)/(\beta\ell_c^2 t)^{1/3}]}{\ell_c^2[\kappa/(\beta\ell_c^2 t)]}[1-\exp(-\kappa)]\frac{d\kappa}{\kappa^{2/3}\beta^{1/3}\ell_c^{2/3}t^{1/3}},
\end{equation}
from which Eq.~(\ref{eq:hscalI}) follows.



In \textbf{stage III} (only observed for narrow defects),
we can still retain just the cubic terms in the exponent and the denominator of Eq.~(\ref{eq:hmesaint}). Since the two sine terms in the second line are now equally important, it is convenient to use the first line, but we can now also assume $kw\ll 1$ and $\sin(kw/2)\approx kw/2$. This gives
\begin{equation}
    h\sim \mathcal{H}\int_0^{\infty}\frac{w}{\ell_c^2 k^2}[1-\exp(-\beta\ell_c^2 k^3 t)]\cos(k\Delta)\,dk,
\end{equation}
and with the same change of variables (\ref{eq:kappa}) we get
\begin{equation}
    h\sim \mathcal{H}\int_0^{\infty}\frac{w}{\ell_c^2[\kappa/(\beta\ell_c^2t)]^{2/3}}[1-\exp(-\kappa)]\cos[\kappa^{1/3}\Delta/(\beta\ell_c^2 t)^{1/3}]\frac{d\kappa}{\kappa^{2/3}\beta^{1/3}\ell_c^{2/3}t^{1/3}},
\end{equation}
which gives Eq.~(\ref{eq:hIII}).


Consider now \textbf{stages II and IV}. In the exponential in Eq.~(\ref{eq:hdotmesa}) now the linear term is dominant, giving
\begin{eqnarray}
    \dot{h}(\Delta,t)&=&\frac{\beta \mathcal{H}}{\pi}\int_0^{\infty}\left\{\sin[k(\Delta+w/2)]-\sin[k(\Delta-w/2)]\right\}\exp[-\beta kt]\, dk\nonumber\\
    &=&\frac{\beta \mathcal{H}}{\pi}\left[\frac{\Delta+w/2}{(\beta t)^2+(\Delta+w/2)^2}-\frac{\Delta-w/2}{(\beta t)^2+(\Delta-w/2)^2}\right].
    \label{eq:hdotII-IV}
\end{eqnarray}
Integrating back in time from the known equilibrium shape [Eq.~(\ref{eq:heq})],
\begin{equation}
    h(\Delta,t)=h_{\rm eq}-\int_t^{\infty}\dot{h}(\Delta,t_1)\, dt_1,
\end{equation}
we get Eq.~(\ref{eq:hII_IV}).
Figure~\ref{fig:profiles_vs_theory} shows that this provides excellent fits in stages II and IV.
Moreover, note that since Eq.~(\ref{eq:hdotII-IV}) has the correct $t\to 0$ limit [cf. Eq.~(\ref{eq:hdottzero})] and the last term of Eq.~(\ref{eq:hII_IV}) is only significantly different from zero in the regions around the edges of the defect of width $\sim\ell_c$ (we will refer to them as the edge regions in what follows), then, aside from these regions, for wide defects Eq.~(\ref{eq:hII_IV}) is also valid in stage I. While, admittedly, in that stage it is the edge regions that are of most interest, this validity is still a remarkable fact and allows us to write, for example, Eq.~(\ref{eq:h0wide}).
%
Note that while in that equation there is a crossover from linear dependence to ``saturation'' at $t\sim w/\beta$ (i.e., on the boundary between stages II and IV), on the boundary between stages I and II there is no similar qualitative change (such changes only occur in the edge regions).

\begin{figure}
    \centering
    \includegraphics[width=0.45\linewidth]{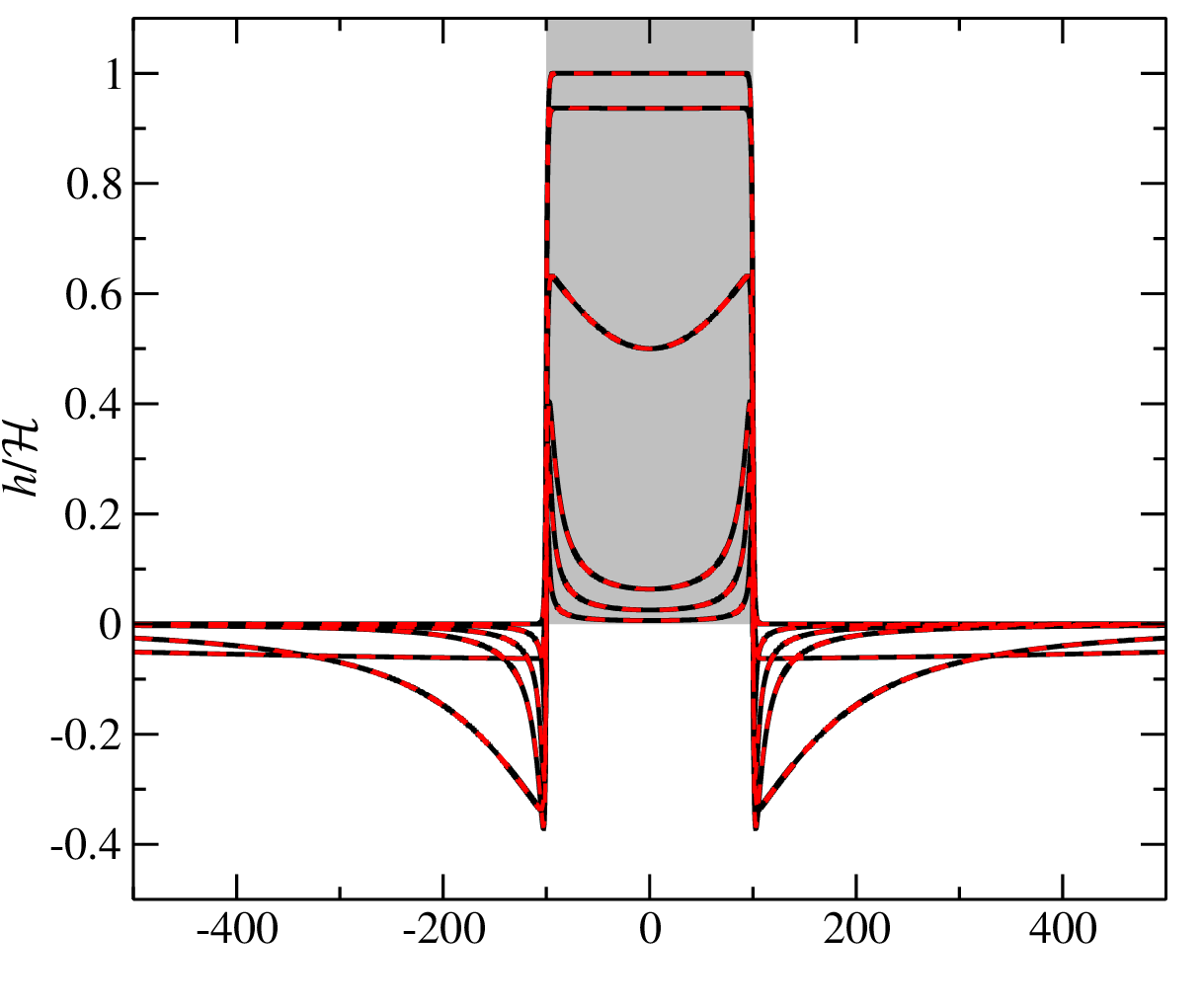}
    \includegraphics[width=0.45\linewidth]{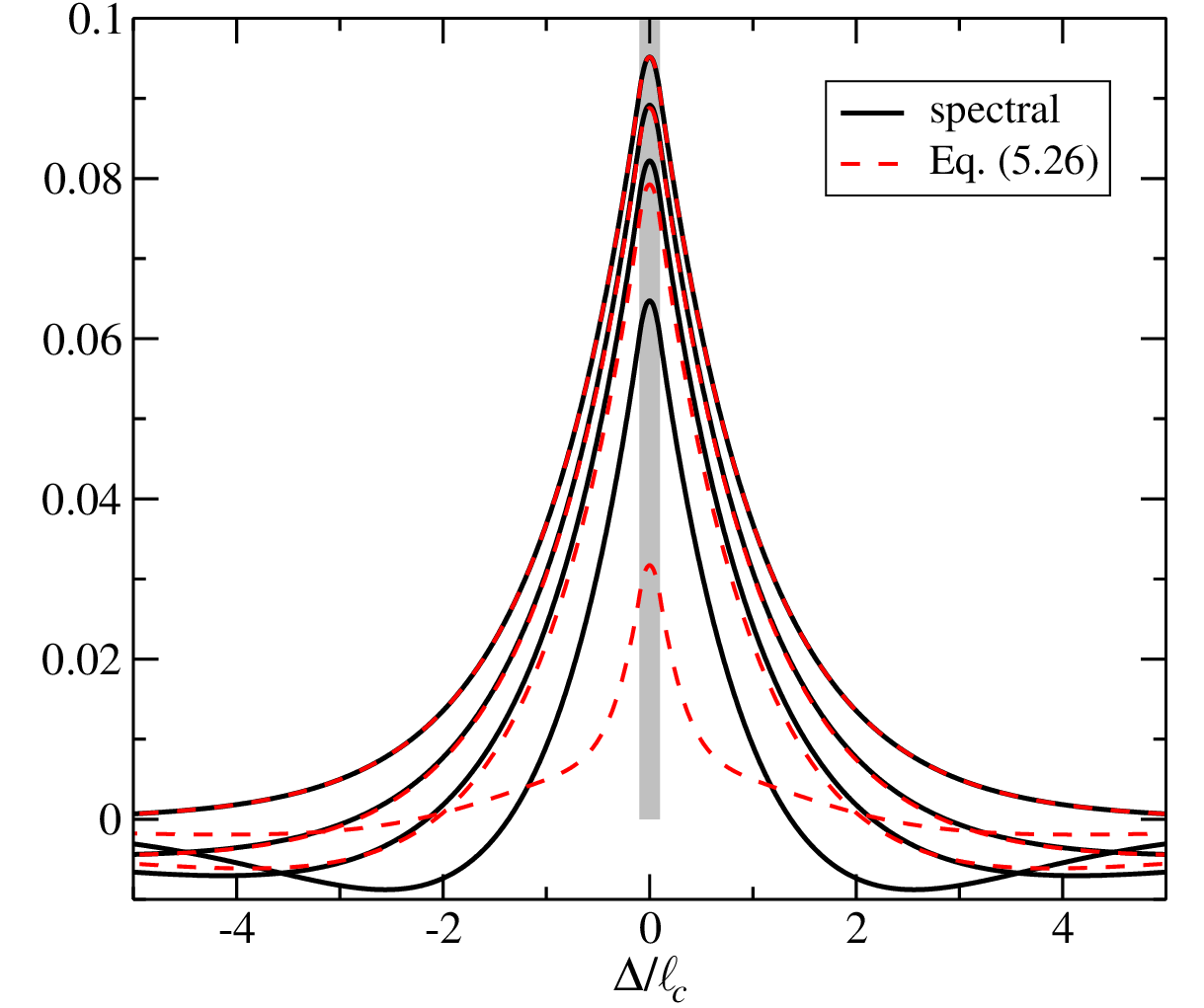}
    \includegraphics[width=0.45\linewidth]{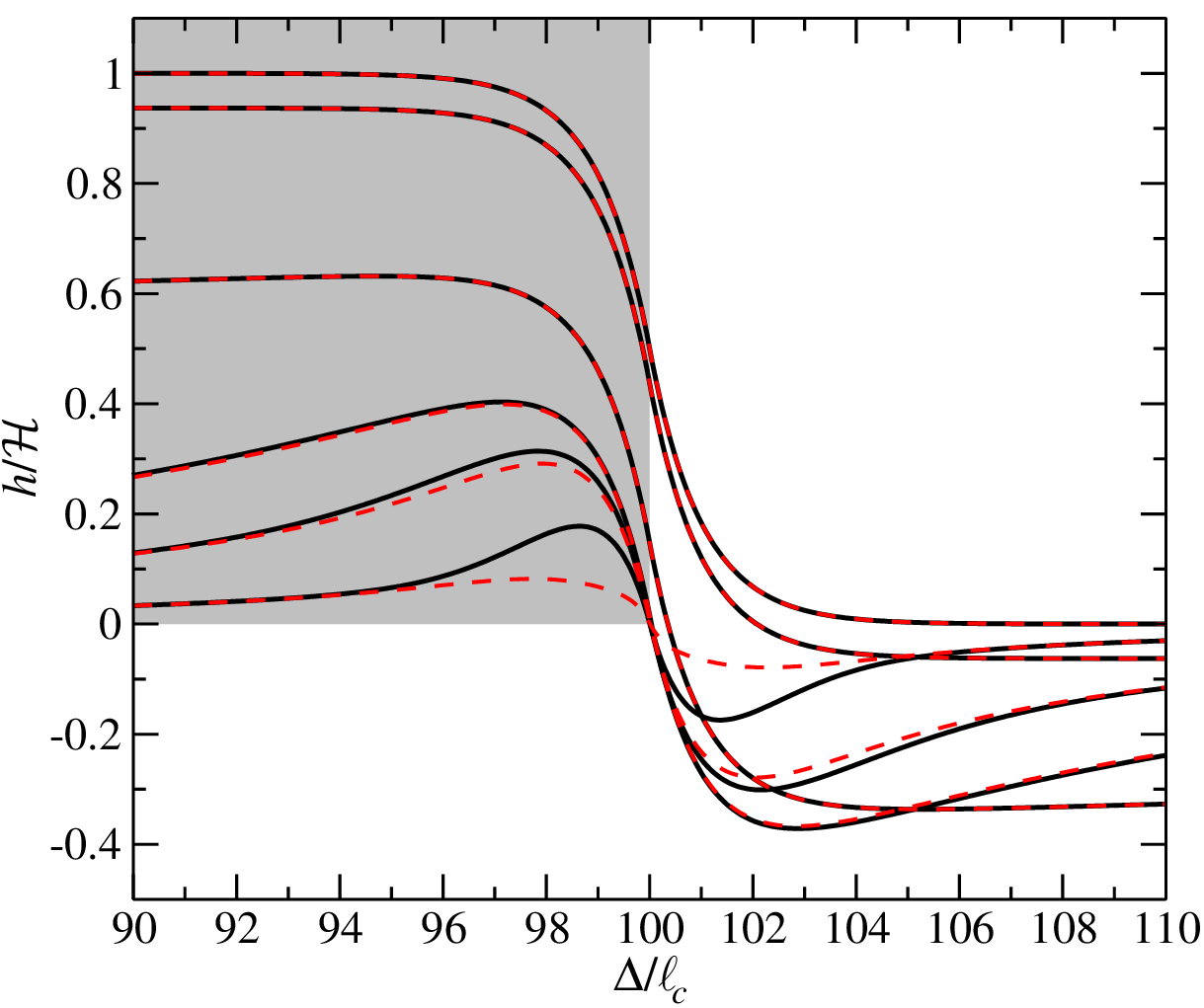}
    \hspace{0.45\linewidth}
    \caption{Interface shapes for Hele-Shaw cells with a single infinitely long mesa defect obtained with the spectral approach (black solid lines), compared to the theoretical result of Eq.~(\ref{eq:hII_IV}) (red dashed lines). For the wide defect ($w/\ell_c=200$; left panels), the times are $\hat{t}=1,4,10,100,1000,\infty$. For the narrow defect ($w/\ell_c=0.2$; right panel), the times are $\hat{t}=1,4,10,\infty$. In both cases, the times increase from bottom to top at $\Delta=0$. Equation~(\ref{eq:hII_IV}) is accurate for $\hat{t}\gg 1$ (stages II and IV); in addition, for the wide defect it is accurate at arbitrarily low times, except in the vicinity of the defect edge.
    }
 \label{fig:profiles_vs_theory}
\end{figure}

Considering now in more detail \textbf{stage II} (only observed for wide defects) and looking specifically at the vicinity of the right edge ($|\Delta-w/2|\ll w$ but not exclusively below $\ell_c$), the second and third terms in Eq.~(\ref{eq:hII_IV}) produce a maximum-minimum pair that is, to a very good approximation (with only negligible exponential corrections) antisymmetric with respect to the defect edge, while the first term  is approximately constant in space and increases slowly in time gradually shifting this perturbation upwards. Expressing this approximation mathematically, we have
\begin{equation}
    h(\Delta,t)\approx
    \begin{cases}
        \frac{\mathcal{H}}{\pi}\arctan\frac{\beta t}{w/2-\Delta}-\frac{\mathcal{H}}{2}\exp\left(-\frac{w/2-\Delta}{\ell_c}\right)+\frac{\beta \mathcal{H}}{\pi w}t, & \Delta-\frac{w}{2}<0,\\
        \frac{\mathcal{H}}{\pi}\arctan\frac{\beta t}{w/2-\Delta}+\frac{\mathcal{H}}{2}\exp\left(-\frac{\Delta-w/2}{\ell_c}\right)+\frac{\beta \mathcal{H}}{\pi w}t, & \Delta-\frac{w}{2}>0.
    \end{cases}
    \label{eq:hII}
\end{equation}
For $|\Delta-w/2|\ll\beta t$,
\begin{equation}
    h(\Delta,t)\approx
    \begin{cases}
        \frac{\mathcal{H}}{2}-\frac{\mathcal{H}}{\pi\beta t}(\frac{w}{2}-\Delta)-\frac{\mathcal{H}}{2}\exp\left(-\frac{w/2-\Delta}{\ell_c}\right)+\frac{\beta \mathcal{H}}{\pi w}t, & \Delta -w/2<0,\\
        -\frac{\mathcal{H}}{2}+\frac{\mathcal{H}}{\pi\beta t}(\Delta-\frac{w}{2})+\frac{\mathcal{H}}{2}\exp\left(-\frac{\Delta-w/2}{\ell_c}\right)+\frac{\beta \mathcal{H}}{\pi w}t, & \Delta -w/2>0,
    \end{cases}
    \label{eq:hIIpeaks}
\end{equation}
which gives Eq.~(\ref{eq:maxII}) for the positions of the extrema and
Eqs.~(\ref{eq:hamplII})--(\ref{eq:hsumII}) for their heights. The tails of the profile extend to distances of order $\beta t$ from the edge and so Eq.~(\ref{eq:hII}), rather than (\ref{eq:hIIpeaks}), must be used to describe them.

Finally, consider \textbf{stage IV} (common for wide and narrow defects). If we are interested in the evolution of the interface in the vicinity of the defect [$|\Delta|\lesssim {\rm max}(w,\ell_c)$], then in stage IV the arguments of the arctangents in Eq.~(\ref{eq:hII_IV}) are large by absolute value and since for $s\to+\infty$ $\arctan s\approx \pi/2-1/s$, we get Eq.~(\ref{eq:hIV}).
On the other hand,
for $|\Delta|\gg{\rm max}(w,\ell_c)$, Eq.~(\ref{eq:hII_IV}) becomes
\begin{equation}
    h(\Delta,t)\approx\frac{\mathcal{H}}{\pi}\left(\arctan\frac{\beta t}{w/2+\Delta}+\arctan\frac{\beta t}{w/2-\Delta}\right)=\frac{\mathcal{H}}{\pi}\arctan\frac{w\beta t}{(w/2)^2-\Delta^2-\beta^2 t^2},
\end{equation}
where an identity for the sum of arctangents was used. Since in stage IV the argument of the arctangent is small, Eq.~(\ref{eq:tail}) is valid.

\subsection{Finite inlet-interface distance}
\label{sec:mesa_details_finite_inlet_distance}

In the above derivations, we have assumed that the mean distance between the interface and the inlet (or the bottom boundary of the cell) can be assumed infinite and is, in particular, large even compared to the width of the cell. This removes the distinction between the pressure-controlled and flow-rate-controlled processes (if in the latter $V=0$) and means that the liquid volume (or, in our approximation, the area under the interface) is preserved. 
If this assumption about the inlet-interface distance cannot be made, then the distinction between the two processes remains and in either case, there is yet another length scale in the problem, which introduces new stages depending on the relation between the new length and other length scales. Still assuming $V=0$ in the flow-rate-controlled case in what follows, if the mean inlet-interface distance $h^{(0)}$ is much smaller than any of the other length scales, then, aside from the shortest time scales, the system behaves as our real-space PDE model discussed in the SI predicts [see Sec.~S2 
in SI; Eq.~(S23) in the flow-rate-controlled case, and there is an additional term on the right-hand side in the pressure-controlled case, as in Eq.~(S6)]. (It is still assumed that the perturbation of the interface is much smaller than the very small $h^{(0)}$.) In essence, the nonlocal hydrodynamic interactions between different parts of the interface are ``screened'' by the lower boundary of the cell. Of all other possible combinations of the lengths, perhaps the most practically interesting one is when $h^{(0)}$ is much larger than both $w$ and $\ell_c$, but may be smaller or larger than the cell width $W$. In this case, Regimes I to III are unaffected. In the flow-rate-controlled case, if $h^{(0)}=h_0\gg W$, there is still, essentially, no effect, as $\tanh(kh_0)\approx 1$ in Eq.~(\ref{eq:omV}) even for the mode with the lowest $k=k_2$. However, if $w,\ell_c\ll h^{(0)}\ll W$, then for the lowest $k$, $\tanh(kh_0)\approx kh_0$ and the dispersion relation of Eq.~(\ref{eq:omV}) becomes quadratic, as for the diffusion equation. Stage IV is split into two, with diffusive spreading of the tail at later times, before crossover to stage V. In the pressure-controlled case, for $k\ll [h^{(0)}]^{-1}\sim h_0^{-1}$ $\omega_p\approx{\rm const}$ [see Eq.~(\ref{eq:omp})]. Thus, for $h^{(0)}\ll W$ stage IV terminates earlier, at $t\sim h^{(0)}/\beta$ instead of $t\sim W/\beta$. At this time, the remaining perturbation starts to decay exponentially, but importantly, this includes the $k=0$ mode that does not conserve the volume. Until this time, volume conservation can be assumed. If $h^{(0)}\gg W$, then, in effect, stage V is split into two. Two modes need to be considered, the area-preserving $k=k_2$ mode with decay time of $W/\beta$ and the area-changing $k=0$ mode with the much longer decay time $h^{(0)}/\beta$. For $W/\beta\ll t \ll h^{(0)}/\beta$, the shape relaxes to equilibrium exponentially, while the area change can be neglected (relative to its total change from the initial state to the equilibrium). After this, at times of order $ h^{(0)}/\beta$, the shape no longer changes, but the interface shifts towards the equilibrium position.

\section{Procedure for generating a rough (disordered) surface (Sec.\ \ref{sec:disorder})}
\label{sec:generating_rough_FR}

We discretise a square domain of size $0<x<W$, $y_0<y<y_0+W$ using an $N\times N$ square grid. We generate a random aperture profile $b(x,y)$, calculating it for all nodes of the grid as
\begin{equation}
b(x_i,y_j)=b_0+\frac{4\sqrt{\pi} L B}{W}\sum_{m,n=0}^{N/2}\frac{\cos(k_{2m}x_i+k_{2n}y_j+\phi_{mn})}{[1+L^2(k_{2m}^2+k_{2n}^2)]^{3/4}},
\end{equation}
where $ k_n={\pi n}/{W}$ [Eq.~(\ref{eq:kn}) in the main text], and $\phi_{mn}$ are random phases uniformly distributed between 0 and $2\pi$. Noting that the average over the random phases is
\begin{equation}
\langle\cos[k_{2m}x+k_{2n}y+\phi_{mn}]\cos[k_{2r}(x+\Delta x)+k_{2s}(y+\Delta y)+\phi_{rs}]\rangle=\frac{1}{2}\cos(k_{2m}\Delta x+k_{2n}\Delta y)\delta_{mr}\delta_{ns},
\end{equation}
where $\delta_{mr}$ and $\delta_{ns}$ are Kronecker's deltas, we get for the correlation function
\begin{eqnarray}
    \langle [b(x,y)-b_0][b(x+\Delta x,y+\Delta y)-b_0]\rangle&=&\frac{8\pi L^2 B^2}{W^2}\sum_{m,n=0}^{N/2}\frac{\cos(k_{2m}\Delta x+k_{2n}\Delta y)}{[1+L^2(k_{2m}^2+k_{2n}^2)]^{3/2}}\nonumber\\
    &=&\frac{2L^2 B^2}{\pi}\int_0^{\infty}dk'\int_0^{\infty}dk''\frac{\cos(k'\Delta x+k''\Delta y)}{[1+L^2(k'^2+k''^2)]^{3/2}}\nonumber\\
    &=&B^2\exp\left[-\frac{(\Delta x^2+\Delta y^2)^{1/2}}{L}\right],
\end{eqnarray}
which is Eq.~(\ref{eq:apercorr}). Thus, correlations decay exponentially with the distance, with correlation length $L$. The cusp in the correlation function at $\Delta x=\Delta y=0$ ensures the roughness of the aperture field, such as that of a real fracture. At any particular point, the deviation from the mean, $b(x,y)-b_0$, being a sum of many uncorrelated random quantities, is normally distributed, with variance $B^2$.


\end{appen}

\bibliographystyle{jfm}
\bibliography{BIB_NEQ_dyn.bib}

\end{document}